\documentclass[rmp,aps,twocolumn,nofootinbib,floatfix]{revtex4}
\usepackage{graphicx,amsfonts,verbatim}

\def\be{\begin{equation}}
\def\ee{\end{equation}}

\def\bea{\begin{eqnarray}}
\def\eea{\end{eqnarray}}

\def\bfig{\begin{figure}[htbp]}
\def\efig{\end{figure}}

\def\btab{\begin{table}[htbp]}
\def\btab{\end{table}}

\def\bcen{\begin{center}}
\def\ecen{\end{center}}

\def\lb{\left[}
\def\rb{\right]}

\def\lbL{\lb\rule{0pt}{2.4ex}}
\def\rbL{\rule{0pt}{2.4ex}\rb}
\def\<{\langle}
\def\>{\rangle}
\def\mymatrix#1{\matrix{#1}}

\newcommand{\ket}[1]{\mbox{$|#1\rangle$}}
\newcommand{\bra}[1]{\mbox{$\langle#1|$}}
\newcommand{\mattwoc}[4]{\left[
        \begin{array}{cc}{#1}&{#2}\\{#3}&{#4}\end{array}\right]}

\begin{document}

\title{NMR Techniques for Quantum Control and Computation}
\author{Lieven M.K. Vandersypen}
\email{lieven@qt.tn.tudelft.nl}
\affiliation{Kavli Institute of NanoScience, Delft University of Technology, Lorentzweg 1, 2628 CJ Delft, The Netherlands}
\author{Isaac L. Chuang}
\email{ichuang@mit.edu}
\affiliation{Center for Bits and Atoms \& Department of Physics,
	Massachusetts Institute of Technology,
	 Cambridge, MA 02139, USA }

\begin{abstract}
Fifty years of developments in nuclear magnetic resonance (NMR) have
resulted in an unrivaled degree of control of the dynamics of coupled
two-level quantum systems.  This coherent control of nuclear spin
dynamics has recently been taken to a new level, motivated by the
interest in quantum information processing.  NMR has been the
workhorse for the experimental implementation of quantum protocols,
allowing exquisite control of systems up to seven qubits in size.
Here, we survey and summarize a broad variety of pulse control and 
tomographic techniques which have been developed for and used in NMR
quantum computation.  Many of these will be useful in other quantum
systems now being considered for implementation of quantum information
processing tasks.
\end{abstract}

\maketitle

\tableofcontents

\section{Introduction}
Precise and complete control of multiple coupled quantum systems is
expected to lead to profound insights in physics as well as to novel
applications, such as quantum
computation~\cite{Bennett00a,Nielsen00b,Galindo02a}. 
Such coherent control is a major goal in atomic
physics~\cite{Leibfried03a,Wieman99a,Science02a}, 
quantum optics~\cite{Zeilinger99a,Science02a} and
condensed matter research~\cite{Makhlin01a,Clark01a,Science02a,Zutic04a}, but 
surprisingly, many of
the leading experimental results are coming from one of the oldest
areas of quantum physics: nuclear magnetic resonance (NMR).

The development of NMR control techniques originated in a strong demand
for precise spectroscopy of complex molecules: NMR is the premier tool
for protein structure determination, and in modern NMR spectroscopy,
often thousands of precisely sequenced and phase controlled pulses are
applied to molecules containing hundreds of nuclear spins.  More
recently, over the past seven years, a wide variety of complex quantum
information processing tasks have been realized using NMR, on systems
ranging from two to seven quantum bits (qubits) in size, on molecules
in
liquid~\cite{Chuang98c,Jones98b,Nielsen98b,Somaroo99a,Knill00a,Vandersypen01a},
liquid crystal~\cite{Yannoni99a}, and solid state
samples~\cite{Zhang98a,Leskowitz03a}.  These demonstrations have been made
possible by application of a menagerie of new and previously existing
control techniques, such as simultaneous and shaped pulses, composite
pulses, refocusing schemes, and effective Hamiltonians.  These allow
control and compensation for a variety of imperfections and
experimental artifacts invariably present in real physical systems,
such as pulse imperfections, Bloch-Siegert shifts, undesired
multiple-spin couplings, field inhomogeneities, and imprecise system
Hamiltonians.

The problem of control of multiple coupled quantum systems is a
signature topic for NMR, and can be summarized as follows: given a
system with Hamiltonian ${\cal H} = {\cal H}_{\rm sys} + {\cal H}_{\rm
control}$, where ${\cal H}_{\rm sys}$ is the Hamiltonian in the
absence of any active control, and ${\cal H}_{\rm control}$ describes
terms which are under external control, how can a desired unitary
transformation $U$ be implemented, in the presence of imperfections,
and using minimal resources?  Similar to other scenarios in which
quantum control is a well-developed idea, such as in laser excitation
of chemical reactions~\cite{Walmsley03a}, ${\cal H}_{\rm control}$
arises from precisely timed sequences of multiple pulses of
electromagnetic radiation, applied phase-coherently, with different
pulse widths, frequencies, phases, and amplitudes.  However,
importantly, in contrast to other areas of quantum control, in NMR
${\cal H}_{\rm sys}$ is composed from multiple distinct physical
pieces, i.e. the individual nuclear spins, providing the tensor
product Hilbert space structure vital to quantum computation.
Furthermore, the NMR systems employed in quantum computation are more well
approximated as being closed, as opposed to open, quantum systems.

Nuclear spins and NMR provide a wonderful model and inspiration for
the advance of coherent control over other coupled quantum systems, as
many of the challenges and solutions are similar across the world of
atomic, molecular, optical, and solid-state systems (see
e.g.~\cite{Steffen03a}).  Here, we review the control techniques
employed in the field of NMR quantum computation, focusing on methods
which are robust under experimental implementation, and including
experimental prescriptions for evaluation of the efficacy of the
techniques.  In contrast to other
reviews~\cite{Cory00a,Jones00a,Vandersypen01c} of and
introductions~\cite{Jones01a,Vandersypen01b,Gershenfeld98a,Steffen01a}
to NMR quantum computation which have appeared in the literature, we
do not assume prior knowledge of, or give specialized descriptions of
quantum computation algorithms, nor do we review NMR quantum computing
experiments.  And although we do not assume prior detailed knowledge
of NMR, a self-contained treatment of several advanced topics, such as
composite pulses, and refocusing, is included.  Finally, as a primary
purpose of this article is to elucidate control techniques which may
generalize beyond NMR, we also assume a regime of operation in which
relaxation and decoherence mechanisms are simple to treat and physical
evolution is dominated by closed systems dynamics.

The organization of this article is as follows. In
section~\ref{sec:hamiltonian}, we briefly review the physics of NMR,
using a Hamiltonian description of single and interacting nuclear
spins-1/2 placed in a static magnetic field, controlled by
radio-frequency fields.  This establishes a foundation for the first
major part of this review, section~\ref{sec:elem_pulse}, which 
discusses the knobs the control Hamiltonian provides to construct 
all the elementary quantum gates, and the limitations that arise from 
the given system and control Hamiltonian, as well as from
instrumental imperfections. The second major part of this review,
section~\ref{sec:adv_pulse}, presents three classes of advanced
techniques for tailoring the control Hamiltonian, which permit accurate
quantum control despite the existing limitations: the methods of 
amplitude and frequency shaped pulses, composite pulses and average 
Hamiltonian theory.  Finally, in section~\ref{sec:evaluation}, we 
conclude by describing a set of standard experiments, derived from 
quantum computation, which demonstrate coherent qubit-control and can 
be used to characterize decoherence.  These include procedures for 
quantum state and process tomography, as well as methods to evaluate 
the fidelity of quantum states and gates.

For further reading on NMR, we recommend the textbooks by
Abragam~\cite{Abragam61a}, Ernst, Bodenhausen and
Wokaun~\cite{Ernst87a} and Slichter~\cite{Slichter96a} for rigorous
discussions of the nuclear spin Hamiltonian and standard pulse
sequences, Freeman~\cite{Freeman97a} for an intuitive explanation of
advanced techniques for control of the spin evolution, and
Levitt~\cite{Levitt01a} for an intuitive understanding of the physics
underlying the spin dynamics. Many useful reviews on specific NMR
techniques are compiled in the Encyclopedia of
NMR~\cite{Encyclopedia_NMR}.

For additional reading on quantum computation, we recommend the book
by Nielsen and Chuang~\cite{Nielsen00b} for the basic theory of
quantum information and computation and
Refs.~\cite{Braunstein00a,Bennett00a} for a review of the state of the
art in experimental quantum information
processing. Ref.~\cite{Lloyd95b} gives a simple introduction to
quantum computation. Excellent presentations of quantum algorithms are
given in Refs.~\cite{Steane98a,Ekert96a}.

The original papers introducing NMR quantum computing are
Refs~\cite{Cory96a,Gershenfeld97a,Cory97a,Cory97b}.
Refs.~\cite{Gershenfeld98a,Steffen01a} give elementary introductions
to NMR quantum computing, and introductions geared towards NMR
spectroscopists are presented in
Refs~\cite{Jones01a,Vandersypen01b}. Summaries of NMR quantum
computing experiments and techniques are given in
Refs.~\cite{Cory00a,Jones00a,Vandersypen01c}.

\section{The NMR system}
\label{sec:hamiltonian}

We begin with a description of the NMR system, based on its system
Hamiltonian and the control Hamiltonian.  The system Hamiltonian gives
the energy of single and coupled spins in a static magnetic field, and
the control Hamiltonian arises from the application of radio-frequency 
pulses to the system at, or near, its resonant frequencies.  A
rotating reference frame is employed, providing a very convenient 
description.

%%%%%%%%%%%%%%%%%%%%%%%%%%%%%%%%%%%%%%%%%%%%%%%%%%%%%%%%%%%%%%%%%%%%%%%%%%%%%
\subsection{The system Hamiltonian}

%%%%%%%%%%%%%%%%%%%%%%%%%%%%%%%%%%%%%%%%
\subsubsection{Single spins}

\label{sec:singlespins}

The time evolution of a spin-1/2 particle (we will not consider higher
order spins in this paper) in a magnetic field $\vec{B}_0$ along
$\hat{z}$ is governed by the Hamiltonian
\be
	{\cal H}_0 
	= -\hbar \gamma B_0 \, I_z 
	= - \hbar \, \omega_0 \; I_z 
	= \left[\matrix{-\hbar \omega_0 /2 & 0 \cr
			 0 & \hbar \omega_0 /2}\right]
\,,
\label{eq:1spin_ham}
\ee
where $\gamma$ is the gyromagnetic ratio of the nucleus, $\omega_0 /
2\pi$ is the Larmor frequency~\footnote{We will sometimes leave the
factor of $2\pi$ implicit and call $\omega_0$ the Larmor frequency.}
and $I_z$ is the angular momentum operator in the $\hat{z}$
direction. $I_z$, $I_x$, and $I_y$ relate to the well-known Pauli
matrices as
\be
	\sigma_x = 2 I_x, 
\quad 	\sigma_y = 2 I_y, 
\quad 	\sigma_z = 2 I_z 
\,,
\ee
where, in matrix notation,
\bea
        \sigma_x  \equiv \mattwoc{0}{1}{1}{0}; \,\,\,\,
        \sigma_y  \equiv \mattwoc{0}{-i}{i}{0}; \,\,\,\,
        \sigma_z  \equiv \mattwoc{1}{0}{0}{-1}
\,.
\label{eq:Pauli_def}
\eea

The interpretation of Eq.~\ref{eq:1spin_ham} is that the $\ket{0}$ or
$\ket{\!\!\uparrow}$ energy (given by $\bra{0}{\cal H}\ket{0}$, the
upper left element of ${\cal H}$) is lower than the $\ket{1}$ or
$\ket{\!\!\downarrow}$ energy ($\bra{1}{\cal H}\ket{1}$) by an amount
$\hbar \omega_0$, as illustrated in the energy diagram of
Fig.~\ref{fig:energy_1spin}.  The energy splitting is known as the
Zeeman splitting.

\bfig
\bcen
\vspace*{1ex}
\includegraphics[width=2cm]{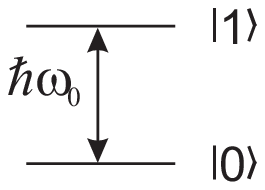} 
\vspace*{-2ex}
\ecen
\caption{Energy diagram for a single spin-1/2.}
\label{fig:energy_1spin}
\efig

We can pictorially understand the time evolution 
$U = e^{-i {\cal H} t/ \hbar}$ under the Hamiltonian of
Eq.~\ref{eq:1spin_ham} as a precessing motion of the Bloch vector
about $\vec{B_0}$, as shown in Fig.~\ref{fig:precession}.  
As is conventional, we define the $\hat{z}$ axis of the Bloch sphere
as the quantization axis of the Hamiltonian, with $\ket{0}$ along
$+\hat{z}$ and $\ket{1}$ along $-\hat{z}$. 

For the
case of liquid-state NMR, which we will largely restrict ourselves to
in this article, typical values of $B_0$ are $5-15$ Tesla, resulting
in precession frequencies $\omega_0$ of a few hundred MHz, the
radio-frequency range.

\bfig
\bcen
\vspace*{1ex}
\includegraphics[height=3.5cm]{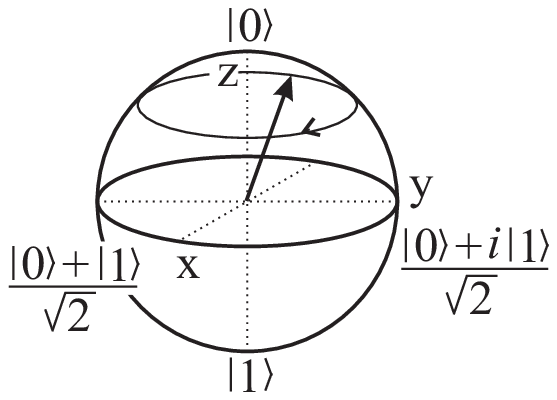} 
\vspace*{-2ex}
\ecen
\caption{Precession
of a spin-1/2 about the axis of a static magnetic field.}
\label{fig:precession}
\end{figure}

Spins of different nuclear species ({\em heteronuclear} spins) can be
easily distinguished spectrally, as they have very distinct values of
$\gamma$ and thus also very different Larmor frequencies
(Table~\ref{tab:larmor_freq}). Spins of the same nuclear
species ({\em homonuclear} spins) which are part of the same molecule
can also have distinct frequencies, by amounts known as their {\em
chemical shifts}, $\tilde{\sigma}_i$.  

The nuclear spin Hamiltonian for a molecule with $n$ uncoupled nuclei with
is thus given by
\be
	{\cal H}_0 
	= - \sum_{i=1}^n \hbar \, 
			(1 - \tilde{\sigma}_i) \gamma_i B_0 \; I_z^i
	= - \sum_{i=1}^n \hbar \, \omega_0^i \;I_z^i 
\,,
\label{eq:ham_0_n}
\ee
where the $i$ superscripts label the nuclei.  

\begin{table}[htbp]
\vspace*{1ex}
\begin{center}
\begin{tabular}{c|ccccccc}
nucleus    & $^1$H & $^2$H & $^{13}$C & $^{15}$N & $^{19}$F & $^{31}$P\\\hline
$\omega_0/2\pi$ & 500   & 77    & 126      & -51      & 470      & 202 
\end{tabular}
\end{center}
\vspace*{-1ex}
\caption{Larmor frequencies [MHz] for some relevant nuclei, at
11.74 Tesla.}
\label{tab:larmor_freq}
\end{table}

\bfig
\bcen
\vspace*{1ex}
\hspace*{-1cm}
\includegraphics[width=5.7cm]{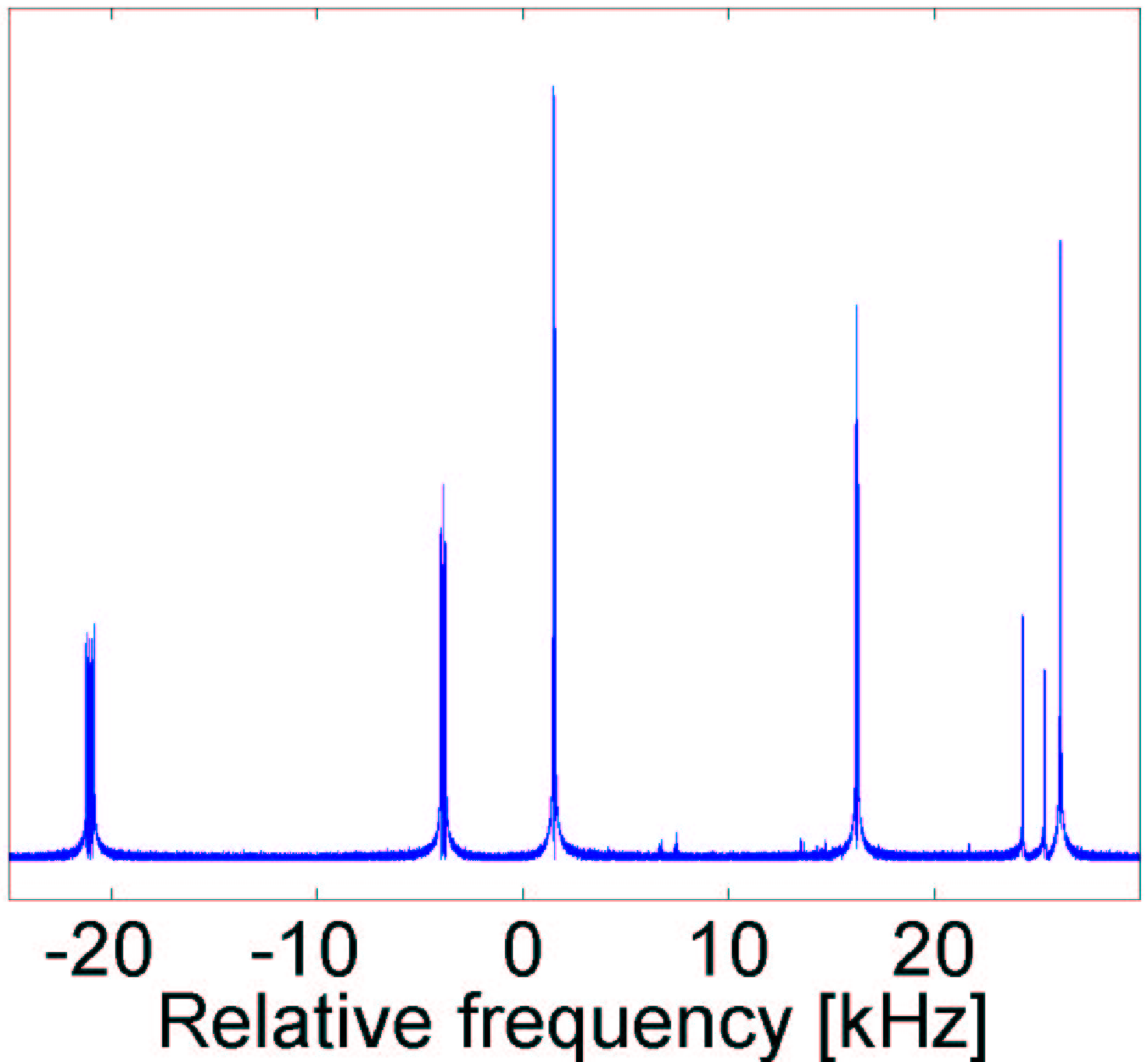}
\raisebox{1cm}{\includegraphics[width=3cm]{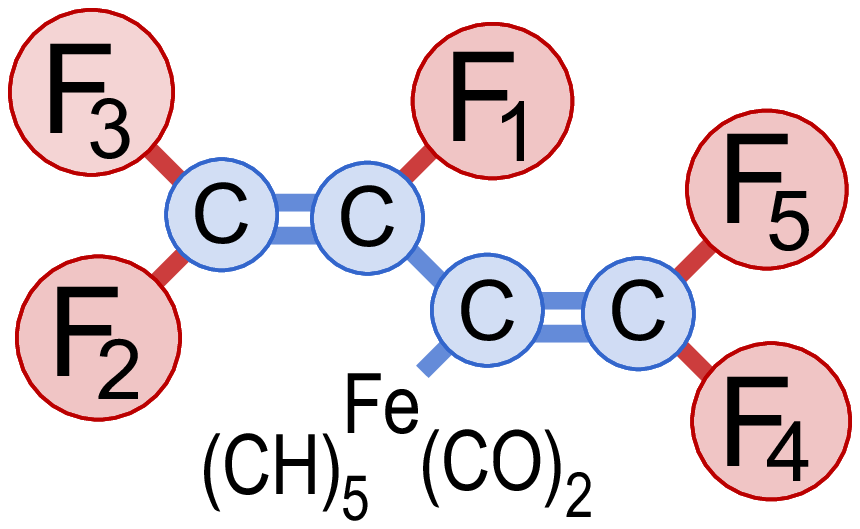}}
\vspace*{-2ex}
\ecen
\caption{Fluorine NMR spectrum (absolute value) centered around 
$\approx 470$ MHz of a specially designed molecule (shown on the 
right). The five main lines in the spectrum
correspond to the five fluorine nuclei in the molecule.  The two small
lines derive from impurities in the sample. The NMR spectra were
acquired by recording the oscillating magnetic field produced by a
large ensemble of precessing spins, and by taking the Fourier
transform of this time-domain signal. The precession motion of
the spins is started by applying a radio-frequency pulse
(section~\protect\ref{sec:rf}) which tips the spins from their
equilibrium position along the $\hat{z}$ axis into the
$\hat{x}-\hat{y}$ plane.}
\label{fig:shor_wide_F}
\efig

The chemical shifts arise from partial shielding of the externally 
applied magnetic field by the electron cloud surrounding the nuclei.
The amount of shielding depends on the electronic environment of each
nucleus, so like nuclei with {\em inequivalent} electronic environments 
have different chemical shifts. Pronounced asymmetries in the molecular
structure generally promote strong chemical shifts.  The range of 
typical chemical shifts $\tilde{\sigma}_i$ varies from nucleus to 
nucleus, e.g. $\approx 10$ parts per million (ppm) for $^1$H, $\approx 
200$ ppm for $^{19}$F and $\approx 200$ ppm for $^{13}$C.  At $B_0 = 10$ 
Tesla, this corresponds to a few kHz to tens of kHz (compared to 
$\omega_0$'s of several hundred MHz).  As an example,
Fig.~\ref{fig:shor_wide_F} shows an experimentally measured spectrum
of a molecule containing five fluorine spins with inequivalent
chemical environments.

In general, the chemical shift can be spatially anisotropic and must
be described by a tensor. In liquid solution, this anisotropy averages
out due to rapid tumbling of the molecules. In solids, the anisotropy
means that the chemical shifts depend on the orientation of the
molecule with respect to $\vec{B}_0$.

%%%%%%%%%%%%%%%%%%%%%%%%%%%%%%%%%%%%%%%%
\subsubsection{Interacting spins}
\label{sec:coupling}
\label{sec:J_coupling}

For nuclear spins in molecules, nature provides two distinct
interaction mechanisms which we now describe, the direct
dipole-dipole interaction, and the electron mediated Fermi contract
interaction known as $J$-coupling.

%%%%%%%%%%%%%%%%%%%%
{\bf Direct coupling}.  The {\em magnetic dipole-dipole} interaction
is similar to the interaction between two bar magnets in each other's
vicinity. It takes place purely {\em through space} --- no medium is
required for this interaction --- and depends on the internuclear 
vector $\vec{r}_{ij}$ connecting the two nuclei $i$ and $j$, as
described by the Hamiltonian
\be
	{\cal H}_D 
	= \sum_{i<j} \frac{\mu_0 \gamma_i \gamma_j \hbar}
		{4\pi |\vec{r}_{ij}|^3} 
	\lbL{ \vec{I}^i \cdot \vec{I}^j - \frac{3}{|\vec{r}_{ij}|^2} 
	(\vec{I}^i \cdot \vec{r}_{ij}) (\vec{I}^j \cdot \vec{r}_{ij}) }\rb
\,,
\label{eq:ham_dipole}
\ee
where $\mu_0$ is the usual magnetic permeability of free space
and $\vec{I}^i$ is the
magnetic moment vector of spin $i$.  This expression can be
progressively simplified as various conditions are met.  These
simplifications rest on averaging effects and can be explained within
the general framework of average-Hamiltonian theory
(section~\ref{sec:average_Ham}).

For large $\omega_0^i = \gamma^i B_0$ (i.e. at high $B_0$), ${\cal
H}_D$ can be approximated as
\be
	{\cal H}_D 
	= \sum_{i<j} \frac{\mu_0 \gamma_i \gamma_j \hbar}
			{8\pi |\vec{r}_{ij}|^3}
		(1-3 \cos^2{\theta_{ij}}) 
			[3 I_z^i I_z^j - \vec{I}^i \cdot \vec{I}^j] 
\,,
\label{eq:ham_dipole2}
\ee
where $\theta_{ij}$ is the angle between $B_0$ and $\vec{r}_{ij}$.
When $|\omega_0^i - \omega_0^j|$ is much larger than the coupling
strength, the
transverse coupling terms can be dropped, so ${\cal H}_D$ simplifies
further to
\be
	{\cal H}_D 
	= \sum_{i<j} \frac{\mu_0 \gamma_i \gamma_j \hbar}
			{4\pi |\vec{r}_{ij}|^3}
		(1-3 \cos^2{\theta_{ij}}) I_z^i I_z^j 
\,,
\label{eq:ham_dipole3}
\ee
which has the same form as the $J$-coupling we describe next 
(Eq.~\ref{eq:ham_J}). 

For molecules in liquid solution, both intramolecular dipolar
couplings (between spins in the same molecule) and intermolecular
dipolar couplings (between spins in different molecules) are averaged
away due to rapid tumbling.  This is the case we shall focus on in
this article.  In solids, similarly simple Hamiltonians can be
obtained by applying multiple-pulse sequences which average out
undesired coupling terms~\cite{Haeberlen68a}, or by physically spinning 
the sample at an angle of $\arccos(1/\sqrt{3})$ (the "magic angle") 
with respect to the magnetic field.

%%%%%%%%%%%%%%%%%%%%
{\bf Indirect coupling}.  The second interaction mechanism between
nuclear spins in a molecule is the {\em $J$-coupling} or {\em scalar
coupling}. This interaction is mediated by the electrons shared in the
chemical bonds between the atoms, and due to the overlap of the shared
electron wavefunction with the two coupled nuclei, a Fermi contact
interaction.  The {\em through-bond} coupling strength $J$ depends on
the respective nuclear species and decreases with the number of
chemical bonds separating the nuclei. Typical values for $J$ are up to
a few hundred Hz for one-bond couplings and down to only a few Hertz
for three- or four-bond couplings.  The Hamiltonian is
\be
	{\cal H}_J 
	= \hbar \sum_{i<j} 2 \pi J_{ij} \vec{I}^i \cdot \vec{I}^j
	= \hbar \sum_{i<j} 2 \pi J_{ij} 
		(I_x^i I_x^j + I_y^i I_y^j + I_z^i I_z^j) 
\,,
\label{eq:ham_bond}
\ee
where $J_{ij}$ is the coupling strength between spins $i$ and $j$.  
Similar to the case of dipolar coupling, Eq.~\ref{eq:ham_bond} 
simplifies to
\be
	{\cal H}_J = \hbar \sum_{i<j}^n 2 \pi J_{ij} I_z^i I_z^j 
\,,
\label{eq:ham_J}
\ee
when $|\omega_i - \omega_j| \gg 2\pi |J_{ij}|$, a condition easily
satisfied for heteronuclear spins and which can also be satisfied for
small homonuclear molecules.

The interpretation of the scalar coupling term of Eq.~\ref{eq:ham_J}
is that a spin ``feels'' a static magnetic field along $ \pm \hat{z}$
produced by the neighboring spins, in addition to the externally
applied $\vec{B}_0$ field. This additional field shifts the energy
levels as in Fig.~\ref{fig:energy_2spins}. As a result, the Larmor
frequency of spin $i$ shifts by $-J_{ij}/2$ if spin $j$ is in
$\ket{0}$ and by $+J_{ij}/2$ if spin $j$ is in $\ket{1}$.

\bfig
\bcen
\vspace*{1ex}
\includegraphics[width=6cm]{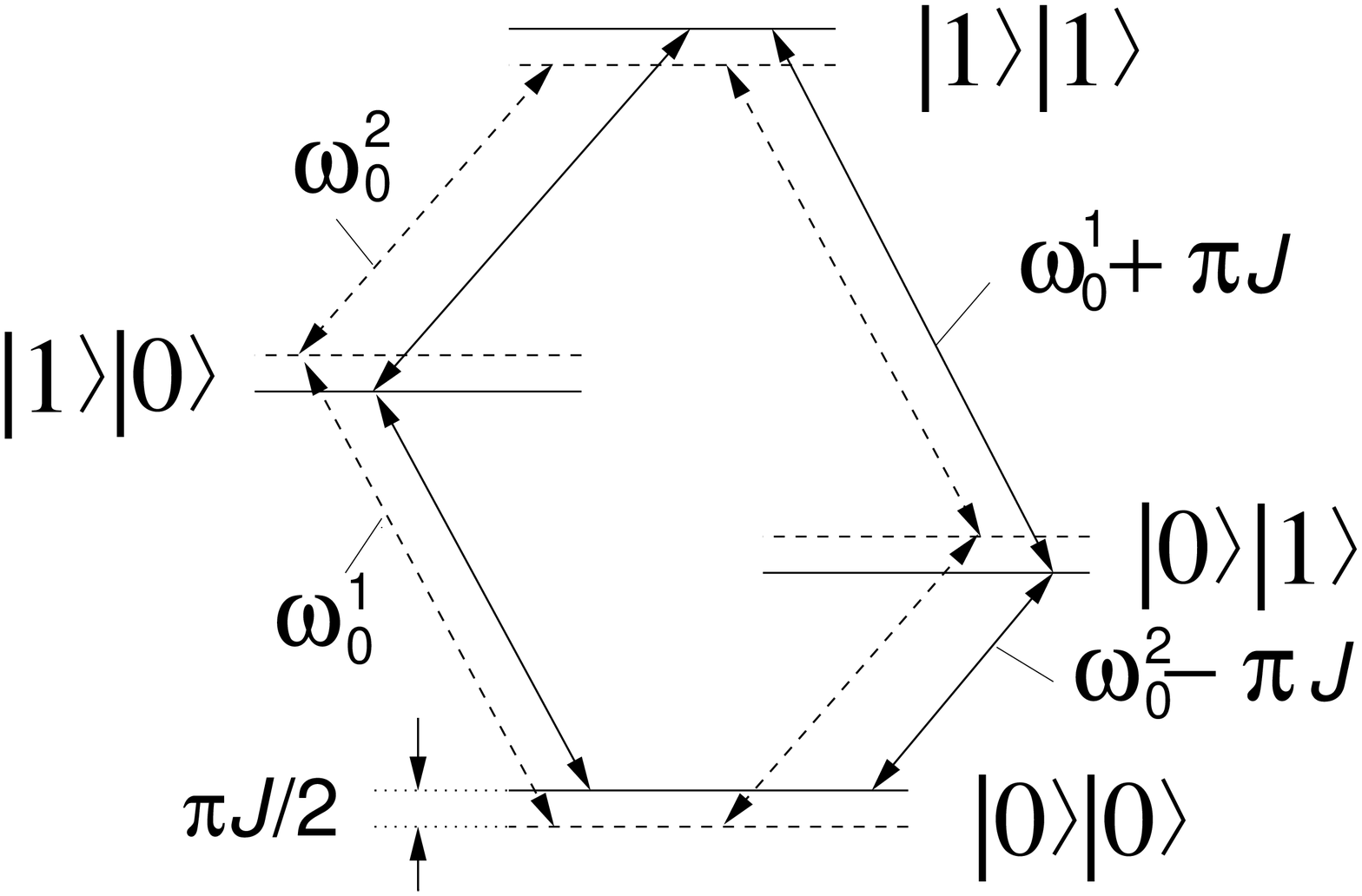} 
\vspace*{-2ex}
\ecen
\caption{Energy level diagram for (dashed lines) two uncoupled spins
and (solid lines) two spins coupled by a Hamiltonian
of the form of Eq.~\ref{eq:ham_dipole3} or Eq.~\ref{eq:ham_J} (in
units of $\hbar$).}
\label{fig:energy_2spins}
\efig

In a system of two coupled spins, the frequency spectrum of spin $i$
therefore actually consists of two lines separated by $J_{ij}$ and
centered around $\omega_0^i$,
each of which can be associated with the state of spin $j$, $\ket{0}$
or $\ket{1}$.  For three pairwise coupled spins, the spectrum of each
spin contains four lines. For every additional spin, the number of
lines per multiplet doubles, provided all the couplings are resolved
and different lines do not lie on top of each other. This is
illustrated for a five spin system in Fig.~\ref{fig:5spin_multiplet}.

\bfig
\bcen
\vspace*{1ex}
\includegraphics[width=6cm]{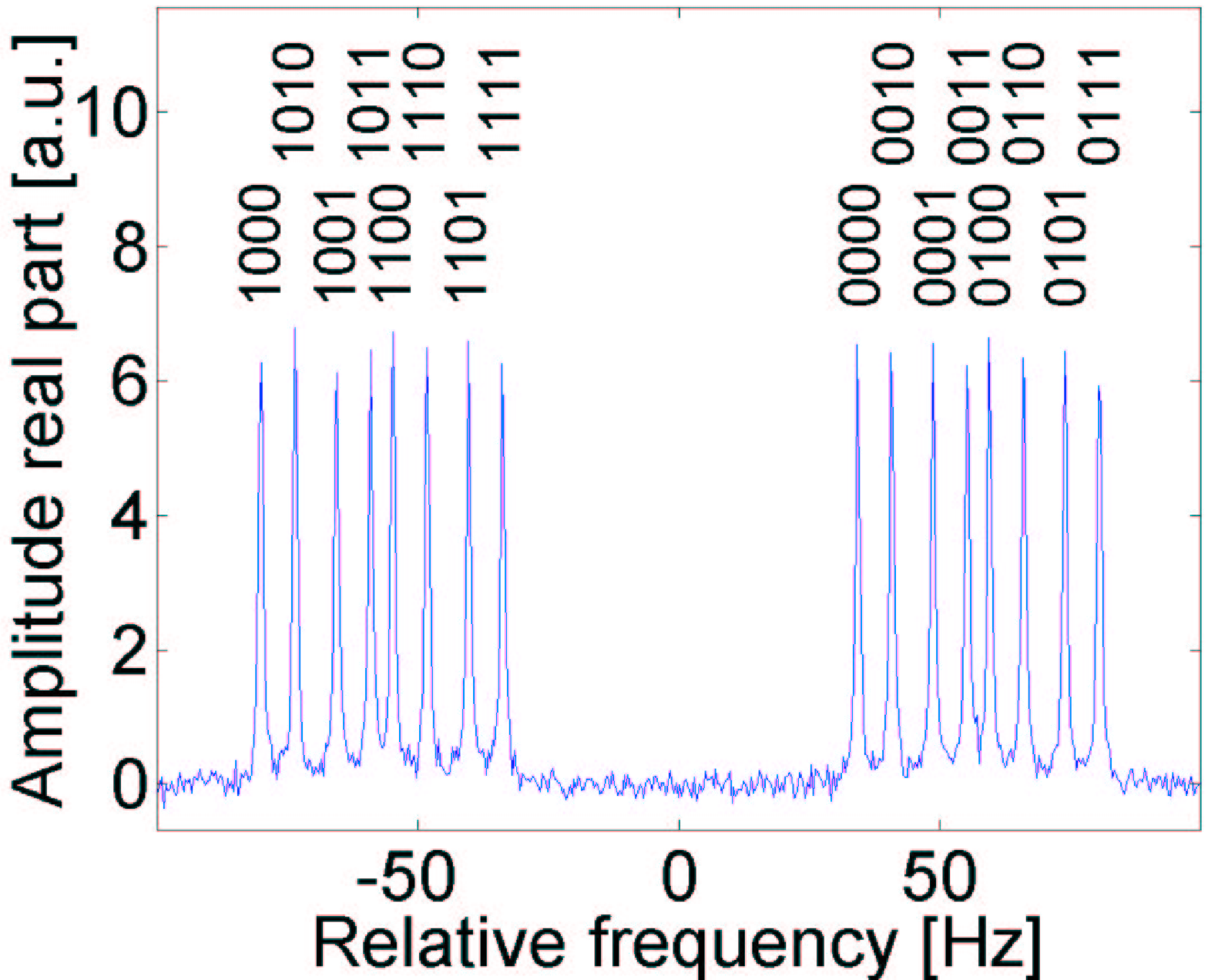} 
\vspace*{-2ex}
\ecen
\caption{The spectrum of spin $F_1$ in the
molecule of Fig.~\protect\ref{fig:shor_wide_F}. This is an expanded view of
the left line in the spectrum of Fig.~\protect\ref{fig:shor_wide_F}.
Frequencies are given with respect to $\omega_0^1$. The state of the
remaining spins is as indicated, based on $J_{12}<0$ and $J_{13},
J_{14}, J_{15} > 0$; furthermore, $|J_{12}| > |J_{13}| > |J_{15}| >
|J_{14}|$.}
\label{fig:5spin_multiplet}
\efig

The magnitude of all the pairwise couplings can be found by looking
for common splittings in the multiplets of different spins. The
relative signs of the $J$ couplings can be determined via appropriate
spin-selective two-pulse sequences, known in NMR as two-dimensional
correlation (soft-COSY) experiments~\cite{Bruschweiler87a} or via
line-selective continuous irradition; both approaches are related to
the {\sc cnot} gate (section~\ref{sec:2qubit_gates}). The signs cannot
be obtained from just the simple spectra.

In summary, the simplest form of the Hamiltonian for a system of $n$
coupled nuclear spins is thus (from Eqs.~\ref{eq:ham_0_n}
and~\ref{eq:ham_J})
\be
	{\cal H}_{\rm sys} 
	= - \sum_{i} \hbar \, \omega_0^i \;I_z^i  
		+ \hbar \sum_{i<j} 2 \pi J_{ij} I_z^i I_z^j 
\,.
\label{eq:ham_iso}
\ee
In almost all NMR quantum computing experiments performed to date, the
system is well described by a Hamiltonian of this form. 

%%%%%%%%%%%%%%%%%%%%%%%%%%%%%%%%%%%%%%%%%%%%%%%%%%%%%%%%%%%%%%%%%%%%%%%%%
\subsection{The control Hamiltonian}

%%%%%%%%%%%%%%%%%%%%%%%%%%%%%%%%%%%%%%%%
\subsubsection{Radio-frequency fields}
\label{sec:rf}

We turn now to physical mechanisms for controlling the NMR system.
The state of a spin-1/2 particle in a static magnetic field $\vec{B}_0$
along $\hat{z}$ can be manipulated by applying an
electromagnetic field $\vec{B}_1(t)$ which rotates in the 
$\hat{x}-\hat{y}$ plane at $\omega_{r\!f}$, at or near the spin
precession frequency $\omega_0$.  The single-spin Hamiltonian
corresponding to the radio-frequency (RF) field is, analogous to
Eq.~\ref{eq:1spin_ham} for the static field
$B_0$,
\be
	{\cal H}_{r\!f} = - \hbar \gamma B_1 
		\lbL{ \cos (\omega_{r\!f} t + \phi) I_x +
			\sin (\omega_{r\!f} t + \phi) I_y }\rb
\,,
\label{eq:ham_rf_lab}
\ee
where $\phi$ is the phase of the RF field, and $B_1$ its amplitude.
Typical values for $\omega_1 = \gamma B_1$ are up to $\approx 50$ kHz
in liquid NMR and up to a few hundred kHz in solid NMR experiments.
For $n$ spins, we have
\be
	{\cal H}_{r\!f} 
	= - \sum_i^n \hbar \gamma_i B_1 
		\lbL{ \cos (\omega_{r\!f} t + \phi) I_x^i +
			\sin (\omega_{r\!f} t + \phi) I_y^i }\rb
\,.
\label{eq:ham_rf_lab_Nspins}
\ee
In practice, a magnetic field is applied which oscillates along a 
fixed axis in the laboratory, perpendicular to the static magnetic 
field.  This oscillating
field can be decomposed into two counter-rotating fields, one of which
rotates at $\omega_{r\!f}$ in the same direction as the spin and so
can be set on or near resonance with the spin. The other component
rotates in the opposite direction and is thus very far off-resonance
(by about $2\omega_0$). As we shall see, its only effect is a 
negligible shift in the Larmor frequency, called the Bloch-Siegert
shift\label{page:counterrot_rf}~\cite{Bloch40a}.

Note that both the amplitude $B_1$ and phase $\phi$ of the RF field
can be varied with time\footnote{For example, the Varian Instruments
Unity Inova 500 NMR spectrometer achieves a phase resolution of
$0.5^\circ$ and has $4095$ linear steps of amplitude control, with a
time-base of $50$ ns. Additional attenuation of the amplitude can be 
done on a logarithmic scale over a range of about $80$ dB, albeit
with a slower timebase.}, unlike the Larmor precession and the
coupling terms.  As we will shortly see, it is the control of the RF
field phases, amplitudes, and frequencies, which lie at the heart of
quantum control of NMR systems.

%%%%%%%%%%%%%%%%%%%%%%%%%%%%%%%%%%%%%%%%
\subsubsection{The rotating frame}
\label{sec:rotating_frame}

The motion of a single nuclear spin subject to both a static and a
rotating magnetic field is rather complex when described in the usual
laboratory coordinate system (the {\em lab frame}). It is much
simplified, however, by describing the motion in a coordinate system
rotating about $\hat{z}$ at $\omega_{r\!f}$ (the {\em rotating
frame}):
\be
	\ket{\psi}^{rot} = \exp(-i \omega_{r\!f} t I_z) \ket{\psi}\,.
\label{eq:def_rot_frame}
\ee
Substitution of $\ket{\psi}$ in the Schr\"odinger equation $
i \hbar \frac{d\ket{\psi}}{dt} = {\cal H} \ket{\psi}$ with
\be
	{\cal H} 
	= - \hbar \, \omega_0 \; I_z     - \hbar \omega_1 
		\lbL{ \cos (\omega_{r\!f} t + \phi) I_x + 
			\sin (\omega_{r\!f} t + \phi) I_y }\rb
\,,
\ee
gives $i \hbar \frac{d \ket{\psi}^{rot}}{d t} = {\cal
H}^{rot} \ket{\psi}^{rot}$, where
\be
	{\cal H}^{rot} 
	= - \hbar \, (\omega_0 - \omega_{r\!f}) \; I_z 
	- \hbar \omega_1 \lbL{ \cos{\phi}\, I_x + \sin{\phi}\, I_y }\rb
\,.
\label{eq:ham_rot}
\ee
Naturally, the RF field lies along a fixed axis in the frame rotating
at $\omega_{r\!f}$. Furthermore, if $\omega_{r\!f} = \omega_0$, the
first term in Eq.~\ref{eq:ham_rot} vanishes. In this case, an observer
in the rotating frame will see the spin simply precess about
$\vec{B}_1$ (Fig.~\ref{fig:nutation}a), a motion called {\em
nutation}. The choice of $\phi$ controls the nutation axis.  An
observer in the lab frame sees the spin spiral down over the surface
of the Bloch sphere (Fig.~\ref{fig:nutation}b).  

\bfig
\bcen
\vspace*{1ex}
\includegraphics[width=7cm]{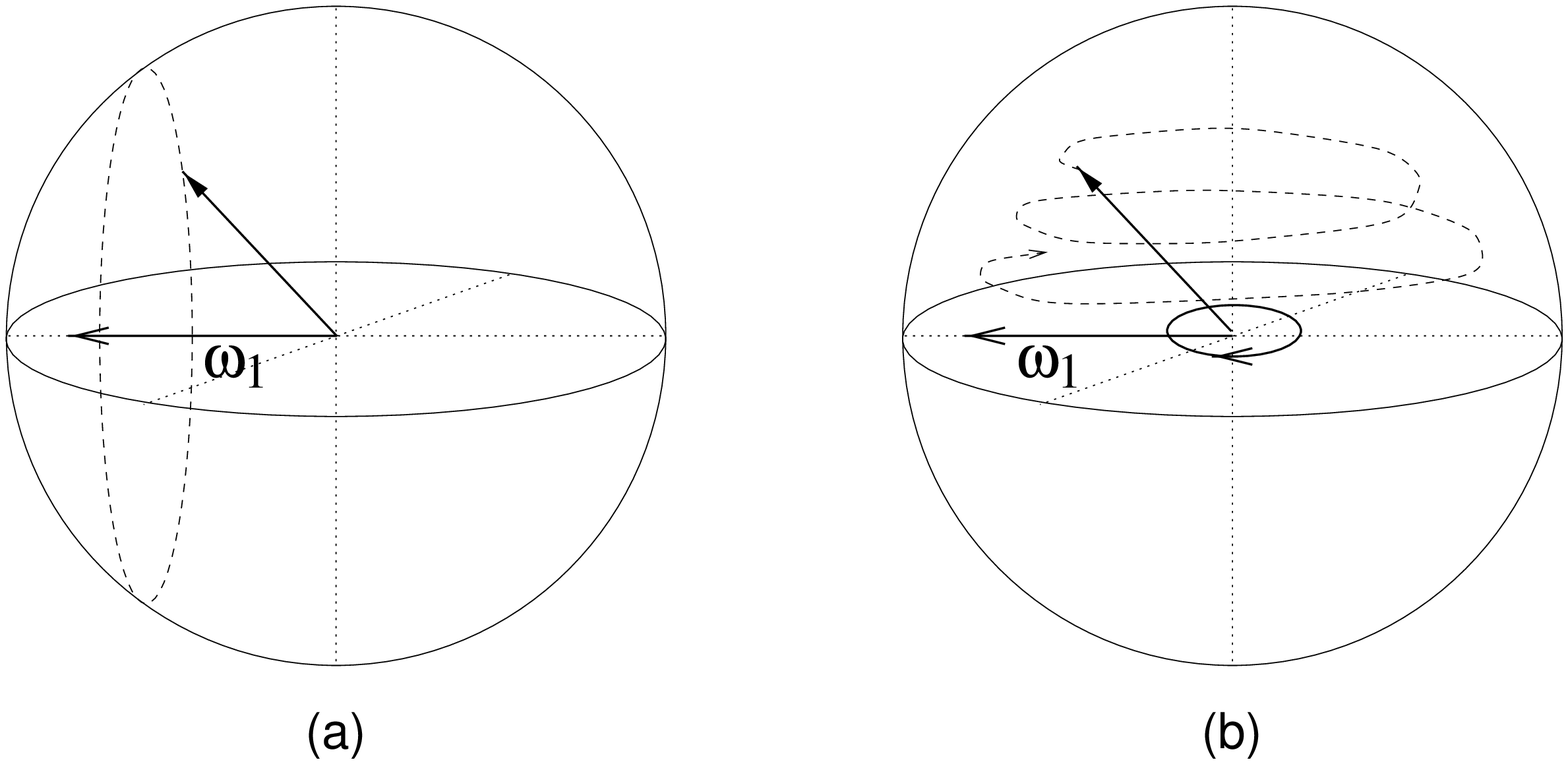} 
\vspace*{-2ex}
\ecen
\caption{Nutation of a spin subject to a transverse RF field (a) 
observed in the rotating frame and (b) observed in the lab frame.}
\label{fig:nutation}
\efig

If the RF field is {\em off-resonance} with respect to the spin
frequency by $\Delta \omega = \omega_0 - \omega_{r\!f}$, 
the spin precesses in the rotating frame about an axis tilted away 
from the $\hat{z}$ axis by an angle
\be 
	\alpha = \mbox{arctan} (\omega_1 / \Delta \omega) 
\,,
\label{eq:axis_offres_rf}
\ee
and with frequency
\be
	\omega_1' = \sqrt{\Delta\omega^2 + \omega_1^2} 
\,,
\label{eq:freq_offres_rf}
\ee
as illustrated in Fig.~\ref{fig:offres_rf}. 

\bfig
\bcen
\vspace*{1ex}
\includegraphics[width=4.5cm]{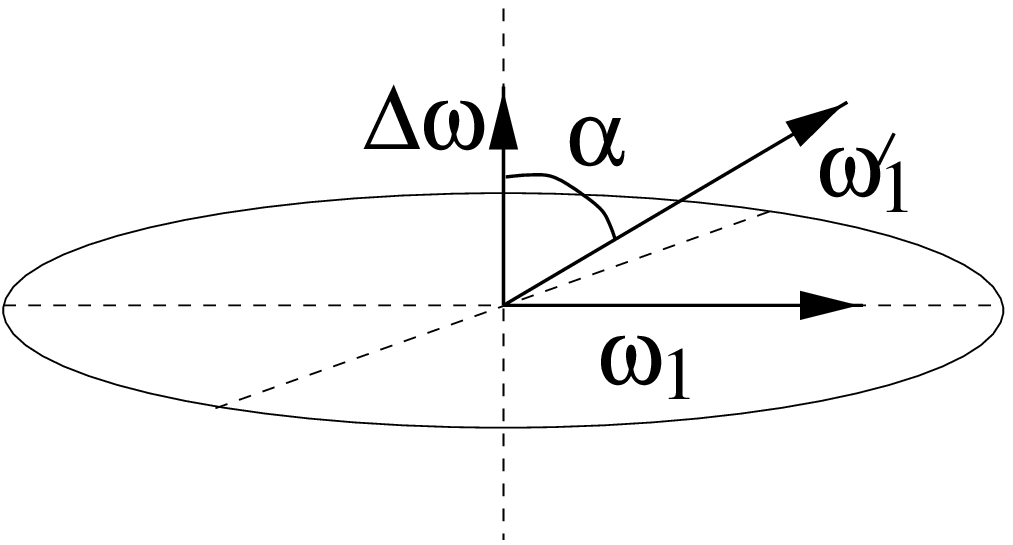} 
\vspace*{-2ex}
\ecen
\caption{Axis of rotation (in the rotating frame) during an 
off-resonant radio-frequency pulse.}
\label{fig:offres_rf}
\efig

It follows that the RF field has virtually no effect on spins which 
are far off-resonance, since $\alpha$ is very small when 
$|\Delta\omega| \gg \omega_1$ (see 
Fig.~\ref{fig:offreson90}).  If all spins have 
well-separated Larmor frequencies, we can thus in principle 
selectively rotate any one qubit without rotating the other spins.  

\bfig
\bcen
\vspace*{1ex}
\includegraphics[width=4.5cm]{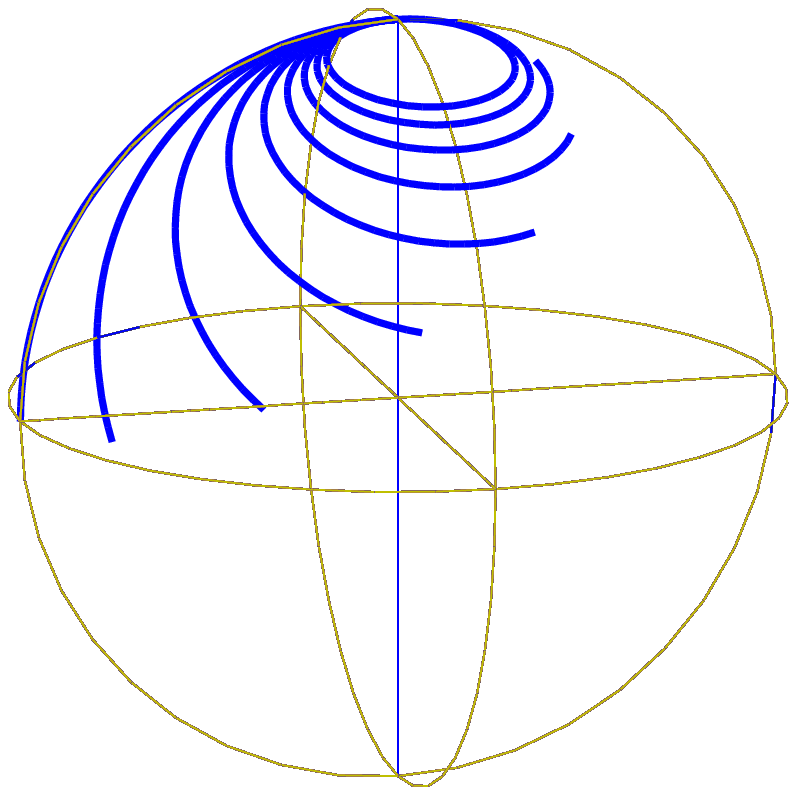}
\vspace*{-2ex}
\ecen
\caption{Trajectory in the Bloch sphere described by a qubit initially
in \ket{0} (along $+\hat{z}$), after a 250 $\mu$s pulse of strength
$\omega_1 = 1$ kHz is applied off-resonance by 0, 0.5, 1, \ldots 
4 kHz. On-resonance, the pulse produces a $90^\circ$ rotation. Far
off-resonance, the qubit is hardly rotated away from $\ket{0}$.}
\label{fig:offreson90}
\efig

Moderately off-resonance pulses ($|\Delta\omega| \approx \omega_1$) do
rotate the spin, but due to the tilted rotation axis, a single such 
pulse cannot, for instance, flip a spin from $\ket{0}$ to $\ket{1}$
(see again Fig.~\ref{fig:offreson90}). Of
course, off-resonance pulses can also be useful, for instance for 
direct implementation of rotations about an axis outside the 
$\hat{x}-\hat{y}$ plane.

We could also choose to work in a frame rotating at $\omega_0$
(instead of $\omega_{r\!f}$), where
\begin{eqnarray}
	{\cal H}^{rot} 
	&=& - \hbar \omega_1 
		\lbL \cos ((\omega_{r\!f} - \omega_0) t + \phi) I_x \right. 
\nonumber \\
	&& ~~~ + \left. \sin((\omega_{r\!f} - \omega_0) t + \phi) I_y \rbL
\,.
\label{eq:ham_rot_rf}
\end{eqnarray}
This transformation does not give a convenient time-independent RF
Hamiltonian (unless $\omega_{r\!f} = \omega_0$), as was the case for
${\cal H}^{rot}$ in Eq.~\ref{eq:ham_rot}. However, it is a natural
starting point for the extension to the case of multiple spins, where
a separate rotating frame can be introduced for each spin:
\be
	\ket{\psi}^{rot} 
	= \lbL{ \prod_{i} \exp(-i \omega_0^i t I_z^i) }\rb \ket{\psi}
\,.
\label{eq:mult_rot_frame}
\ee
In the presence of multiple RF fields indexed $r$, the RF Hamiltonian 
in this multiply rotating frame is 
\begin{eqnarray}
	{\cal H}^{rot} 
	&=& \sum_{i,r}^{} - \hbar \omega_1^r
	\lbL \cos ((\omega_{r\!f}^r - \omega_0^i) t + \phi^r) I_x^i \right. 
\nonumber \\
	&& \left. ~~~ + \sin((\omega_{r\!f}^r - \omega_0^i) t + \phi^r)
		I_y^i \rbL
\,,
\label{eq:mult_rot_frame1}
\end{eqnarray}
where the amplitudes $\omega^r_1$ and phases $\phi^r$ are under user
control.

The system Hamiltonian of Eq.~\ref{eq:ham_iso} is simplified, in the
rotating frame of Eq.~\ref{eq:mult_rot_frame}; the $I_z^i$ terms drop
out leaving just the $J_{ij} I_z^i I_z^j$ couplings, which remain
invariant.  Note that coupling terms of the form $\vec{I}^i \cdot
\vec{I}^j$ do not transform cleanly under Eq.~\ref{eq:mult_rot_frame}.

Summarizing, in the multiply rotating frame, the NMR Hamiltonian
${\cal H} = {\cal H}_{\rm sys} + {\cal H}_{\rm control}$ takes the
form
\bea
	{\cal H}_{\rm sys} 
	&=& %- \sum_{i} \hbar \, \omega_0^i \;I_z^i +
		  \hbar \sum_{i<j} 2 \pi J_{ij} I_z^i I_z^j 
\label{eq:nmr_sys_ham}
\\
	{\cal H}_{\rm control} 
	&=& \sum_{i,r}^{} - \hbar \omega_1^r
	\lbL \cos ((\omega_{r\!f}^r - \omega_0^i) t + \phi^r) I_x^i \right. 
\nonumber \\
	&& ~~~~~ \left. + \sin((\omega_{r\!f}^r - \omega_0^i) t + \phi^r)
		I_y^i \rbL
\,.
\label{eq:nmr_control_ham}
\eea

%%%%%%%%%%%%%%%%%%%%%%%%%%%%%%%%%%%%%%%%%%%%%%%%%%%%%%%%%%%%%%%%%%%%%%%%%
\subsection{Relaxation and decoherence}
\label{sec:decoherence}

One of the strengths of nuclear spins as quantum bits is precisely the
fact that the system is very well isolated from the environment,
allowing coherence times to be long compared with the dynamical
timescales of the system.  Thus, our discussion here focuses on closed
system dynamics, and it is important to be aware of the limits of this
approximation.

The coupling of the NMR system to the environment may be described by
an additional Hamiltonian term ${\cal H}_{\rm env}$, whose magnitude
is small compared to that of ${\cal H}_{\rm sys}$ or ${\cal H}_{\rm
control}$.  It is this coupling which leads to decoherence, the loss
of quantum information, which is traditionally parameterized by two
rates: $T_1$, the energy relaxation rate, and $T_2$, the phase 
randomization rate 
(see also Sections~\ref{sec:echo} and~\ref{sec:recovery}).

$T_2$ originates from spin-spin couplings which are imperfectly
averaged away, or unaccounted for in the system Hamiltonian.  For
example, in molecules in liquid solution, spins on one molecule may
have a long range, weak interaction with spins on another molecule.
Fluctuating magnetic fields, caused by spatial anisotropy of the
chemical shift, local paramagnetic ions, or unstable laboratory
fields, also contribute to $T_2$.  Nevertheless, in well prepared
samples and in a good experimental apparatus at reasonably high
magnetic fields, the $T_2$ for molecules in solution is easily on the
order of one second or more.  This decoherence mechanism can be
identified with elastic scattering in other physical systems; it
does not lead to loss of energy from the system.

$T_1$ originates from couplings between the spins and the ``lattice,''
that is, excitation modes which can carry away energy
quanta on the scale of the Larmor frequency.  For example, these may
be vibrational quanta, paramagnetic ions, chemical reactions such as
ions exchanging with the solvent, or spins with higher order magnetic
moments (such as $^{2}$H, $^{17}$Cl, or $^{35}$Br) which relax quickly
due to their quadrupolar moments interacting with electric field
gradients.  In well chosen molecules and liquid samples with good
solvents, $T_1$ can easily be tens of seconds, while isolated nuclei
embedded in solid samples with a spin-zero host crystal matrix (such
as $^{31}$P in $^{28}$Si) can have $T_1$ times of days.  
This mechanism is analogous to inelastic scattering in other
physical systems.

The description of relaxation in terms of only two parameters is known
to be an oversimplification of reality, particularly for coupled spin
systems, in which coupled relaxation mechanisms appear
\cite{Redfield57a,Jeener82a}. Nevertheless, the 
independent spin decoherence model is useful for its simplicity and 
because it can capture well the main effects of decoherence on simple 
NMR quantum computations~\cite{Vandersypen01a}, which are typically 
designed as pulse sequences shorter in time than $T_2$.

\section{Elementary pulse techniques}
\label{sec:elem_pulse}

This section begins our discussion of the main subject of this
article, a review of the control techniques developed in NMR quantum
computation for coupled two-level quantum systems.  We begin with a
quick overview of the language of quantum circuits and its important
universality theorems, then connect this with the language of pulse
sequences as used in NMR, and indicate how pulse sequences can be
simplified.  The main approximations employed in this section are that
pulses can be strong compared with the system Hamiltonian while
selectively addressing only one qubit at a time, and can be perfectly
implemented. The limits of these approximations are discussed in the
last part of the section.

%%%%%%%%%%%%%%%%%%%%%%%%%%%%%%%%%%%%%%%%%%%%%%%%%%%%%%%%%%%%%%%%%%%%%%%%%%%%%%
\subsection{Quantum control, quantum circuits, and pulses}

The goal of quantum control, in the context of quantum computation, is
the implementation of a unitary transformation $U$, specified in terms
of a sequence $U = U_k U_{k-1} \cdots U_2 U_1$ of standard ``quantum
gates'' $U_i$, which act locally (usually on one or two qubits) 
and are simple to implement.
As is conventional for unitary operations, the $U_i$ are ordered in
time from right to left.

%%%%%%%%%%%%%%%%%%%%%%%%%%%%%%%%%%%%%%%%
\subsubsection{Quantum gates and circuits}

The basic single-qubit quantum gates are rotations, defined as
\be
	R_{\hat n}(\theta) 
	= \exp \lbL{ - \frac{i \theta \hat{n}\cdot \vec{\sigma}}{2}}\rb
\,,
\label{eq:rotdef}
\ee
where $\hat{n}$ is a (three-dimensional) vector specifying the axis of
the rotation, $\theta$ is the angle of rotation, and $\vec{\sigma} =
\sigma_x \hat{x} + \sigma_y \hat{y} + \sigma_z \hat{z}$ is a vector of
Pauli matrices.  It is also convenient to define the Pauli matrices 
(see Eq.~\ref{eq:Pauli_def})
themselves as logic gates, in terms of which $\sigma_x$ can be
understood as being analogous to the classical {\sc not} gate, which
flips $|0\>$ to $|1\>$ and vice versa.  
In addition, the {\em Hadamard} gate $H$ and $\pi/8$ gate $T$
\bea
	H = \frac{1}{\sqrt{2}}\mattwoc{1}{1}{1}{-1}
~~,~~	T = \mattwoc{1}{0}{0}{\exp(i\pi/4)}
\eea
are useful and widely employed.  These, and any other single qubit
transformation $U$ can be realized using a sequence of rotations
about just two axes, according to Bloch's theorem: for any 
single-qubit $U$, there exist real numbers $\alpha,\beta,\gamma$ and 
$\delta$ such that 
\be
        U = e^{i\alpha} R_x(\beta) R_y(\gamma) R_x(\delta)
\,.  
\label{thm:X-Y_decomp}
\end{equation}

The basic two-qubit quantum gate is a controlled-{\sc not} ({\sc
cnot}) gate
\be
    U_{\mbox{\sc cnot}} = \left[
        \mymatrix{
                1 & 0 & 0 & 0  \cr
                0 & 1 & 0 & 0   \cr
                0 & 0 & 0 & 1  \cr
                0 & 0 & 1 & 0   \cr
        } \right ]
\ee
where the basis elements in this notation are $|00\>$, $|01\>$,
$|10\>$, and $|11\>$ from left to right and top to bottom. 
$U_{\mbox{\sc cnot}}$ flips the second qubit (the target) if and
only if the first qubit (the control) is $\ket{1}$. This gate
is the analogue of the classical exclusive-{\sc or} gate, since
$U_{\mbox{\sc cnot}} |x,y\> = |x,x\oplus y\>$, for $x,y\in \{0,1\}$ 
and where $\oplus$ denotes addition modulo two.

A basic theorem of quantum computation is that up to an irrelevant
overall phase, any $U$ acting on $n$
qubits can be composed from $U_{\mbox{\sc cnot}}$ and 
$R_{\hat n}(\theta)$ gates~\cite{Nielsen00b}. 
Thus, the problem of quantum control can be reduced to implementing 
$U_{\mbox{\sc cnot}}$ and single qubit rotations, where at least two 
non-trivial rotations are required.  Other such sets of universal 
gates are known, but this is the one which has been employed in NMR.

These gates and sequences of such gates may be conveniently
represented using quantum circuit diagrams, employing standard symbols.  
We shall use a notation commonly employed in the
literature~\cite{Nielsen00b} in this article.

%%%%%%%%%%%%%%%%%%%%%%%%%%%%%%%%%%%%%%%%
\subsubsection{Implementation of single qubit gates}
\label{sec:1qubit_gates}

Rotations on single qubits may be implemented directly in the rotating
frame using RF
pulses.  From the control Hamiltonian, Eq.~\ref{eq:nmr_control_ham}, 
it follows that when an RF field of amplitude $\omega_1$ is applied 
to a single-spin system at $\omega_\textit{rf} = \omega_0$, the spin 
evolves under the transformation
\be
	U = \exp \lbL{i \omega_1 ( \cos{\phi}\, I_x + \sin{\phi}\, I_y ) 
			t_{pw} }\rb
\,,
\label{eq:spin_ideal_rot}
\ee
where $t_{pw}$ is the {\em pulse width} (or pulse length), the time
duration of the RF pulse.  $U$ describes a rotation in the Bloch
sphere over an angle $\theta$ proportional to the product of $t_{pw}$
and $\omega_1=\gamma B_1$, and about an axis in the $\hat{x}-\hat{y}$
plane determined by the phase $\phi$.  

Thus, a pulse with phase $\phi=\pi$ and $\omega_1 t_{pw} = \pi/2$
will perform $R_x(90)$ (see Eq.~\ref{eq:rotdef}), which is a 
90$^\circ$ rotation about $\hat{x}$, denoted for short as
$X$. A similar pulse but twice as long realizes a $R_x(180)$ rotation,
written for short as $X^2$.  By changing the phase of the RF pulse to
$\phi=-\pi/2$, $Y$ and $Y^2$ pulses can similarly be implemented. 
For $\phi=0$, a negative rotation about $\hat{x}$, denoted
$R_x(-90)$ or $\bar{X}$, is obtained, and similarly $\phi = \pi/2$ 
gives $\bar{Y}$.
For multi-qubit systems, subscripts are used to indicate on which
qubit the operation acts, e.g.  $\bar{Z}^2_3$ is a $180^\circ$
rotation of qubit $3$ about $-\hat{z}$.

%\begin{comment}
It is thus not necessary to apply the RF field along different 
spatial axis in the lab frame
to perform $\hat{x}$ and $\hat{y}$ rotations. Rather,
the {\em phase} of the RF field determines the nutation axis in the 
rotating frame. Furthermore, note that only the {\em relative} 
phase between pulses applied to the same spin matters.  The absolute 
phase of the first pulse on any given spin does not matter in itself. 
It just establishes a phase reference against which the phases of all
subsequent pulses on that same spin, as well as the read-out of that
spin, should be compared.
%\end{comment}

We noted earlier that the ability to implement arbitrary rotations 
about $\hat{x}$ and $\hat{y}$ is sufficient for performing arbitrary 
single-qubit rotations (Eq.~\ref{thm:X-Y_decomp}).  Since $\hat{z}$ 
rotations are very common, two useful explicit decompositions of 
$R_z(\theta)$ in terms of $\hat{x}$ and $\hat{y}$ rotations are:
\be
	R_z(\theta) = X R_y(\theta) \bar{X} = Y R_x(-\theta) \bar{Y} 
\,.
\label{eq:composite_z}
\ee

%%%%%%%%%%%%%%%%%%%%%%%%%%%%%%%%%%%%%%%%
\subsubsection{Implementation of two-qubit gates}
\label{sec:2qubit_gates}

\begin{figure}
\bcen
\vspace*{1ex}
\includegraphics*[width=8cm]{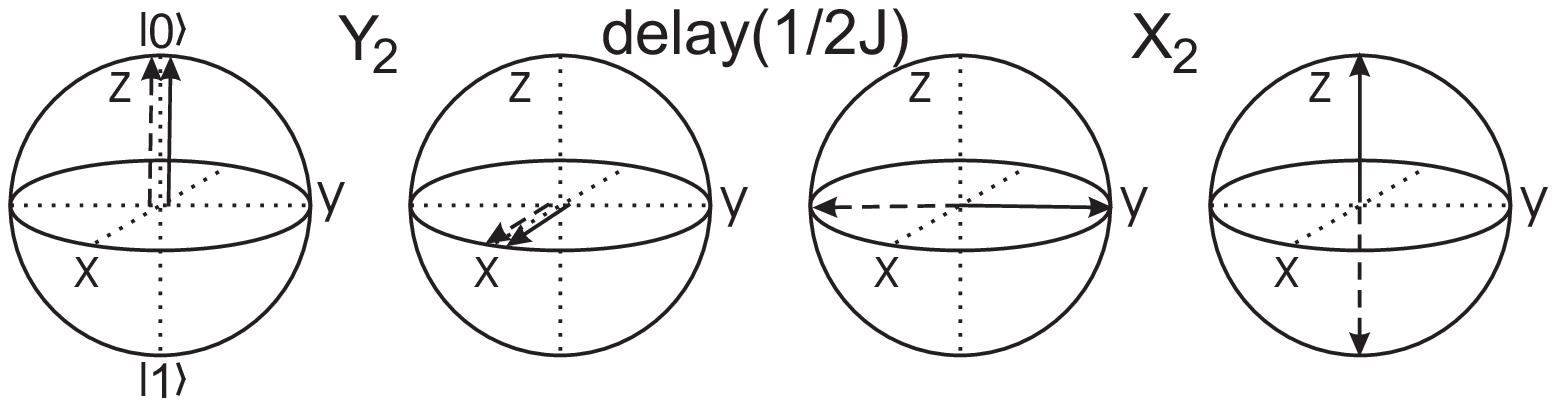}
\vspace*{-2ex}
\ecen
\vspace*{-1.0ex}
\caption{Bloch-sphere representation of the operation of the 
{\sc CNOT}$_{12}$ gate between two qubits $1$ and $2$ coupled by
$\hbar 2\pi J I_z^1 I_z^2$. 
Here, qubit $2$ starts off in $\ket{0}$ (along $\hat{z}$) and is 
depicted in a reference frame rotating about
$\hat{z}$ at $\omega_0^2/2\pi$. Solid and dashed arrows correspond to 
the case where qubit $1$ is $\ket{0}$ and \ket{1} respectively.}
\label{fig:inept}
\end{figure}

The most natural two-qubit gate is the one generated directly by
the spin-spin coupling 
Hamiltonian. For nuclear spins in a molecule in liquid solution, the
coupling Hamiltonian is given by Eq.~\ref{eq:ham_J} (in the lab frame
as well as in the rotating frame), from which we obtain the time
evolution operator $U_J(t) = \exp[-i 2\pi J I_z^1 I_z^2 t]$, or in
matrix form
\be
U_J(t) =
\left[\matrix{	e^{-i \pi J t/2} & 0 & 0 & 0 \cr
		0 & e^{+i \pi J t/2} & 0 & 0 \cr
		0 & 0 & e^{+i \pi J t/2} & 0 \cr
		0 & 0 & 0 & e^{-i \pi J t/2} }\right] \,.
\ee
Allowing this evolution to occur for time $t=1/2J$ gives a
transformation known as the {\em controlled
phase gate}, up to a $90^\circ$ phase shift on each qubit and an
overall (and thus irrelevant) phase:
\be
	U_{\mbox{\sc cphase}} =
	\sqrt{-i} \bar{Z_1} \bar{Z_2} U_J(1/2J)  
	=
	\left[\matrix{ 1 & 0 & 0 & 0 \cr
			0 & 1 & 0 & 0 \cr
			0 & 0 & 1 & 0 \cr
			0 & 0 & 0 & -1 }\right] 
\,.
\ee
This gate is equivalent to the well-known {\sc cnot} gate up to a
basis change of the target qubit and a phase shift on the control
qubit:
\begin{eqnarray}
\label{eq:cnot}
	U_{\mbox{\sc cnot}} &=& i Z_1^2 \bar{Y}_2 U_{\mbox{\sc cphase}} Y_2  
\nonumber \\
	&=& i Z_1^2 \bar{Y}_2 \left[\sqrt{-i} \bar{Z_1} 
		\bar{Z_2} U_J(1/2J)\right] Y_2 
\nonumber \\
	&=& \sqrt{i} Z_1 \bar{Z}_2 X_2 U_J(1/2J) Y_2 
\nonumber \\
	&=&
	\left[\matrix{ 1 & 0 & 0 & 0 \cr
			0 & 1 & 0 & 0 \cr
			0 & 0 & 0 & 1 \cr
			0 & 0 & 1 & 0 }\right] 
\,.
\end{eqnarray}
The core of this sequence, $X_2 U_J(1/2J) Y_2$,
can be graphically understood via 
Fig.~\ref{fig:inept}~\cite{Gershenfeld97a}, assuming the spins start
along $\pm \hat{z}$. First, a 
spin-selective pulse on spin $2$
about $\hat{y}$ (an rf pulse centered at $\omega_0^2/2\pi$ and of a
spectral bandwidth such that it covers the frequency range
$\omega_0^2/2\pi \pm J_{12}/2$ but not 
$\omega_0^1/2\pi \pm J_{12}/2$,
rotates spin 2 from $\hat{z}$ to $\hat{x}$.  Next, the spin system is
allowed to freely evolve for a duration of $1/2J_{12}$ seconds.
Because the precession frequency of spin 2 is shifted by $\pm
J_{12}/2$ depending on whether spin 1 is in $\ket{1}$ or $\ket{0}$
(see Fig.~\ref{fig:energy_2spins}),
spin 2 will arrive in $1/2J$ seconds at either $+\hat{y}$ or 
$-\hat{y}$, depending on the state of spin 1.  Finally, a 90$^\circ$
pulse on spin 2 about the $\hat{x}$ axis rotates spin 2 back to
$+\hat{z}$ if spin 1 is $\ket{0}$, or to $-\hat{z}$ if spin 1 is in
$\ket{1}$.

The net result is that spin 2 is flipped if and only if spin 1 is in
$\ket{1}$, which corresponds exactly to the classical truth table for
the {\sc cnot}.  The extra $\hat{z}$ rotations in Eq.~\ref{eq:cnot}
are needed to give all elements in $U_{\mbox{\sc cnot}}$ the same
phase, so the sequence works also for superposition input states.

An alternative implementation of the {\sc cnot} gate, up to a relative
phase factor, consists of applying a line-selective 180$^\circ$ pulse
at $\omega_0^2 + J_{12}/2$ (see Fig.~\ref{fig:energy_2spins}). This
pulse inverts spin 2 (the target qubit) if and only if spin 1 (the
control) is $\ket{1}$~\cite{Cory97b}. In general, if a spin is coupled
to more than one other spin, half the lines in the multiplet must be
selectively inverted in order to realize a {\sc cnot}.  Extensions to
doubly-controlled {\sc not}s are straightforward: in a three-qubit
system for example, this can be realized through inversion of one out
of the eight lines~\cite{Freeman98a}. As long as all the lines are 
resolved, it is in
principle possible to invert any subset of the lines. Demonstrations
using very long multi-frequency pulses have been performed with up to
five qubits~\cite{Khitrin02a}.  However, this approach cannot be used
whenever the relevant lines in the multiplet fall on top of each
other.

If the spin-spin interaction Hamiltonian is not of the form $I_z^i
I_z^j$ but contains also transverse components (as in
Eqs.~\ref{eq:ham_dipole}, \ref{eq:ham_dipole2} and~\ref{eq:ham_bond}),
other sequences of pulses are needed to perform the {\sc cphase} and
{\sc cnot} gates. These sequences are somewhat more
complicated~\cite{Bremner02a}.
 
If two spins are not directly coupled two each other, it is still 
possible to
perform a {\sc cnot} gate between them, as long as there exists a
network of couplings that connects the two qubits. For example,
suppose we want to perform a {\sc cnot} gate with qubit 1 as the
control and qubit 3 as the target, {\sc cnot}$_{13}$, but $1$ and $3$
are not coupled to each other. If both are coupled to qubit $2$, as in
the coupling network of Fig.~\ref{fig:coupling_networks} (b), we can
first swap the state of qubits $1$ and $2$ (via the sequence {\sc
cnot}$_{12}$ {\sc cnot}$_{21}$ {\sc cnot}$_{12}$), then perform a {\sc
cnot}$_{23}$, and finally swap qubits $1$ and $2$ again (or relabel
the qubits without swapping back).  The net
effect is {\sc cnot}$_{13}$.  By extension, at most $O(n)$ {\sc swap}
operations are required to perform a {\sc cnot} between any pair of
qubits in a chain of $n$ spins with just nearest-neighbor couplings
(Fig.~\ref{fig:coupling_networks}b). {\sc swap} operations can also 
be used to perform two-qubit gates between any two qubits which are 
coupled to a common ``bus" qubit (Fig.~\ref{fig:coupling_networks}c).

\bfig
\bcen
\vspace*{1ex}
\includegraphics*[width=8cm]{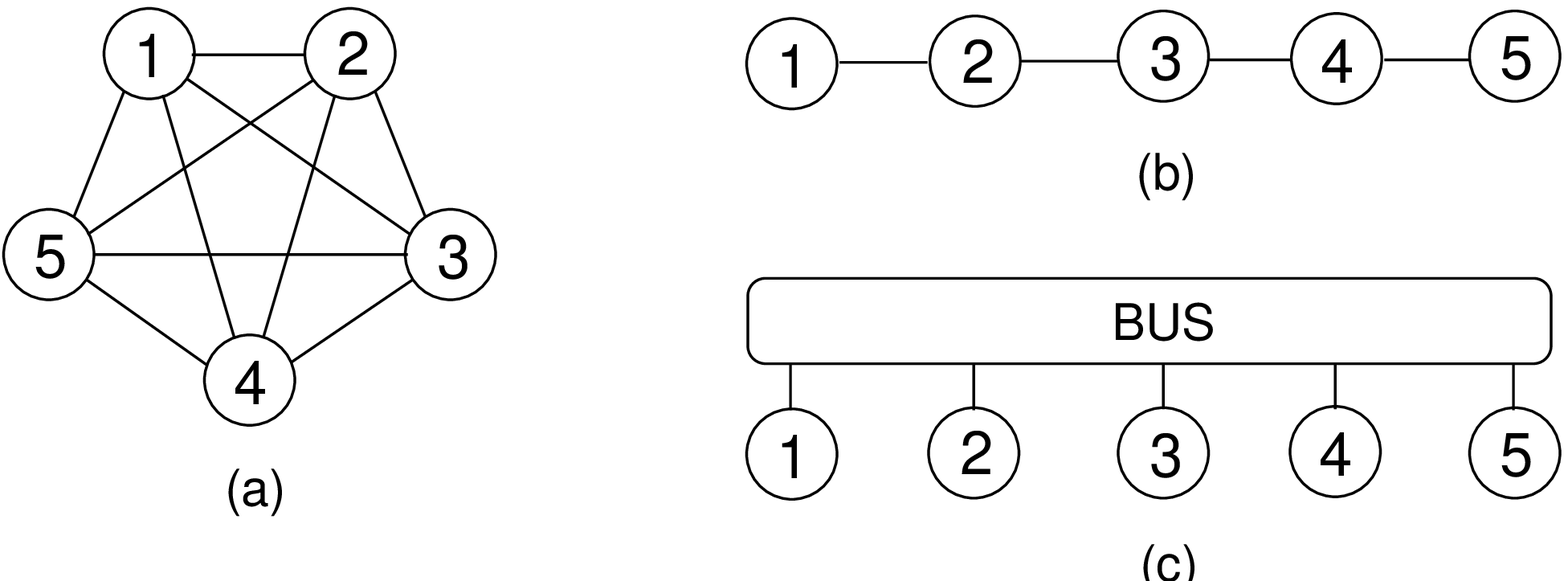} 
\vspace*{-2ex}
\ecen
\caption{Three possible coupling networks between five qubits. (a) A
full coupling network. Such networks will in practice always be
limited in size, as physical interactions tend to decrease with
distance. (b) A nearest-neighbor coupling network. Such linear chains
with nearest-neighbor couplings, or two-dimensional variants are used
in many solid-state proposals. (c) Coupling via a ``bus''. This is the
case of ion-trap schemes for example. Similar to case (a), the bus
degree of freedom will in reality couple well to only a finite number 
of qubits.}
\label{fig:coupling_networks}
\efig

Conversely, if a qubit is coupled to many other qubits 
(Fig.~\ref{fig:coupling_networks}a) and we want to
perform a {\sc cnot} between just two of them, we must remove the effect
of the remaining couplings. This can accomplished using the technique
of refocusing, which has been widely adopted in a variety of NMR
experiments.

\subsubsection{Refocusing: turning off undesired $I_z^i I_z^j$ couplings}
\label{sec:refocusing}

The effect of coupling terms during a time interval of free evolution
can be removed via so-called ``refocusing'' techniques.  For coupling
Hamiltonians of the form $I_z^i I_z^j$, as is often the case in liquid
NMR experiments (see Eq.~\ref{eq:ham_J}), the refocusing
mechanism can be understood at a very intuitive level. Reversal of the
effect of coupling Hamiltonians of other forms, such as in
Eqs.~\ref{eq:ham_dipole}, \ref{eq:ham_dipole2} and~\ref{eq:ham_bond},
is less intuitive, but can be understood within the framework of
average Hamiltonian theory (section~\ref{sec:average_Ham}).

Let us first look at two ways of undoing $I_z^i I_z^j$ in a two-qubit
system. In Fig.~\ref{fig:refocusing2}a, the evolution of qubit 1 in
the first time interval $\tau$ is reversed in the second time
interval, due to the $180^\circ$ pulse on qubit $2$. In
Fig.~\ref{fig:refocusing2}b, qubit 1 continues to evolve in the same
direction all the time, but the first $180^\circ$ pulse causes the
two components of qubit 1 to be refocused by the end of the second time
interval. The second $180^\circ$ pulse ensures that both qubits always
return to their initial state.

\bfig
\bcen
\vspace*{1ex}
\includegraphics*[width=8cm]{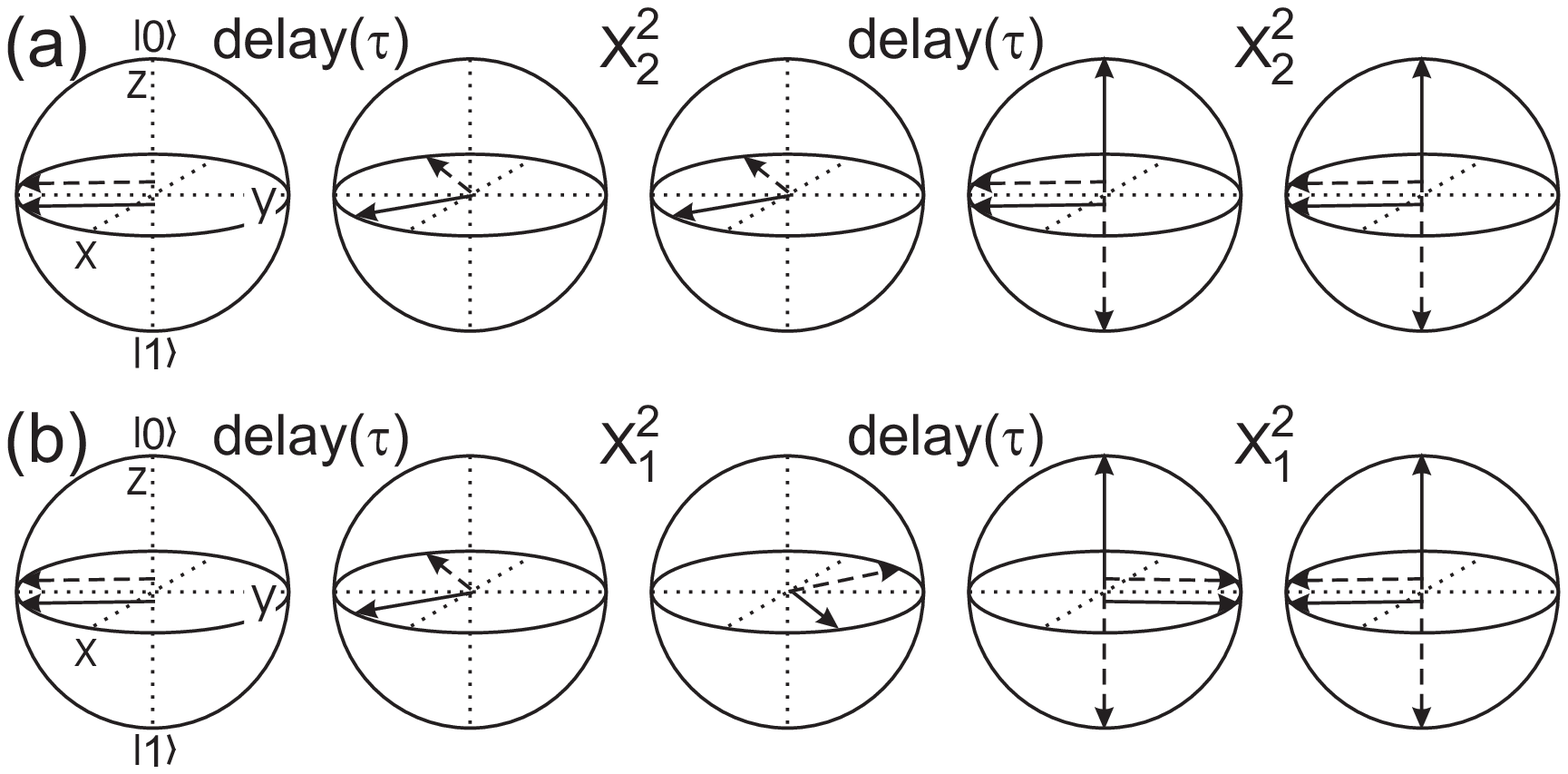} 
\vspace*{-2ex}
\ecen
\vspace*{-1.0ex}
\caption{Bloch-sphere representation of the operation of two simple 
schemes to refocus the coupling between two coupled qubits. The diagram
shows the evolution of qubit $1$ (in the rotating frame) initially
along $-\hat{y}$, when qubit 2 is in $\ket{0}$ (solid) or in $\ket{1}$
(dashed). The refocusing pulse can be applied to either (a) qubit 2
or (b) qubit 1.}
\label{fig:refocusing2}
\efig

Mathematically, we can see how refocusing of $J$ couplings works using
the fact that for all $\tau$
\be
X_1^2 \, U_J(\tau) \, X_1^2 
= U_J(-\tau) =
X_2^2 \, U_J(\tau) \, X_2^2 \,,
\ee
which leads to
\be
X_1^2\, U_J(\tau) \, 
X_1^2\, U_J(\tau)  = I =
X_2^2\, U_J(\tau)\, 
X_2^2\, U_J(\tau) \;.
\ee
Replacing all $X_i^2$ with $Y_i^2$, the sequence works just the same.
However, if we use sometimes $X_i^2$ and sometimes $Y_i^2$, we get the
identity matrix only up to some phase shifts. Also, if we applied
pulses on both qubits simultaneously, e.g. $X_1^2 X_2^2 \, U_J(\tau)
\, X_1^2 X_2^2 \, U_J(\tau)$, the coupling would not be removed.

Fig.~\ref{fig:refocusing3} gives insight in refocusing techniques in a
multi-qubit system. Specifically, this scheme preserves the effect of
$J_{12}$, while effectively inactivating all the other couplings. The
underlying idea is that a coupling between spins $i$ and $j$ acts
``forward'' during intervals where both spins have the same sign in
the diagram, and acts ``in reverse'' whenever the spins have opposite
signs. Whenever a coupling acts forward and in reverse for the same
duration, it has no net effect.
 
\bfig
\bcen
\vspace*{1ex}
\includegraphics*[width=6cm]{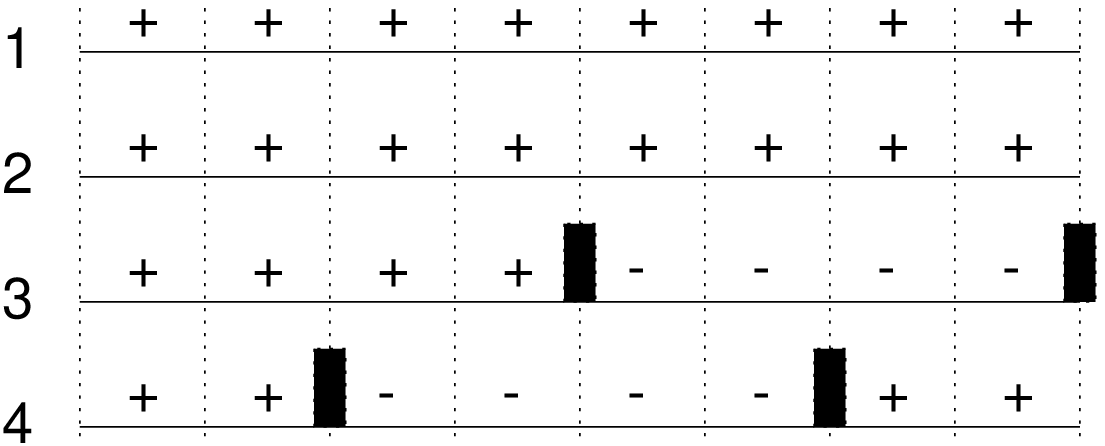} 
\vspace*{-2ex}
\ecen
\vspace*{-1.0ex}
\caption{Refocusing scheme for a four spin system, designed to leave 
$J_{12}$ active the whole time but to neutralize the effect of the other 
$J_{ij}$.
The interval is divided into slices of equal duration, and the ``$+$''
and ``$-$'' signs indicate whether a spin is still in its original
position, or upside down. The black rectangles represent $180^\circ$ 
pulses, which flip the corresponding spin.}
\label{fig:refocusing3}
\efig

Systematic methods for designing refocusing schemes for multi-qubit
systems have been developed specifically for the purpose of quantum
computing. The most compact scheme is based on Hadamard 
matrices~\cite{Leung00a,Jones99a}. A Hadamard matrix of order $n$, denoted 
by $H(n)$, is an $n \times n$ matrix with entries $\pm 1$, such that
\be
H(n) H(n)^T=nI \,.
\ee
The rows are thus pairwise orthogonal, and any two rows agree in exactly 
half of the entries. Identifying $+1$ and $-1$ with $+$ and $-$ as in the 
diagram of Fig.~\ref{fig:refocusing3}, we see that $H(n)$ gives a valid 
decoupling scheme for $n$ spins using only $n$ time intervals. 
An example of $H(12)$ is
\be
\left[
\begin{array}{cccccccccccc}
+ & + & + & + & + & + & + & + & + & + & + & + \\
+ & + & + & - & - & + & - & - & + & - & - & + \\
+ & + & + & + & - & - & - & + & - & + & - & - \\
+ & - & + & + & + & - & - & - & + & - & + & - \\
+ & - & - & + & + & + & - & - & - & + & - & + \\
+ & + & - & - & + & + & - & + & - & - & + & - \\
+ & - & - & - & - & - & - & + & + & + & + & + \\
+ & - & + & - & - & + & + & - & - & + & + & - \\
+ & + & - & + & - & - & + & - & - & - & + & + \\
+ & - & + & - & + & - & + & + & - & - & - & + \\
+ & - & - & + & - & + & + & + & + & - & - & - \\
+ & + & - & - & + & - & + & - & + & + & - & - 
\end{array}
\right]
\ee
If we want 
the coupling between one pair of qubits to remain active while removing 
the effect of all other couplings, we can simply use the same row of $H(n)$ 
for those two qubits.

$H(n)$ does not exist for all $n$, but we can always find a 
decoupling sequence for $n$ qubits by taking the first $n$ rows of 
$H(\bar{n})$, with $\bar{n}$ the smallest integer that satisfies 
$\bar{n} \ge n$ with known $H(\bar{n})$. From the properties of Hadamard 
matrices, we can show that $\bar{n}/n$ is always close to 1~\cite{Leung00a}.
So decoupling schemes for $n$ spins require $\bar{n}$ time intervals and no
more than $n \bar{n}$ $180^\circ$ pulses.

Another systematic approach to refocusing sequences is illustrated via the 
following 4-qubit scheme~\cite{Linden99c}:
\be
\left[
\begin{array}{cccccccccccc}
+ & + & + & + & + & + & + & + \\
+ & + & + & + & - & - & - & - \\
+ & + & - & - & - & - & + & + \\
+ & - & - & + & + & - & - & + 
\end{array}
\right] \,.
\ee
For every additional qubit, the number of time intervals is doubled,
and $180^\circ$ pulses are applied to this qubit after the first,
third, fifth, $\ldots$ time interval. The advantage of this scheme over
schemes based on Hadamard matrices, is that it does not require
simultaneous rotations of multiple qubits. The main drawback is that
the number of time intervals increases exponentially.

We end this subsection with three additional remarks.  First, each
qubit will generally be coupled to no more than a fixed number of other
qubits, since coupling strengths tend to decrease with distance. In
this case, all refocusing schemes can be greatly
simplified~\cite{Linden99c,Leung00a,Jones99a}.

Second, if the forward and reverse evolutions under $J_{ij}$ are not
equal in duration, a net coupled evolution takes place corresponding
to the excess forward or reverse evolution. In principle, therefore,
we can organize any refocusing scheme such that it incorporates any
desired amount of coupled evolution for each pair of qubits.

Third, refocusing sequences can also be used to remove the effect of 
$I_z^i$ terms in the Hamiltonian. Of course, these 
terms vanish in principle if we work in the multiply rotating frame
(see Eq.~\ref{eq:nmr_sys_ham}). However, there may be some spread in 
the Larmor frequencies, for instance due to magnetic field 
inhomogeneities. This effect can then be reversed using refocusing 
pulses, as is routinely accomplished in spin-echo experiments
(section~\ref{sec:echo}).

%%%%%%%%%%%%%%%%%%%%%%%%%%%%%%%%%%%%%%%%%%%%%%%%%%%%%%%%%%%%%%%%%%%%%%%%%%%%%
\subsubsection{Pulse sequence simplification}
\label{sec:simplif}

There are many possible pulse sequences which in an ideal world
result in exactly the
same unitary transformation. Good pulse sequence design therefore
attempts to find the {\em shortest and most effective} pulse sequence
that implements the desired transformations. In 
section~\ref{sec:adv_pulse}, we will see that the use of more complex
pulses or pulse sequences may sometimes increase the degree of quantum
control. Here, we look at three levels of pulse sequence simplification.
%as well as a mathematical approach to finding the shortest possible
%pulse sequence that implements a given quantum gate.

At the most abstract level of pulse sequence simplification, careful
study of a quantum algorithm can give insight in how to reduce the
resources needed. For example, a key step in both the modified
Deutsch-Jozsa algorithm~\cite{Cleve98a} and the Grover
algorithm~\cite{Grover97a} can be described as 
the transformation $\ket{x}\ket{y} \rightarrow \ket{x}\ket{x \oplus
y}$, where $\ket{y}$ is set to $(\ket{0}-\ket{1})/\sqrt{2}$, so that the
transformation in effect is $\ket{x}(\ket{0}-\ket{1})/\sqrt{2}
\rightarrow (-1)^{f(x)} \ket{x}(\ket{0}-\ket{1})/\sqrt{2}$. We might 
thus as well leave out the last qubit as it is never changed.

At the next level, that of quantum circuits, we can use 
simplification rules such as those illustrated in
Fig.~\ref{fig:simplify_circuits}. In this process, we can fully take
advantage of commutation rules to move building blocks around, as
illustrated in Fig.~\ref{fig:commutations}. Furthermore, gates which 
commute with each other can be executed simultaneously.  Finally, we 
can take advantage of the fact that most building blocks have many
equivalent implementations, as shown for instance in
Fig.~\ref{fig:equivalent_circuits}.

\begin{figure}[htbp]
\bcen
\vspace*{1ex}
\includegraphics*[width=8.5cm]{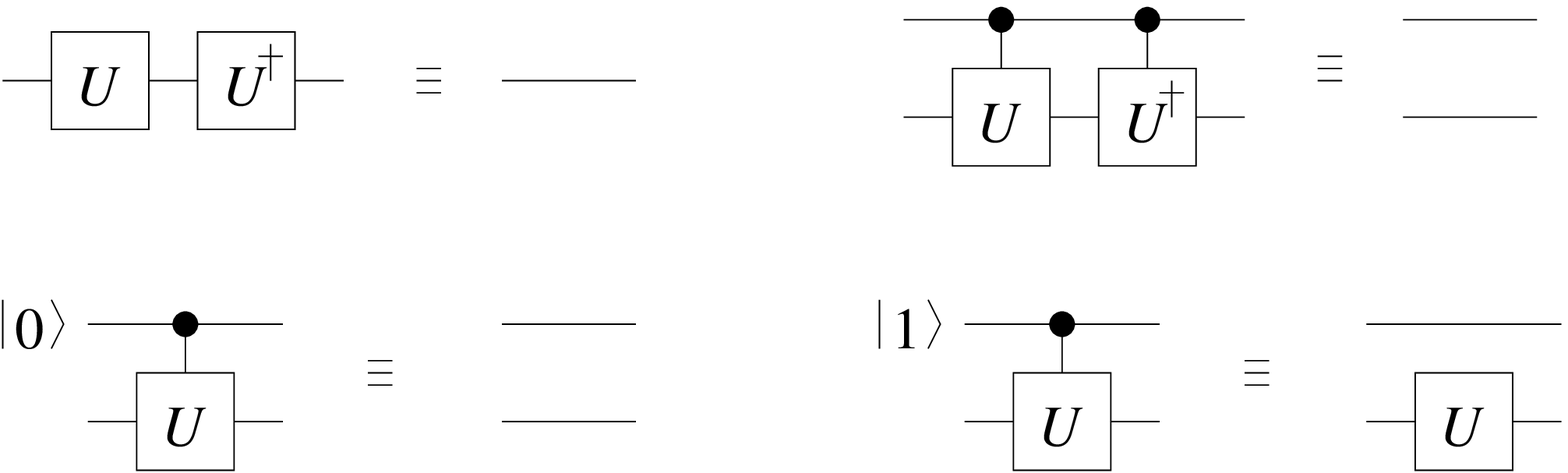} 
\vspace*{-2ex}
\ecen
\caption{Simplification rules for quantum circuits, drawn using 
         standard quantum gate symbols, where time goes from left to
	 right, each wire represents a qubit, boxes represent simple
	 gates, and solid black dots indicate control terminals.}
\label{fig:simplify_circuits}
\efig

\bfig
\bcen
\vspace*{1ex}
\includegraphics*[width=8.5cm]{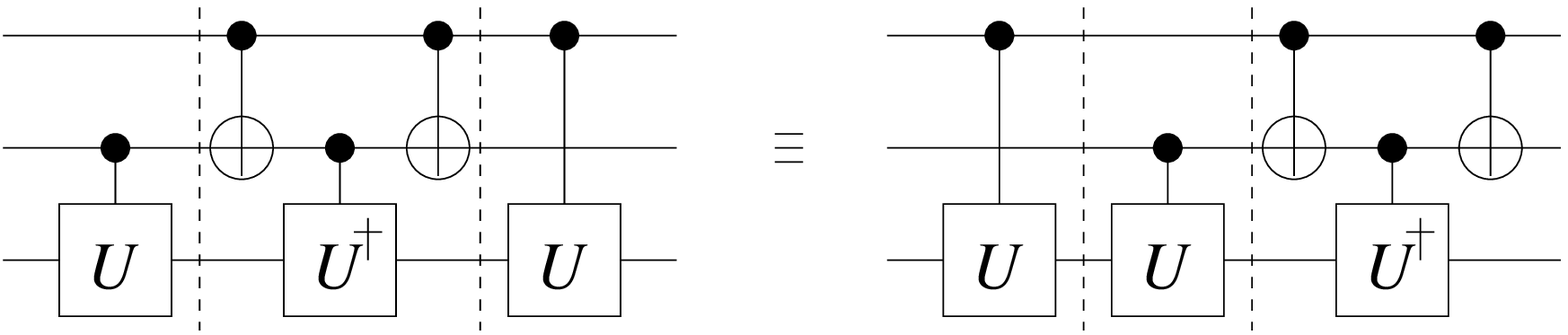} 
\vspace*{-2ex}
\ecen
\caption{Commutation of unitary operators can help simplify quantum 
circuits by moving building blocks around such that cancellation of
operations as in Fig.~\protect\ref{fig:simplify_circuits} become
possible. For example, the three segments (separated by dashed
lines) in these two equivalent realizations of the {\sc toffoli} gate
(doubly-controlled {\sc not}) 
commute with each other and can thus be executed in any order.}
\label{fig:commutations}
\efig

\bfig
\bcen
\vspace*{1ex}
\includegraphics*[width=8.5cm]{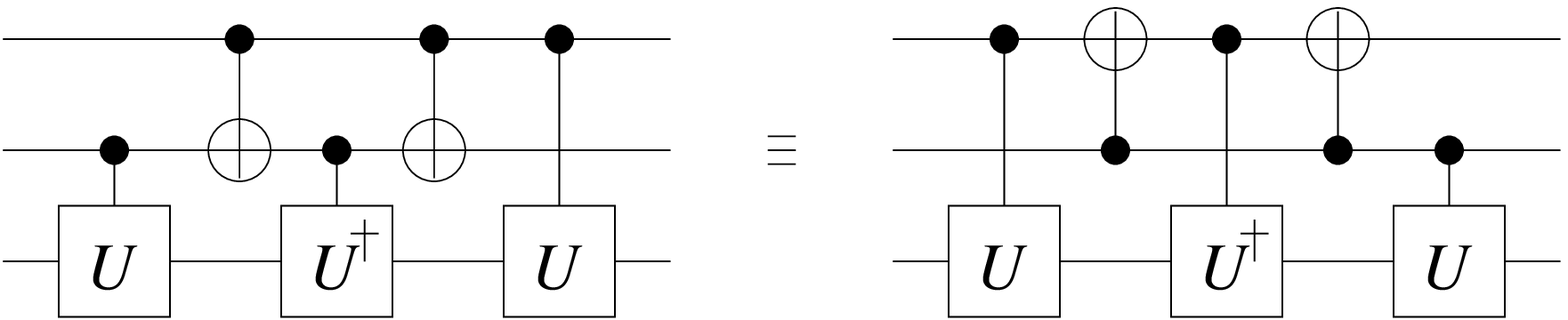} 
\vspace*{-2ex}
\ecen
\caption{Choosing one of several equivalent implementations can help 
simplify quantum circuits, again by enabling cancellation of
operations as in Fig.~\protect\ref{fig:simplify_circuits}. For instance,
the two control qubits in the {\sc toffoli} gate play equivalent roles, 
so they can be interchanged.}
\label{fig:equivalent_circuits}
\efig

Sometimes, a quantum gate may be replaced by another quantum gate, which 
is easier to implement. For instance, refocusing sequences 
(section~\ref{sec:refocusing}) can be kept simple by examining which 
couplings really need to be refocused. Early on in a pulse sequence, 
several qubits may still be along $\pm \hat{z}$, in which case their 
mutual $I_z^i I_z^j$ 
couplings have no effect and thus need not be refocused. 
Similarly, if a subset of the qubits can be traced out at some point in 
the sequence, the mutual interaction between these qubits does not matter 
anymore, so only their coupling with the remaining qubits must be 
refocused. Fig.~\ref{fig:refocusing4} gives an example of such a 
simplified refocusing scheme for five coupled spins.

\bfig
\bcen
\vspace*{1ex}
\includegraphics*[width=6cm]{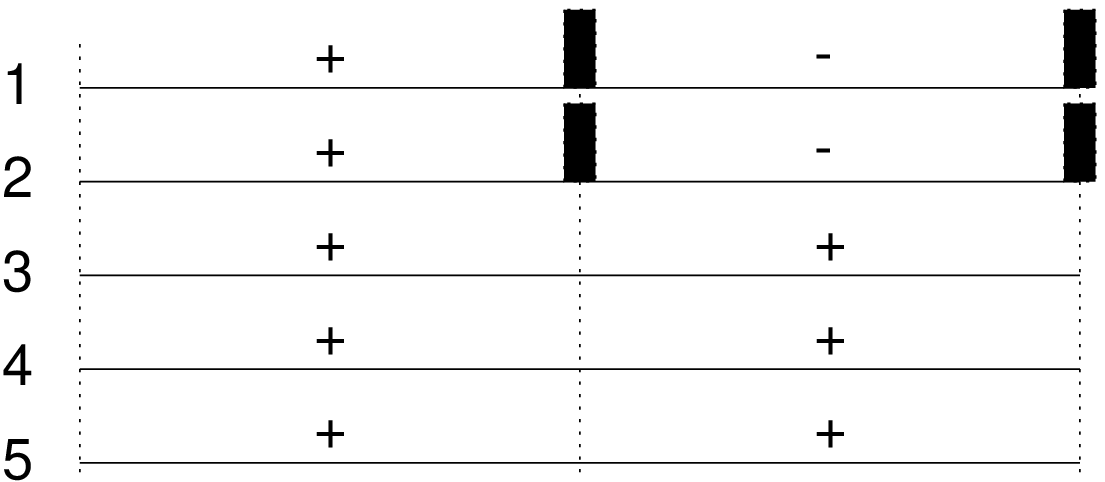} 
\vspace*{-2ex}
\ecen
\vspace*{-1.0ex}
\caption{Simplified refocusing scheme for five spins, designed such that
the coupling of qubits $1$-$2$ with qubits $3-5$ is switched off, i.e.
$J_{13},$ $J_{14},$ $J_{15},$ $J_{23},$ $J_{24}$ and $J_{25}$ are inactive
whereas $J_{12},$ $J_{34},$ $J_{35}$ and $J_{45}$ are active.}
\label{fig:refocusing4}
\efig

More generally, the relative phases between the entries in the 
unitary matrix describing a quantum gate are irrelevant when the gate acts
on a diagonal density matrix. In this case, we can for instance implement 
a {\sc cnot} simply as $X_2 \; U_J(1/2J) \;Y_2$ rather than the sequence
of Eq.~\ref{eq:cnot}.

At the lowest level, that of pulses and delay times, further
simplification is possible by taking out adjacent pulses which cancel
out, such as $X$ and $\bar{X}$ (an instance of the first simplification
rule of Fig.~\ref{fig:simplify_circuits}), and by converting 
``difficult'' operations to ``easy'' operations. 

\def\Uj{U_J\left(\frac{1}{2J}\right)}

Cancellation of adjacent pulses can be maximized by properly choosing 
the pulse sequences for subsequent quantum gates. For this purpose, it 
is convenient to have a library of equivalent implementations for the 
most commonly used quantum gates. For example, two equivalent 
decompositions of a {\sc cnot}$_{12}$ gate (with $J_{12} > 0$) are
\be
	Z_1 \, \bar Z_2 \, X_2 \; \Uj \;Y_2 
\,,
\label{eq:cnot_decomposition}
\ee
as in Eq.~\ref{eq:cnot}, and
\be
\bar Z_1 \, \bar Z_2 \, \bar X_2 \; \Uj \; \bar Y_2 \,.
\ee
Similarly, two equivalent implementations of the {\sc hadamard} gate on
qubit 2 are
\be
X_2^2 \, Y_2
\ee
and 
\be
\bar Y_2 \, X_2^2 \,.
\label{eq:had_decomposition}
\ee
Thus, if we need to perform a {\sc hadamard} operation on qubit 2
followed by a {\sc cnot}$_{12}$ gate, it is best to choose the
decompositions of Eqs.~\ref{eq:cnot_decomposition}
and~\ref{eq:had_decomposition}, such that the resulting pulse
sequence,
\be
Z_1 \, \bar Z_2 \, X_2 \; \Uj \; Y_2 \quad \bar Y_2 \, X_2^2
\ee
simplifies to
\be
Z_1 \, \bar Z_2 \, X_2 \; \Uj \; X_2^2 \,.
\ee

An example of a set of operations which is easy to perform is the 
rotations about $\hat{z}$. While the implementation of $\hat{z}$ rotations 
in the form of three RF pulses (Eq.~\ref{eq:composite_z}) takes more work 
than a rotation about $\hat{x}$ or $\hat{y}$, rotations about $\hat{z}$ need 
in fact not be executed at all, provided the coupling Hamiltonian is of the 
form $I_z^i I_z^j$, as in Eq.~\ref{eq:nmr_sys_ham}. In this case, 
$\hat{z}$ rotations commute with free evolution under the system 
Hamiltonian, so we can interchange the order of $\hat{z}$ rotations and 
time intervals of free evolution. Using equalities such as
\be
	Z \bar{Y} = X Y \bar{X} \bar{Y} = X Z \,,
\ee
we can also move $\hat{z}$ rotations across $\hat{x}$ and $\hat{y}$ 
rotations, and gather all $\hat{z}$ rotations at the end or the beginning
of a pulse sequence. At the end, $\hat{z}$ rotations do not affect the 
outcome of measurements in the usual $|0\>$, $|1\>$ ``computational'' 
basis. Similarly, $\hat{z}$ rotations at the start of a pulse sequence
have no effect on the usually diagonal initial state. In either case,
$Z$ rotations do then not require any physical pulses and are in a
sense ``for free'' and perfectly executed.  Indeed, $Z$ rotations
simply define the reference frame for $\hat{x}$ and $\hat{y}$ and can
be implemented by changing the phase of the reference frame throughout 
the pulse sequence.

It is thus advantageous to convert as many $X$ and $Y$ rotations as 
possible into $Z$ rotations, using identities similar to 
Eq.~\ref{eq:composite_z}, for example
\be
X Y = X Y \bar{X} X = Z X \,.
\ee

A key point in pulse sequence simplification of any kind is
that the simplification process must itself be {\em efficient}. For
example, suppose an algorithm acts on five qubits with initial state
$\ket{00000}$ and outputs the final state
$(\ket{01000}+\ket{01100}/\sqrt{2}$. The overall result of the
algorithm is thus that qubit $2$ is flipped and that qubit $3$ is
placed in an equal superposition of $\ket{0}$ and $\ket{1}$. This net
transformation can obviously be obtained immediately by the sequence
$X^2_2 Y_3$. However, the effort needed to compute the overall
input-output transformation generally increases exponentially with the
problem size, so such extreme simplifications are not practical.

%%%%%%%%%%%%%%%%%%%%%%%%%%%

\subsubsection{Time-optimal pulse sequences}
\label{sec:time-optimal}

Next to the widely used but rather naive set of pulse sequence 
simplification rules of the previous subsection, there exist powerful 
mathematical techniques for determing the \emph{minimum} time needed to 
implement a quantum gate, using a given system and control Hamiltonian, 
as well as for finding time-optimal pulse sequences~\cite{Khaneja01a}. 
These methods build on earlier optimization procedures for mapping an 
initial operator onto a final operator via unitary 
transformations~\cite{Sorenson89a,Glaser98a}, as in coherence or 
polarization transfer experiments, common tasks in NMR spectroscopy.

The pulse sequence optimization technique expresses pulse sequence
design as a geometric problem in the space of all possible unitary
transformations. The goal is to find the shortest path between the 
identity transformation, $I$, and the point in the space corresponding 
to the desired quantum gate, $U$, while travelling only in directions 
allowed by the given system and control Hamiltonian. 
Let us call $K$ the set of all unitaries $k$ that can be produced
using the control Hamiltonian only. Next we assume that the terms in 
the control Hamiltonian are much stronger than the system Hamiltonian 
(as we shall see in section~\ref{sec:coupled_evolution}, this assumption 
is valid in NMR only when using so-called hard, high-power pulses). 
Then, starting from $I$, any point in $K$ can be reached in a negligibly 
short time, and similarly, $U$ can be reached in no time from any point 
in the coset $K U$, defined by $\{k U | k \in K  \}$. Evolution
under the system Hamiltonian for a finite amount of time is required to 
reach the coset $KU$ starting from $K$. Finding a time-optimal sequence
for $U$ thus comes down to finding the shortest path from $K$ to $KU$, 
allowed by the system Hamiltonian. 

Such optimization problems have been extensively studied in 
mathematics~\cite{Brockett81a}, and have been solved explicitly for 
elementary quantum gates on two coupled spins~\cite{Khaneja01a} and a 
three-spin chain with nearest-neighbour couplings~\cite{Khaneja02a}. 
For example, a sequence was found for producing the trilinear propagator 
$\exp(-i 2\pi I_z^1 I_z^2 I_z^3)$ from the system Hamiltonian 
$\hbar 2 \pi J (I_z^1 I_z^2 + I_z^2 I_z^3)$ in a time $\sqrt{3}/2J$,
the shortest possible time~\cite{Khaneja02a}. This propagator is the 
starting point for useful quantum gates such as the doubly-controlled 
{\sc not} or {\sc toffoli} gate. The standard quantum circuit approach, 
in comparison, would yield a sequence of duration $3/2J$ (it uses only 
one coupling at a time while refocusing the other coupling), and the 
common NMR pulse sequence has duration $1/J$. 

Clearly, the time needed to {\em find} a time-optimal pulse sequence
increases exponentially with the number of qubits, $n$, involved in the 
transformation, since the unitary matrices involved are of size 
$2^n \times 2^n$. Therefore, the main use of the techniques presented 
here lies in finding efficient pulse sequences for building blocks
acting on only a few qubits at a time, which can then be incorporated 
in more complex sequences acting on many qubits by adding appropriate 
refocusing pulses to remove the couplings with the remaining qubits.
While the examples given here are for the typical NMR system and 
control Hamiltonian, the approach is completely general and may be
useful for other qubit systems too.

%%%%%%%%%%%%%%%%%%%%%%%%%%%%%%%%%%%%%%%%%%%%%%%%%%%%%%%%%%%%%%%%%%%%%%%%%%%%%%
\subsection{Experimental Limitations}
\label{sec:limitations}

Many years of experience have taught NMR spectroscopists that while
the ideal control techniques described above are theoretically
attractive, they neglect important experimental artifacts and
undesired Hamiltonian terms which must be addressed in any actual
implementation.  First, a pulse intended to selectively rotate one
spin will to some extent also affect the other spins.  Second, the
coupling terms $2 \pi J_{ij} I_z^i I_z^j$ cannot be switched
off in NMR. During time intervals of free evolution under the system
Hamiltonian, the effect of these coupling terms can easily be removed
using refocusing techniques (section~\ref{sec:refocusing}), so long as
the single-qubit rotations are perfect and instantaneous. However,
during RF pulses of finite duration, the coupling terms also distort
the single-qubit rotations.  In addition to these two limitations
arising from the NMR system and control Hamiltonian, a number of
instrumental imperfections cause additional deviations from the
intended transformations.

%%%%%%%%%%%%%%%%%%%%%%%%%%%%%%%%%%%%%%%%

\subsubsection{Cross-talk}
\label{sec:cross_talk}

Throughout the discussion of single- and two-qubit gates, we have 
assumed that we can selectively address each qubit. Experimentally,
qubit selectivity could be accomplished if the qubits are 
well-separated in space or, as in NMR, in frequency. In practice, 
there will usually be some cross-talk, which causes an RF pulse
applied on resonance with one qubit to slightly rotate another 
qubit, or shift its phase. Cross-talk effects are even more complex
when two or more pulse are applied simultaneously.

The frequency bandwidth over which qubits are rotated by a pulse of 
length $t_{pw}$ is roughly speaking of order $1/t_{pw}$. Yet, since 
the qubit response to an RF field is not linear (it is sinusoidal in 
$\omega_1 t_{pw}$), the exact frequency response cannot be 
computed using Fourier theory.

For a constant amplitude (rectangular) pulse, the unitary
transformation as a function of the detuning $\Delta \omega$ is easy 
to derive analytically from Eqs.~\ref{eq:axis_offres_rf}
and~\ref{eq:freq_offres_rf}. Alternatively, we can exponentiate the
Hamiltonian of Eq.~\ref{eq:ham_rot} to get $U$ directly.  An example
of a qubit response to a rectangular pulse
is shown in Fig.~\ref{fig:hard180_freq}.

\bfig
\bcen
\vspace*{1ex}
\includegraphics*[width=6cm]{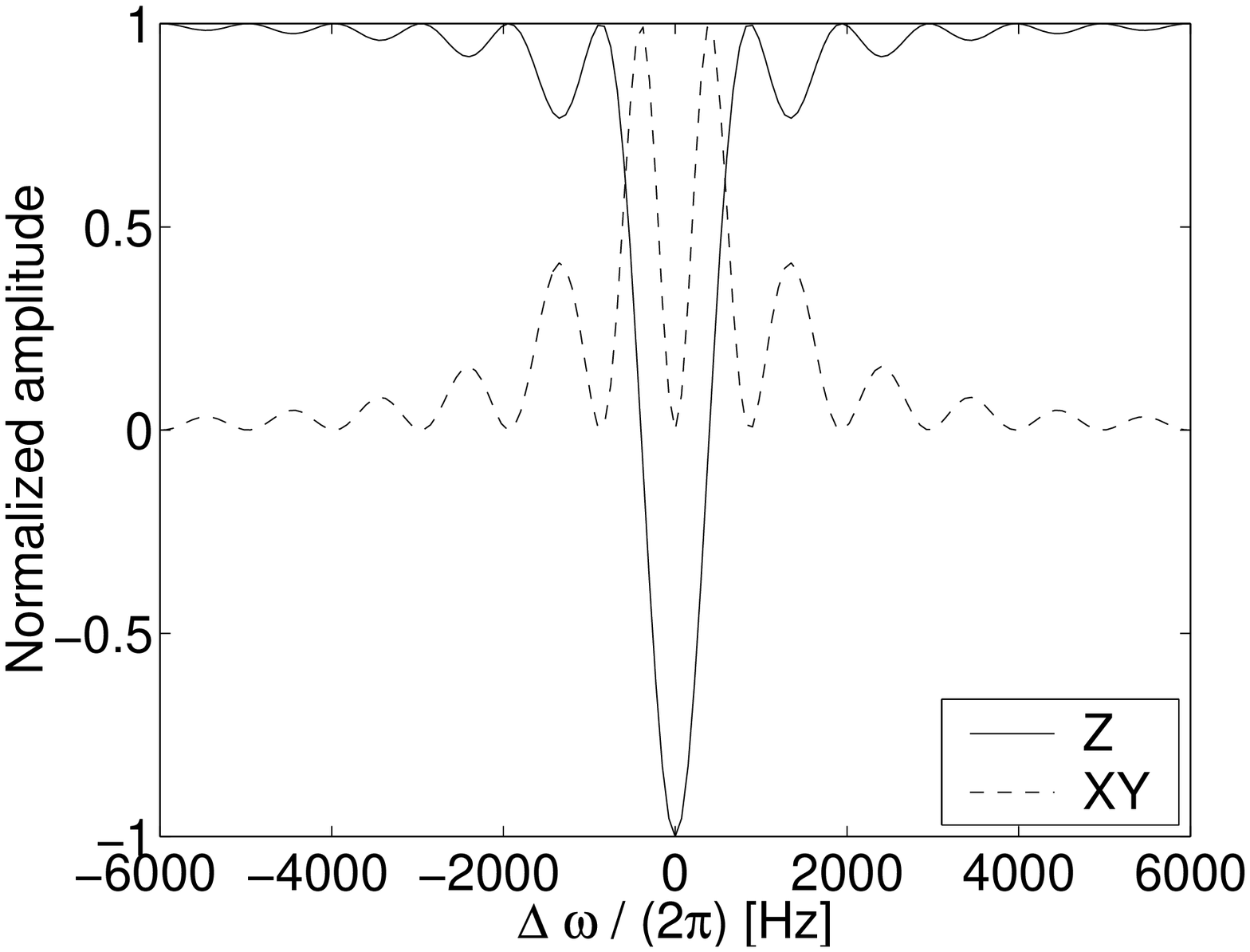} 
\vspace*{-2ex}
\ecen
\caption{Simulation of the spin response to a 1 ms constant amplitude 
RF pulse 
as a function of the frequency offset $\Delta \omega$ between $\omega_0$ 
and $\omega_{rf}$. The spin starts off in $\ket{0}$ (along $+\hat{z}$ in 
the Bloch sphere) and $\omega_1/ 2\pi = 500$ Hz 
is chosen such that the rotation angle amounts to 
$180^\circ$ for an on-resonance pulse.}
\label{fig:hard180_freq}
\efig

It is evident from Fig.~\ref{fig:hard180_freq} that short rectangular
pulses (known as ``hard'' pulses) excite spins over a very wide
frequency range. The frequency selectivity of a pulse can of course be
increased by increasing $t_{pw}$ while lowering $B_1$ accordingly
(thus creating what is known as a ``soft'' pulse), but decoherence
effects become more severe as the pulses get longer. Fortunately, as
we will see in sections~\ref{sec:shaped_pulses}
and~\ref{sec:composite_pulses}, the use of shaped and composite pulses
can dramatically improve the frequency selectivity of the RF excitation.

Even if a pulse is designed not to produce any net $\hat{x}$ or
$\hat{y}$ rotations of spins outside a specified frequency window, 
the presence of RF irradiation
during the pulse still causes a shift $\Delta \omega_{BS}^i$ in
the precession frequency of spins $i$ at frequencies well outside the
excitation frequency window~\cite{Emsley90a}.  As a result, each spin
accumulates a spurious phase shift during RF pulses applied to spins
at nearby frequencies.

This effect is related to the Bloch-Siegert shift mentioned in 
section~\ref{sec:rf}, and is known as the transient generalized 
Bloch-Siegert shift in the NMR community. It is related to the AC Stark 
effect in atomic physics. At a deeper level, the acquired phase can be 
understood as an instance of Berry's phase~\cite{Berry84a}: the spin 
describes a closed trajectory on the surface of the Bloch sphere and thus 
returns to its initial position, but it acquires a phase shift proportional 
to the area enclosed by its trajectory.

The frequency shift is given by
\be
\Delta \omega_{BS} \approx \frac{\omega_1^2}{2(\omega_0 - \omega_{r\!f})}
\label{eq:BS}
\ee
(provided $\omega_1 \ll |\omega_0 - \omega_{r\!f}|$), where $\omega_0/2\pi$ 
is the original Larmor frequency (in the absence of the RF field). 
In typical NMR experiments, 
the frequency shifts can easily reach several hundred Hz in magnitude. 
We see from Eq.~\ref{eq:BS} that the Larmor frequency shifts up if
$\omega_0 > \omega_{r\!f}$ and shifts down if $\omega_0 < \omega_{r\!f}$.

Fortunately, the resulting phase shifts can be easily computed in advance 
for each possible spin-pulse combination, if all the frequency 
separations, pulse amplitude profiles and pulse lengths are known. The unintended phase shifts $R_z(\theta)$ can then be compensated for during 
the execution of a pulse sequence by inserting appropriate
$R_z(-\theta)$, which can be executed at no cost, as we saw in
section~\ref{sec:simplif}.

Cross-talk effects are aggravated during simultaneous pulses, applied to 
two or more spins with nearby frequencies $\omega_0^1$ and
$\omega_0^2$ (say $\omega_0^1 < \omega_0^2$).  The pulse at
$\omega_0^1$ then temporarily shifts the frequency of spin $2$ to
$\omega_0^2 + \Delta \omega_{BS}$. As a result, the pulse on spin $2$,
if applied at $\omega_0^2$, will be off-resonance by an amount
$-\Delta \omega_{BS}$. Analogously, the pulse at $\omega_0^1$ is now
off the resonance of spin 1 by $\Delta \omega_{BS}$. The resulting
rotations of the spins deviate significantly from the intended
rotations.

The detrimental effect of the Bloch Siegert shifts during simultaneous
pulses is illustrated in Fig.~\ref{fig:simwithoutbs}, which shows the 
simulated inversion profile for
a spin subject to two simultaneous 180$^{\circ}$ pulses separated by
3273 Hz. The centers of the inverted regions have shifted away from
the intended frequencies and the inversion is incomplete, which can be
seen most clearly from the substantial residual
$\hat{x}-\hat{y}$-magnetization ($> 30\%$) 
over the whole region intended to be inverted. Note also that 
since the frequencies of the applied pulses are off the spin resonance
frequencies, complete inversion cannot be achieved no matter what tip
angle is chosen (see section~\ref{sec:rotating_frame}).

In practice, simultaneous soft pulses at nearby frequencies have been
avoided in NMR~\cite{Linden99a} or the poor quality of the spin
rotations was accepted. Pushed by the stringent requirements of quantum 
computation, several techniques have meanwhile been invented to generate 
accurate simultaneous rotations of spins at nearby frequencies 
(sections~\ref{sec:phase_ramping} and~\ref{sec:strongly_mod_pulses}).

\begin{figure}[htbp]
\begin{center}
\includegraphics*[width=6cm]{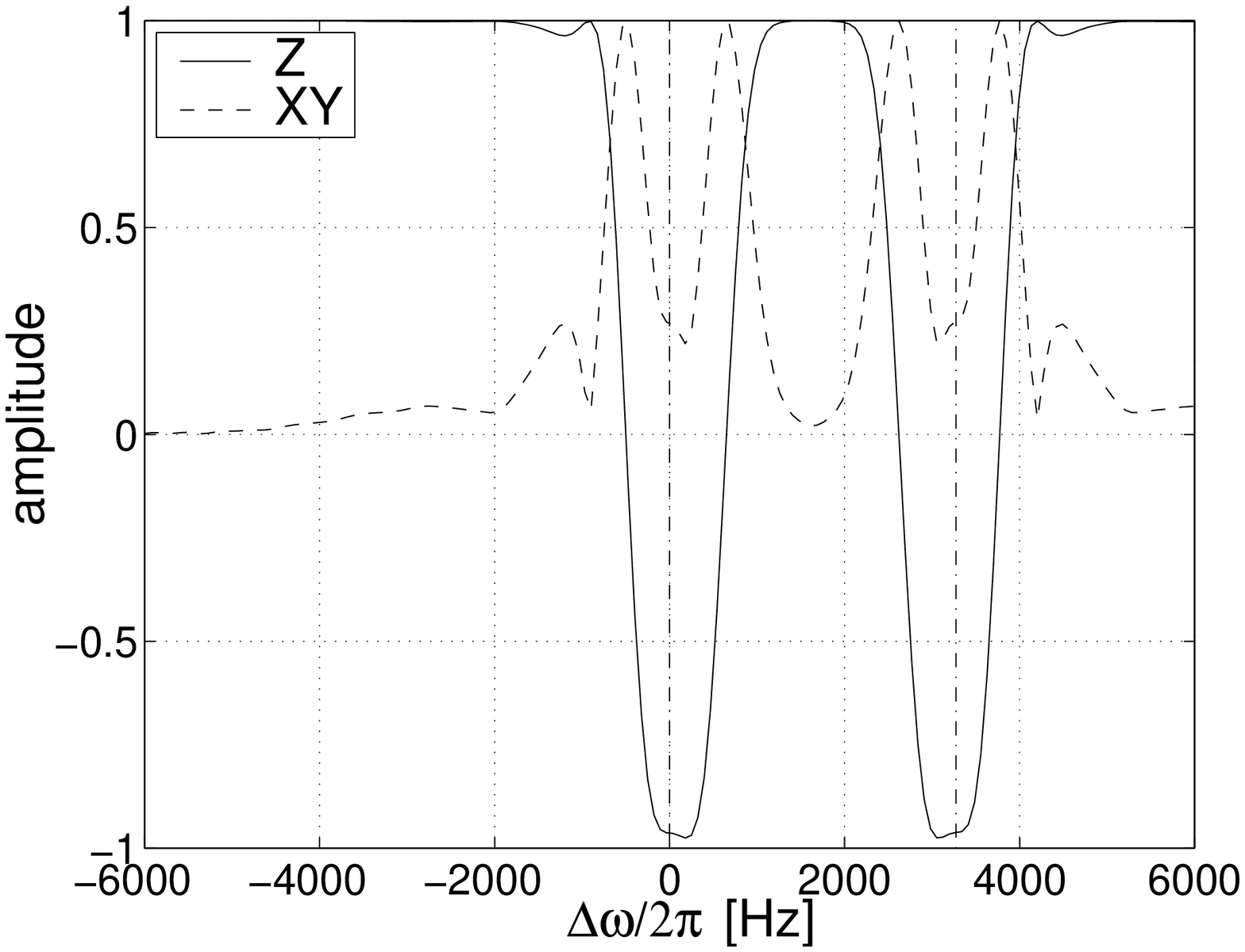} 
\end{center}
\vspace*{-2.0ex}
\caption{Simulation of the spin response to two simultaneous pulses
with carrier frequencies at 0 Hz and 3273 Hz (vertical dashed lines)
away from the spin resonance frequency, with a calibrated pulse length
of 2650$\mu$s (as for an ideal 180$^{\circ}$). The amplitude profile
of the pulses is Hermite shaped (Section~\ref{sec:shaped_pulses}) in
order to obtain a smooth spin response.  For ideal inversion, the
solid line should be $-1$ at the two frequencies, and the dashed line
should be zero.}
\label{fig:simwithoutbs}
\end{figure}

%%%%%%%%%%%%%%%%%%%%%%%%%%%%%%%%%%%%%%%%
\subsubsection{Coupled evolution}
\label{sec:coupled_evolution}

The spin-spin couplings in a molecule are essential for the
implementation of two-qubit gates (section~\ref{sec:2qubit_gates}),
but they cannot be turned off and are thus also active during the RF
pulses, which are intended to be just single-qubit transformations.
Unless $\omega_1$ is much stronger than the coupling strength, the
interactions strongly affect the intended nutation.  For couplings of
the form $J I_z^i I_z^j$, the effect is similar to the off-resonance
effects illustrated in Fig.~\ref{fig:offres_rf}: the coupling to
another spin shifts the spin frequency to $\omega_0/2\pi \pm J/2$, so
a pulse sent at $\omega_0/2\pi$ hits the spin off-resonance by $\mp
J/2$.

In practice, $J$-coupling terms can only be neglected for short,
high-power pulses used in heteronuclear spin systems: typically $J <
300$ Hz while $\omega_1$ is up to $\approx 50$ kHz.  For low-power
pulses, often used in homonuclear spin systems, $\omega_1$ can be of
the same order as $J$ and coupling effects become prominent.
The coupling terms also lead to additional complications when two qubits 
are pulsed simultaneously. In general, the qubits become partially
entangled~\cite{Kupce95a}.

As was the case for cross-talk, NMR spectroscopists have developed 
special shaped and composite pulses to compensate for coupling effects 
during RF pulses while performing spin-selective rotations. In recent 
years, the use of such pulses has been extended and perfected for 
quantum computing experiments (sections~\ref{sec:shaped_pulses} and~\ref{sec:composite_pulses}).

%%%%%%%%%%%%%%%%%%%%%%%%%%%%%%%%%%%%%%%%
\subsubsection{Instrumental errors}
\label{sec:instrum_errors}

A number of experimental imperfections lead to errors in the quantum
gates.  In NMR, the most common imperfections are inhomogeneities in
the static and RF magnetic field, pulse length calibration errors, 
frequency offsets, and pulse timing and phase imperfections.

The static field $B_0$ in modern NMR magnets can be made homogeneous
over the sample volume (a cylinder $5$ mm in diameter and $1.5$ cm
long) to better than $1$ part in $10^9$. This amazing homogeneity is
obtained by meticulously adjusting the current through a set of
so-called ``shim'' coils, which compensate for the inhomogeneities
produced by the large solenoid. At $\omega_0 = 500 \cdot 2\pi$ MHz,
linewidths of $0.5$ Hz can thus be obtained, corresponding to a
dephasing time constant $T_2^*$ (see Section~\ref{sec:larmor}) of
$1/(2\pi 0.5) = 0.32$ s.  In the course of long pulse sequences (of
order $0.1 - 1$ s), even the tiny remaining inhomogeneity would
therefore have a large effect, so its effect must be reversed using
refocusing sequences (section~\ref{sec:refocusing})

The RF field homogeneity is typically very poor, due to constraints on 
the geometry of the RF coils: the envelope of Rabi oscillations 
(section~\ref{sec:rabi}) often  decays by as much as $5\%$ per $90^\circ$ 
rotation, corresponding to a quality factor of only $\approx 5$. In 
sequences containing only a few pulses, this is not problematic, but in 
multiple-pulse experiments, the RF field inhomogeneity is often the 
dominant source of errors and signal loss.

Imperfect pulse length calibration has an effect similar to $B_1$ 
inhomogeneity: the qubit rotation angle is different than was intended.
Only the correlation time for the error is different. Miscalibrations are
constant throughout an experiment, whereas the RF field experienced by any
given molecule changes on the timescale of diffusion through the sample 
volume.

Frequency offsets occur in different contexts. In traditional NMR
experiments, the Larmor frequencies are often not known in advance. RF
pulses are then expected to rotate the spins over a wide range of
frequencies, quite the opposite case of quantum computing, where the Larmor
frequencies are precisely known and rotations should be spin-selective. 
However, we have seen earlier that $I_z^i I_z^j$ coupling terms act as
a frequency offset of one spin, which depends on the state of the other 
spin. 
Qubit-selective rotations of qubit $i$ thus require a uniform rotation 
over a range $\omega_0^i \pm \sum_{j \neq i} |J_{ij}|/2$.

Various approaches have been developed to reduce the sensitivity of RF
pulses and pulse sequences to these instrumental errors, sometimes in 
combination with solutions to cross-talk and coupling artifacts. These 
advanced techniques are the subject of the next section.

\section{Advanced pulse techniques}
\label{sec:adv_pulse}

The accuracy of quantum gates that can be achieved using the simple pulse
techniques of the previous section is unsatisfactory when applied to
multi-spin systems, where the given NMR system and control Hamiltonian
lead to undesired cross-talk and coupling effects.
In addition, the available instrumentation can only imperfectly 
approximate ideal pulse amplitudes, timings, and phases, for realistic 
sample geometries and coil configurations, and any real molecule includes 
additional Hamiltonian terms such as couplings to the environment, which 
are undesired. Nevertheless, extremely precise control can be achieved
despite these imperfections, and this is accomplished using the art of
{\em shaped pulses}, {\em composite pulses} and {\em average Hamiltonian
theory}, the subject of this second major section of this review.

These advanced techniques are based on the assumption that errors are, 
at least on some accessible timescale, {\em systematic}, rather than 
{\em random}. This assumption clearly holds for the terms in the ideal 
NMR Hamiltonian of Eqs.~\ref{eq:nmr_sys_ham} and~\ref{eq:nmr_control_ham}, 
and applies also to most instrumental errors. Then, by using the special 
properties of evolution in unitary groups, such as the $SU(2^n)$ which 
describes the space of operators acting on $n$ qubits, the systematic 
errors can in principle be canceled out. 

%%%%%%%%%%%%%%%%%%%%%%%%%%%%%%%%%%%%%%%%%%%%%%%%%%%%%%%%%%%%%%%%%%%%%%%%%%%%%%
\subsection{Shaped pulses}
\label{sec:shaped_pulses}

The amplitude and phase profile of RF pulses can be specially
tailored in order to ease the cross-talk and coupling effects
discussed in sections~\ref{sec:cross_talk}
and~\ref{sec:coupled_evolution}.
In practice, the pulse is divided in a few tens to many hundreds of
discrete time slices; to achieve an arbitrarily shaped pulse, it
suffices to control the amplitude and phase of the slices separately.
Furthermore, 
multiple shaped pulses applied at various frequencies can be combined 
into a single pulse shape, since a linear vector sum of pulse slices 
also results in a valid pulse. Here, we consider simple amplitude and 
phase shaped pulses.

%%%%%%%%%%%%%%%%%%%%%%%%%%%%%%%%%%%%%%%%
\subsubsection{Amplitude profiles}
\label{sec:amplitude_profiles}

The frequency selectivity of RF pulses can be much improved compared
to standard rectangular pulses with sharp edges, by using pulse shapes 
which smoothly modulate the pulse amplitude with time.  Such pulses 
are typically specially designed to excite or invert spins over a 
limited frequency region, while minimizing $\hat{x}$ and $\hat{y}$ 
rotations for spins outside this region~\cite{Freeman98a,Freeman97a}.

Furthermore, specialized pulse shapes exist which minimize the effect
of couplings during the pulses.
Such self-refocusing pulses~\cite{Geen91a} take a spin over a
complicated trajectory in the Bloch sphere, in such a way that the net
effect of couplings between the selected and non-selected spins is
reduced (Fig.~\ref{fig:iburp1_offres}). 
It is as if those couplings are only in part or even not at
all active during the pulse (couplings between pairs of non-selected
spins will still be fully active but their effect can be removed using
standard refocusing techniques~\ref{sec:refocusing}).  
As a general rule, it is relatively 
easy to make $180^\circ$ pulses self-refocusing, but much harder to do 
so for $90^\circ$ pulses. 

\begin{figure}[htbp]
\begin{center}
\includegraphics*[width=4cm]{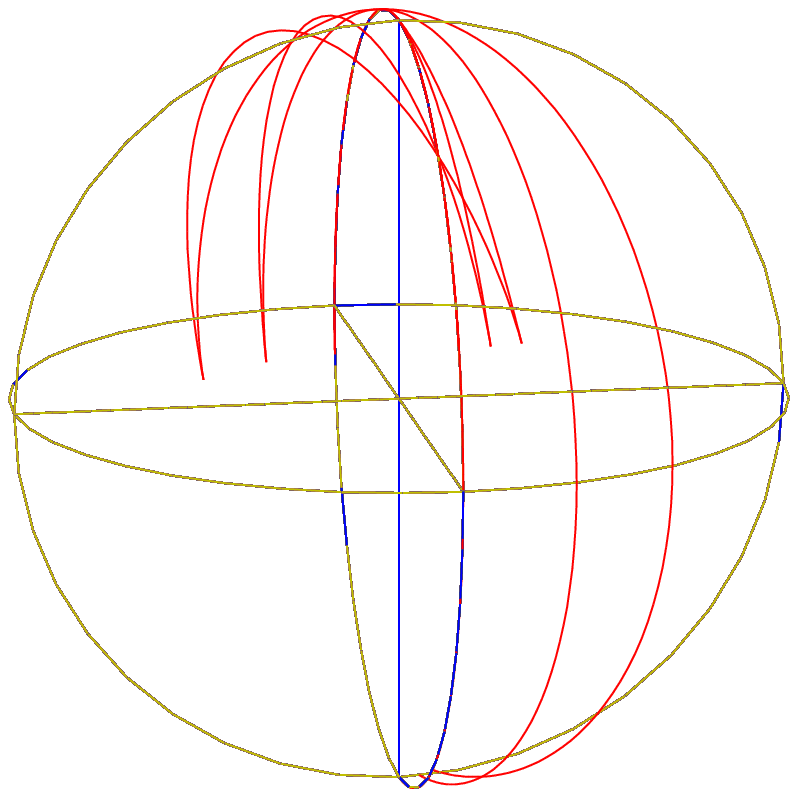}
\end{center}
\vspace*{-2.0ex}
\caption{Trajectory on the Bloch sphere of a qubit initially in 
$\ket{0}$, when a so-called {\sc iburp1} pulse ~\cite{Geen91a} is
applied, of duration 1 ms and $\omega_1 = 3342$ Hz, with a frequency
offset (analogous to $I_z^i I_z^j$ coupling) of 0, 100 and 200 Hz.
This pulse is intended to rotate the qubit from $\ket{0}$ ($+\hat{z}$)
to $\ket{1}$ ($-\hat{z}$).
We see that the effect of the frequency offset is largely removed by 
the specially designed pulse shape; all three trajectories terminate 
near $-\hat{z}$.}
\label{fig:iburp1_offres}
\end{figure}

The self-refocusing behavior of certain shaped pulses can be intuitively 
understood to some degree. Nevertheless, many actual pulse shapes have been 
the result of numerical optimizations. Often, the pulse shape is expressed 
in a basis of several functions, for instance a Fourier 
series~\cite{Geen91a},
\be
\omega_1(t) = \left\{ A_0 
+ \sum_n \left[ A_n \cos\left(\! n \frac{2\pi}{t_{pw}} t\right)
+ B_n \sin\left(\! n \frac{2\pi}{t_{pw}} t\right) \right] \!\right\} \,,
\ee
and the weights of the basis functions, $A_n$ and $B_n$, are optimized 
using numerical routines such as simulated annealing.

Comparison of the performance of various pulse shapes is facilitated by 
computing the corresponding spin responses. This is
most easily done by concatenating the unitary operators of each time slice 
of the shaped pulse, as the Hamiltonian is time-independent within each 
time slice.
Fig.~\ref{fig:profile_pulses} presents the amplitude profile and pulse 
response for three standard pulse shapes of equal duration, illustrating 
that different pulse shapes produce strikingly different spin response 
profiles.

\bfig
\begin{center}
\includegraphics{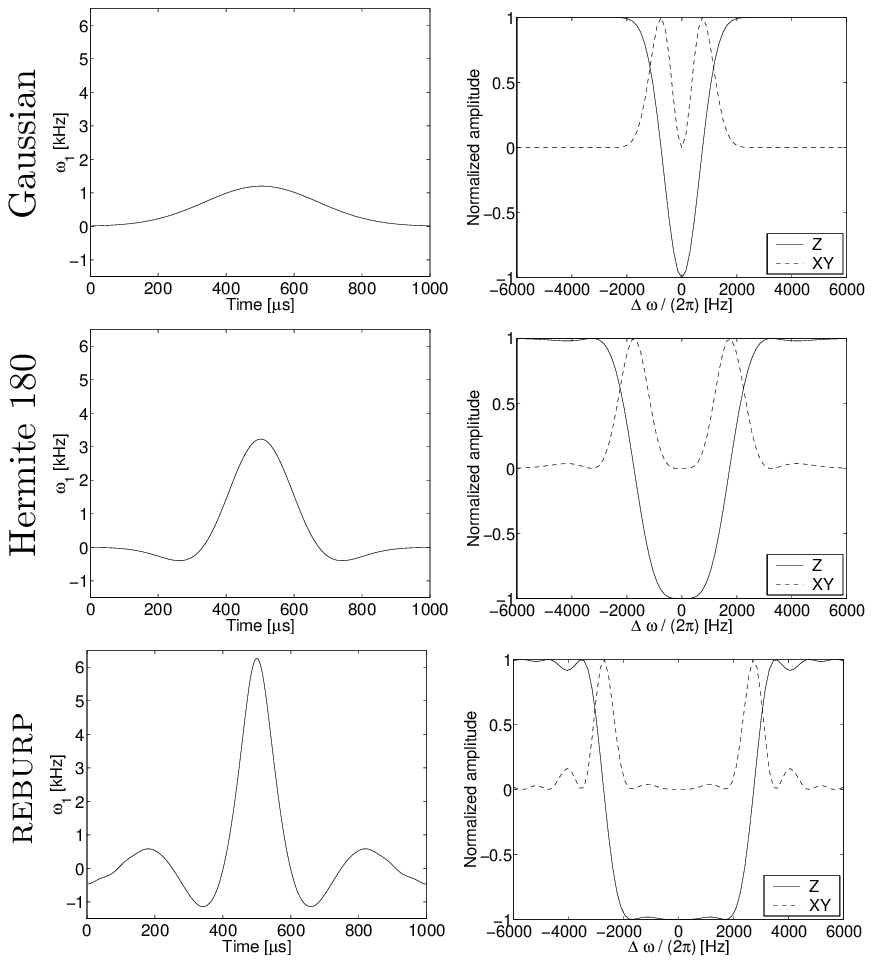}
\end{center}
\caption{(Left) Time profile and (Right) frequency response, displaying 
the $\hat{z}$ and $\hat{x}-\hat{y}$ components of the Bloch vector
after a pulse when the Bloch vector is along $+\hat{z}$ before the
pulse for three relevant pulse shapes.}
\label{fig:profile_pulses}
\efig

Properties relevant for choosing a pulse shape include:
\begin{itemize}
\item frequency selectivity: product of excitation bandwidth and 
pulse length (lower is more selective),
\item transition range: the width of the transition region between 
      the selected and non-selected frequency region,
\item power: the peak power required for a given pulse length and tip
angle (lower is less demanding),
\item self-refocusing behavior: degree to which the $J$
      coupling between the selected spin and other spins are refocused (the
      signature for self-refocusing behavior is a flat top in the excitation
      profile),
\item robustness to
      experimental imperfections such as pulse length errors,
\item universality: whether the pulse performs the intended rotation 
      for arbitrary input states or only for specific input states.
\end{itemize}

Table~\ref{tab:shaped_pulses} summarizes these properties for a
selection of widely used pulse shapes. Only universal pulses (also known
as general-rotation pulses) are included
in the table, since quantum computations must work for any input state.

\begin{table}[h]
\begin{center}
\begin{tabular}{l|c|c|c|c|c}
	    & selec- &transition&       &  self-   & robust- \\
	    & tivity &range     & power &refocusing& ness\\\hline
Rectangular & poor   & very wide&minimal& no       & good    \\
Gauss 90    &excellent&wide     & low    & fair     & good    \\
Gauss 180   &excellent&wide     & low    & fair     & good    \\
Hrm 90      &moderate&moderate  &average& good     & fair    \\
Hrm 180     & good   &moderate  &average&very good & fair    \\
{\sc uburp} 90    & poor   & narrow   & high  &excellent & poor    \\
{\sc reburp}180   & poor   & narrow   & high  &excellent & poor    \\
{\sc av} 90       & fair   &moderate  &average& good   & fair    \\
\end{tabular}
\end{center}
\caption{Properties of relevant pulse shapes. The Gaussian~\cite{Bauer84a} 
and Hermite~\cite{Warren84a} shapes are described by analytical functions 
and have been identified early on. The {\sc burp} family of 
pulses~\cite{Geen91a} resulted from numerical optimization routines. 
Continued work in this area has produced a large number of additional 
pulse shapes, such as the {\sc av} 90~\cite{Abramovich93a}.}
\label{tab:shaped_pulses}
\end{table}

Obviously, no single pulse shape optimizes for all
properties simultaneously, so pulse shape design consists of finding
the optimal trade-off for the desired application. For quantum computing 
experiments, we can select molecules with large chemical
shifts, so sharp transition regions are not so important. Furthermore,
the probe and spectrometer can deal with relatively high powers. The
crucial parameters are the self-refocusing behavior, the
selectivity (short, selective pulses minimize decoherence) and to some
extent the robustness.

It is also possible to start from a desired frequency response, and
invert the transformation to find the pulse shape that produces this
response.  Again, given the non-linear nature of the response, the
inverse transformation is not given by a Fourier transform, but it can
nevertheless be computed directly~\cite{Pauly91a}.

Even self-refocusing shaped pulses do generally not remove the
coupling terms completely. Furthermore, when two spins are
pulsed simultaneously with self-refocusing pulses, the refocusing
effects are often destroyed~\cite{Kupce95a}.  In both cases, the 
remaining coupled evolution that takes place during the pulses must 
be reversed at an earlier and/or later stage in the pulse sequence.

If we could decompose the evolution during an actual pulse into an
idealized, instantaneous $X$ or $Y$ rotation with no coupling present, 
followed and/or preceded by a time interval of free evolution, we could 
compensate for the coupling effects simply by adjusting the appropriate 
time intervals of free evolution in between the pulses
(section~\ref{sec:refocusing}).
However, ${\cal H}_{r\!f}$ and ${\cal H}_J$ do not commute, so 
such a decomposition is not possible.

Nevertheless, the coupled evolution can still be unwound to first
order~\cite{Vandersypen01a,Knill00a}, 
when a time interval of reverse evolution both before and after the 
pulse is used:
\be
e^{+i {\cal H}_J\, \tau /\hbar} e^{-i ({\cal H}_{r\!f} + {\cal
H}_J)\, t_{pw} \,/ \hbar} e^{+i {\cal H}_J\, \tau / \hbar}
\approx e^{-i {\cal H}_{r\!f}\, t_{pw} \,/ \hbar} \,,
\label{eq:unwind_J_symm}
\ee
where $\tau$ is chosen such that the approximations are as good as
possible according to some distance or fidelity measure (see
section~\ref{sec:fidelity}). The optimal $\tau$ is usually close but
not equal to $t_{pw} / 2$.  In comparison, a negative time interval
only before {\em or} after the pulse,
\begin{eqnarray}
	e^{+i {\cal H}_J\, \tau/\hbar}
	e^{-i ({\cal H}_{r\!f} + {\cal H}_J)\, t_{pw}\,/\hbar}
	\approx e^{-i {\cal H}_{r\!f} \,t_{pw}\,/\hbar} \nonumber 
\\
	\approx
	e^{-i ({\cal H}_{r\!f} + {\cal H}_J) \,t_{pw}\,/\hbar}
	e^{+i {\cal H}_J\, \tau/\hbar} \,,
	\end{eqnarray}
is much less effective.

%%%%%%%%%%%%%%%%%%%%%%%%%%%%%%%%%%%%%%%%
\subsubsection{Phase profiles}
\label{sec:phase_ramping}

An alternative to amplitude shaping that is often useful is frequency
or phase shaping.  One specific phase shaping method utilizes fixed,
small increments $\Delta \phi$ to the phase of successive slices of a
pulse to achieve an excitation profile which is centered at a
frequency which differs from the RF carrier frequency $\omega_{r\!f}$
by $\Delta \phi/\Delta t$, where $\Delta t$ is the duration of each
time slice.  This technique for shifting the RF frequency is known as
phase-ramping~\cite{Patt91a}.  We can express the effect of phase
ramping mathematically by replacing Eq.~\ref{eq:ham_rf_lab} by
\begin{eqnarray}
{\cal H}_{r\!f} / \hbar \omega_1 && =  
\cos \left[\omega_{r\!f} t + 
	\left(\phi_0 + \frac{\Delta\phi}{\Delta t} t\right) \right] I_x 
\hspace*{0cm} \nonumber \\
&& ~~~+ \hspace*{0cm}
      %\right. \nonumber \\
	%& &   \left.  
	\sin \left[\omega_{r\!f} t + 
	\left(\phi_0 + \frac{\Delta\phi}{\Delta t} t\right) \right] I_y 
	    %\right\}    
    \nonumber \\
&& = 
    \cos \left[\left(\omega_{r\!f} +
	   	\frac{\Delta\phi}{\Delta t}\right) t + \phi_0 \right] I_x   \nonumber \\
&& ~~~+	 \sin \left[\left(\omega_{r\!f} + 
		\frac{\Delta\phi}{\Delta t}\right) t + \phi_0 \right] I_y 
     \,.
\end{eqnarray}
The use of phase shifts thus permits us to obtain an RF field at a
different frequency than is generated by the signal generator.
Furthermore, the displaced frequency can be chosen different for every
pulse, and can even be varied in the course of a pulse.\\

A useful application of phase ramping lies in compensation for 
Bloch-Siegert effects during simultaneous pulses, where the RF applied 
at $\omega_0^i$ shifts the resonance frequency of spin $j$ to 
$\omega_0^j + \Delta \Omega_{BS}$ (Section~\ref{sec:cross_talk}).  
The rotations of both spins can be
significantly improved simply by shifting the RF excitation frequencies
via phase ramping such that they track the shifts of the corresponding 
spin frequencies~\cite{Steffen00a}. In this way, the pulses are always
applied on-resonance with the respective spins. The calculation of
the frequency shift throughout a shaped pulse is straightforward and
needs to be done only once, at the start of a series of experiments.

Fig.~\ref{fig:simwithbs} shows the simulated inversion profiles for
the same conditions as in Fig.~\ref{fig:simwithoutbs}, but this time
using the frequency shift corrected scheme. The inversion profiles are
clearly much improved and there is very little left-over
$\hat{x}-\hat{y}$ magnetization.  Simulations of the inversion
profiles for a variety of pulse shapes, pulse widths and frequency
separations, confirm that the same technique can be used to correct
the frequency offsets caused by three or more simultaneous soft pulses
at nearby frequencies.  The improvement is particularly pronounced
when the frequency window of the shaped pulse is two to eight times
the frequency separation between the pulses~\cite{Steffen00a}.

\begin{figure}[htbp]
\begin{center}
\includegraphics*[width=6cm]{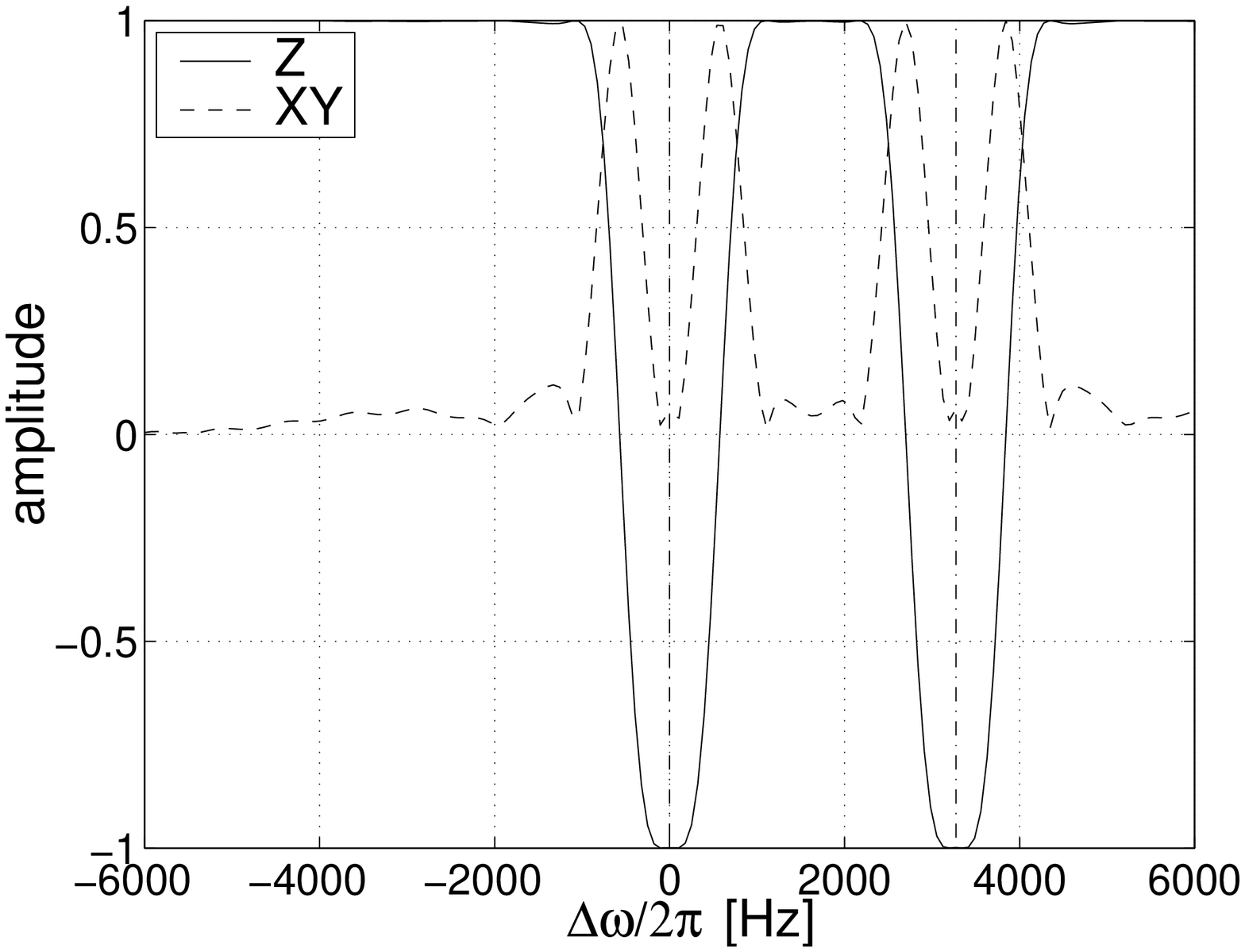} 
\end{center}
\vspace*{-2.0ex}
\caption{Similar to Fig.~\ref{fig:simwithoutbs} but with 
frequency shift correction.}
\label{fig:simwithbs}
\end{figure}

%%%%%%%%%%%%%%%%%%%%%%%%%%%%%%%%%%%%%%%%%%%%%%%%%%%%%%%%%%%%%%%%%%%%%%%%%%%%%%
\subsection{Composite pulses}
\label{sec:composite_pulses}

Another practical method for compensating systematic control errors in
NMR experiments is the application of a sequence of pulses instead of
a single pulse.  This method of {\em composite} pulses arises from the
observation that concatenation of several pulses can produce more
accurate rotations than is possible using just a single pulse, due to
strategic cancellation of systematic errors and other unwanted
systematic effects.  Composite pulses work particularly well for
compensating errors arising from the RF field inhomogeneity, frequency
offsets, imperfect pulse length calibration, and other instrumental
artifacts introduced in Section~\ref{sec:instrum_errors}.  They
leverage the ability to control one parameter precisely to compensate
for the inability to control another parameter well.  We describe
two approaches to construction of composite pulses: an analytical
method, and one employing numerical optimization.

%%%%%%%%%%%%%%%%%%%%%%%%%%%%%%%%%%%%%%%%
\subsubsection{Analytical approach}

The three parameters which characterize a hard pulse are its frequency
offset $\Delta \omega$, phase $\phi$, and area $\hbar \gamma B_1
t_{pw}$, given by the product of the pulse amplitude $B_1$ and pulse
duration $t_{pw}$ (Section~\ref{sec:rf}).  In terms of qubit
operations, errors in these parameters translate directly into errors
in the axis $\hat n$ and angle $\theta$ of rotation, such that the
actual operation applied is not the ideal $R_{\hat n}(\theta)$ of
Eq.(\ref{eq:rotdef}), but rather,
\be
	\tilde{R}_{\hat n}(\theta) 
	= \exp \lbL{ - i \frac{f(\theta,\hat{n})\cdot \vec{\sigma}}{2}}\rb
\,,
\label{eq:roterr}
\ee
where $f(\theta,\hat{n})$ is a function which characterizes the
systematic error.  For example, under and over-rotation errors caused
by pulse amplitude miscalibration or RF field inhomogeneity may be
described by $f(\theta,\hat{n}) = \theta(1+\epsilon) \hat{n}$, while
RF phase errors may be described by
$f(\theta,\hat{n}) = \theta [\hat{n}_x \cos\epsilon + \hat{n}_y \sin\epsilon,
\hat{n}_y \cos\epsilon - \hat{n}_x \sin\epsilon,\hat{n}_z]$, where
$\epsilon$ is a fixed, but unknown parameter.  The essence of the
composite pulses technique is that a number of erroneous operations are
concatenated, varying $\hat n$ and $\theta$, to obtain a final
operation which is as independent of $\epsilon$ as possible.  This is
done without knowing $\epsilon$.  

This technique can be illustrated by considering the specific case of
linear amplitude errors, in which
\be
	\tilde{R}_{\hat n}(\theta) 
	= \exp \lbL{ - \frac{i \theta (1 + \epsilon) 
				\hat{n}\cdot \vec{\sigma}}{2}}\rb 
\,.
\ee
Let the goal be to obtain $R_x(\pi/2)$.  Using as a measure of error
the average gate fidelity, defined in Eq.(\ref{eq:gatefavg}),
we find that $\bar{F}(R_x(\pi/2),\tilde{R}_x(\pi/2)) =
(2+\cos(\epsilon \pi/2))/3 \approx 1 - \pi^2 \epsilon^2/24$, so the
error is quadratic in $\epsilon$ for small $\epsilon$.  Consider, in
contrast, the sequence
\be
	BB1_{\theta} = 
		   \tilde{R}_{\phi}(\pi)
		   \tilde{R}_{3\phi}(2\pi)
		   \tilde{R}_{\phi}(\pi)
               \tilde{R}_x(\theta) 
\,,
\label{eq:bb1def}
\ee
where $\tilde{R}_{\phi}(\cdot)$ denotes a rotation about the axis
$[\cos\phi,\sin\phi,0]$, and the choice $\phi =
\cos^{-1}(-\theta/4\pi)$ is made.  This sequence gives average gate fidelity
$\bar{F}(R_x(\pi/2),BB1_{90}) \approx 1-21 \pi^6 \epsilon^6/16384$,
which is much better than for the single pulse, even for relatively
large values of $\epsilon$, as shown in
Figure~\ref{fig:plotbb1-distance}.
The operation of the $BB1_{90}$ sequence is illustrated graphically in 
Fig.~\ref{fig:plotbb1}.

\begin{figure}[htbp]
\begin{center}
\includegraphics*[width=6cm]{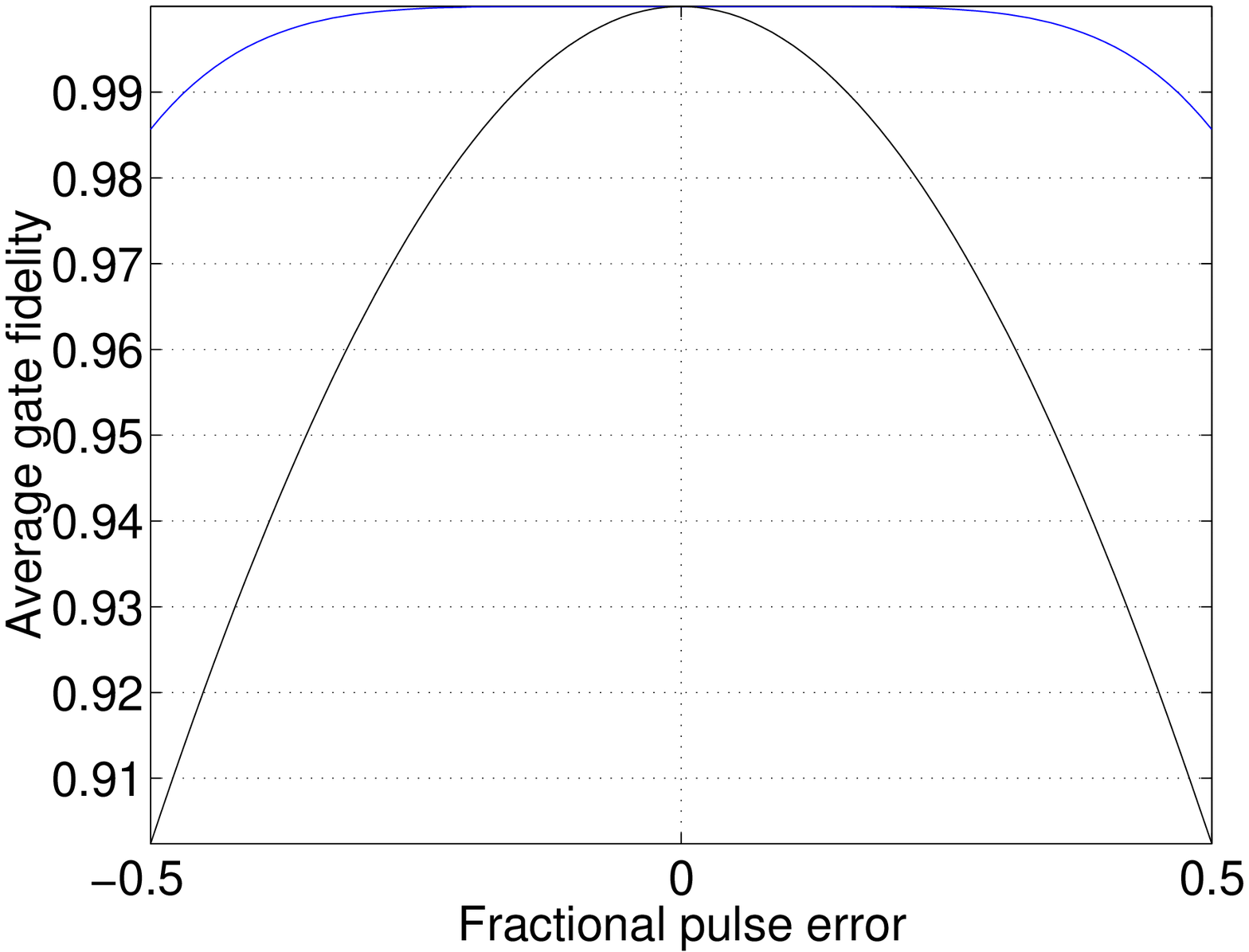} 
\end{center}
\caption{Plot of the average gate fidelity between the ideal $R_x(\pi/2)$ and
    actual unitary transforms $\tilde{R}_x(\pi/2)$ (black line) and
	between the ideal $R_x(\pi/2)$ and the composite sequence
	$BB1_{90}$ (blue line), as a function of the fraction of
	over-rotation error $\epsilon$.  Note how much higher fidelity
	the $BB1$ sequence has (the best possible fidelity is $1$),
	over a wide range of errors.}
\label{fig:plotbb1-distance}
\end{figure}

\begin{figure}[htbp]
\begin{center}
\includegraphics*[width=4cm]{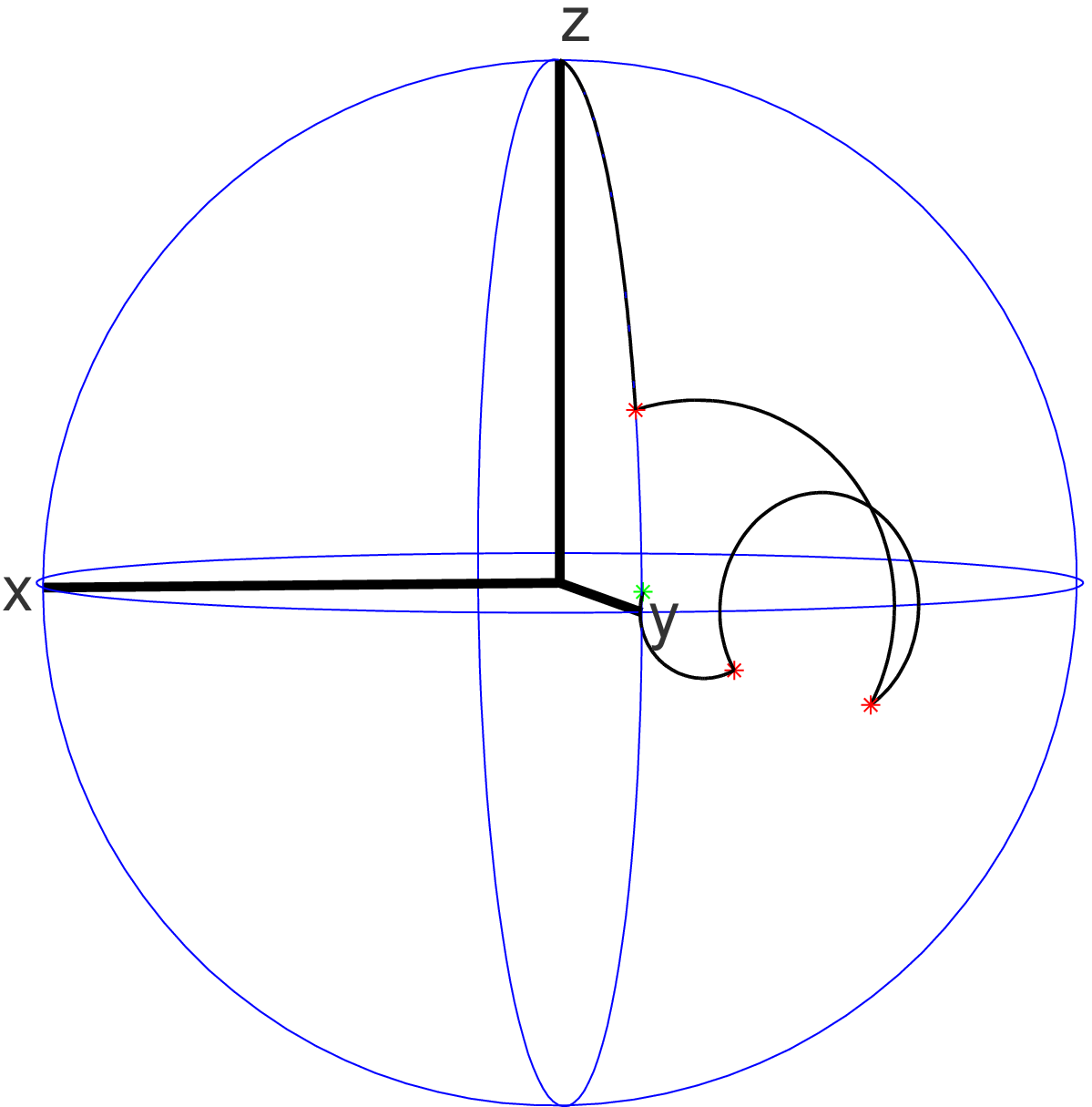} \\
\includegraphics*[width=4cm]{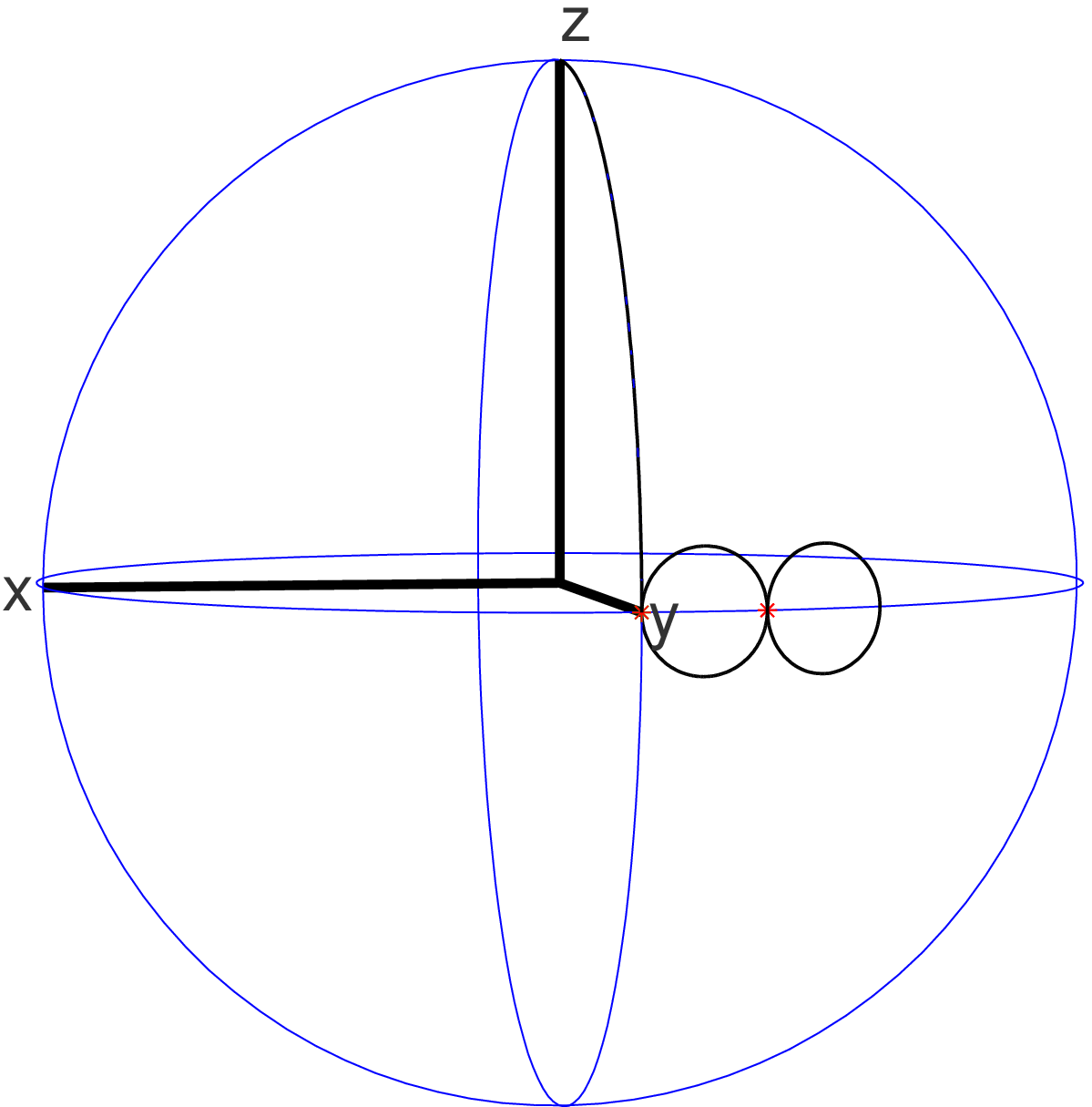} \\
\includegraphics*[width=4cm]{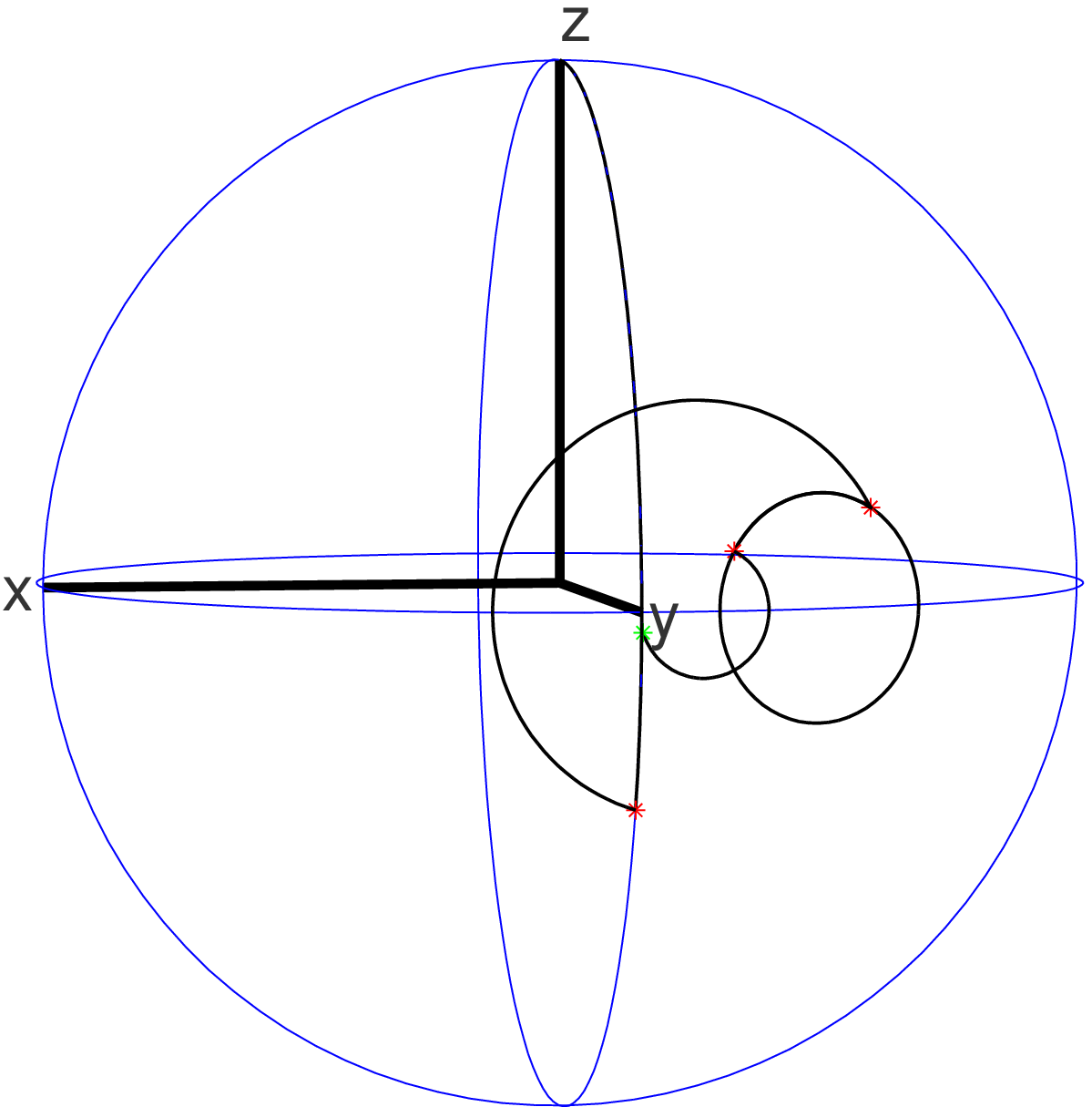} 
\end{center}
\caption{Illustration of the trajectories of a spin as it transforms
    under the $BB1$ pulse sequence of Eq.(\ref{eq:bb1def}), starting
    initially in the $|0\>$ state. Three trajectories are shown, in
    which the error is $50\%$ under-rotation (top), zero (middle), and
    $50\%$ over-rotation (bottom).  Plotted symbols denote the
    endpoints of each pulse in the sequences.
}
\label{fig:plotbb1}
\end{figure}

A few comments about this result are in order.  This result is the
best which has been presented in the literature
to-date~\cite{Cummins00a,Jones03b,Wimperis94a}; currently, no pulse
sequence which cancels out errors to higher order (for all possible
initial states) has yet been published.  It is also fairly general;
$BB1_{\theta}$ approximates $R_x(\theta)$.  Also, while composite
pulses have been widely studied and employed in the art of NMR, this
sequence is special in that it is {\em universal} (also termed
fully-compensating or class $A$): the amount of error cancellation is
independent of the starting state of the
spin~\cite{Tycko83a,Tycko85a}.  Other examples of such universal
composite pulses are the sequence
\be
	\tilde{R}_{60}(180) \tilde{R}_{300}(180) \tilde{R}_{60}(180)
\,,
\ee
which performs a $X^2$ rotation with compensation for pulse length
errors, and  
\be
	\tilde{R}_y(385) \tilde{R}_y(-320) \tilde{R}_y(25)
\,,
\ee
which performs a $Y$ rotation compensating for off-resonance errors
and to some extent for pulse length errors as well.

Earlier, in the
original work which introduced the concept of composite pulses into
NMR~\cite{Levitt79a,Levitt86a}, only limited pulse sequences were
known, which only work for particular initial states; for example,
there is the common $\tilde{R}_x(\pi/2)
\tilde{R}_{-y}(\pi) \tilde{R}_x(\pi/2)$, used to
approximate $R_x(\pi)$.  Figure~\ref{fig:compos180} illustrates
how this simple sequence removes the effect of errors in either the
rotation angle or the rotation axis.

\begin{figure}[htbp]
\begin{center}
\includegraphics*[width=4cm]{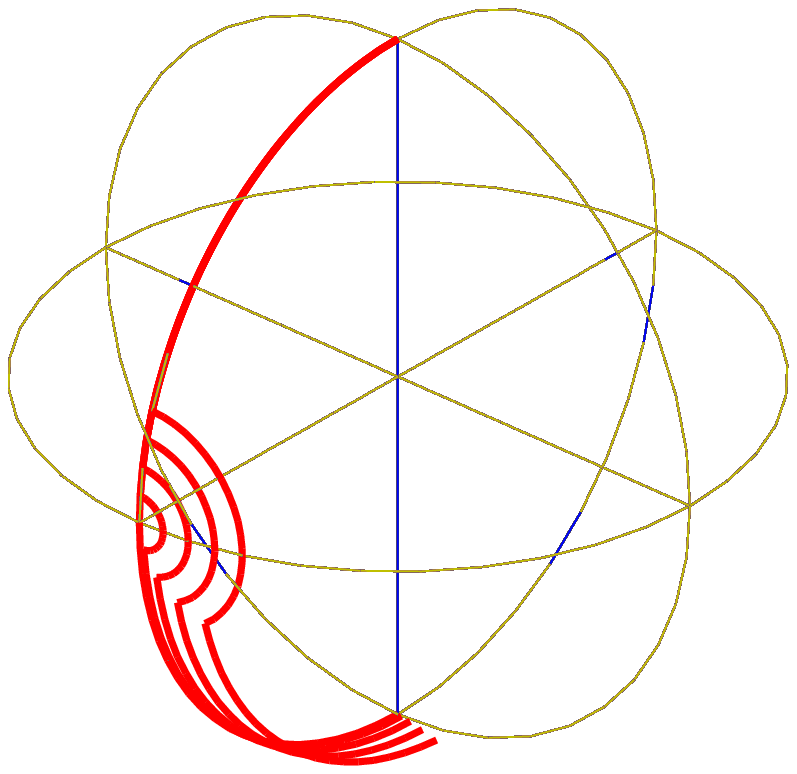}
\includegraphics*[width=4cm]{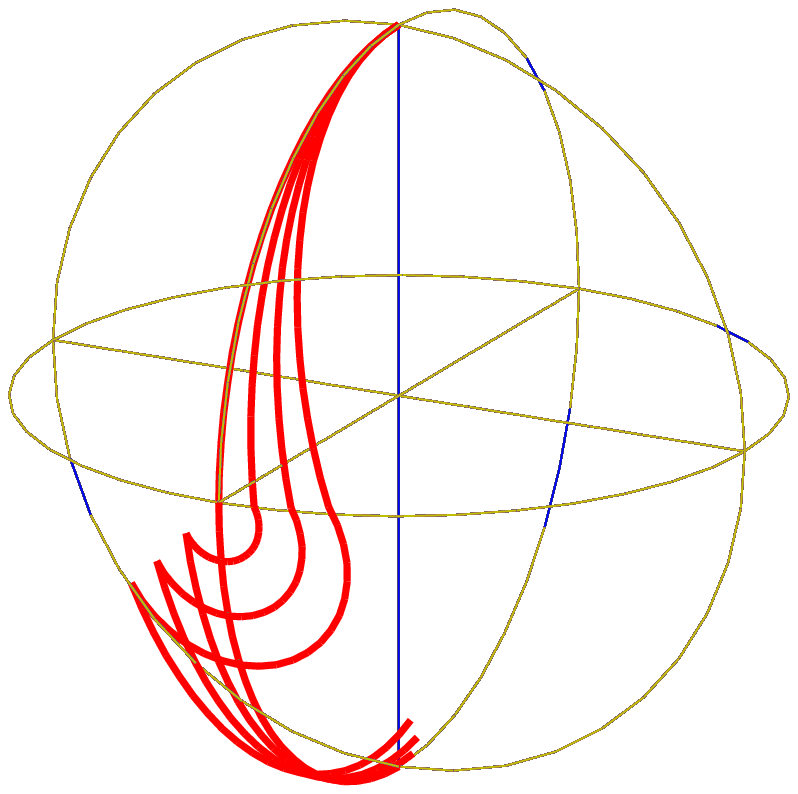}
\end{center}
\caption{Trajectory in the Bloch sphere described by a qubit 
initially in $\ket{0}$, when a composite $180^\circ$ rotation is
applied, consisting of three imperfect rotations, 
$\tilde{R}_x(\pi/2) \tilde{R}_{-y}(\pi) \tilde{R}_x(\pi/2)$. 
(Left) The tip angles are set $0, 5, \ldots 20\%$ too short. 
(Right) The pulse is applied off-resonance, with $(\omega_0 -
\omega_{r\!f})/ \omega_1 = 0, 0.05, \ldots 0.20$. In both cases, the
effect of the errors in the individual pulses is largely removed by 
the composite pulse.}
\label{fig:compos180}
\end{figure}

Systematic errors in the coupling strengths can also be tackled using
composite rotations, in order to obtain accurate two-qubit gates.
This was shown explicitly for the case of Ising
couplings~\cite{Jones03a}.

Similar compensation of slowly-fluctuating errors can be achieved during 
a train of pulses, separated by time intervals of free evolution.
The simplest instance of such a pulse train uses only
$180^\circ$ pulses. Off-resonance effects in such pulses can be largely
reversed by properly choosing the phases of the pulses. For instance, 
and at first sight surprisingly, the errors from off-resonant pulses
$X^2 \bar{X^2}$ roughly add up, while they largely compensate each other 
in $X^2 X^2$. This cancellation can be appreciated via a simple Bloch
sphere picture (Fig.~\ref{fig:two180}). 
The remaining errors are further reduced for a properly
chosen train of four pulses, $X^2 X^2 \bar{X^2} \bar{X^2}$, which 
performs markedly better than 
$X^2 \bar{X^2} X^2 \bar{X^2}$~\cite{Levitt82a}. Further reduction of
the effect of off-resonance errors can be obtained by using even longer
trains of $180^\circ$ pulses~\cite{Levitt82a}.

\bfig
\bcen
\vspace*{1ex}
\includegraphics*[width=4cm]{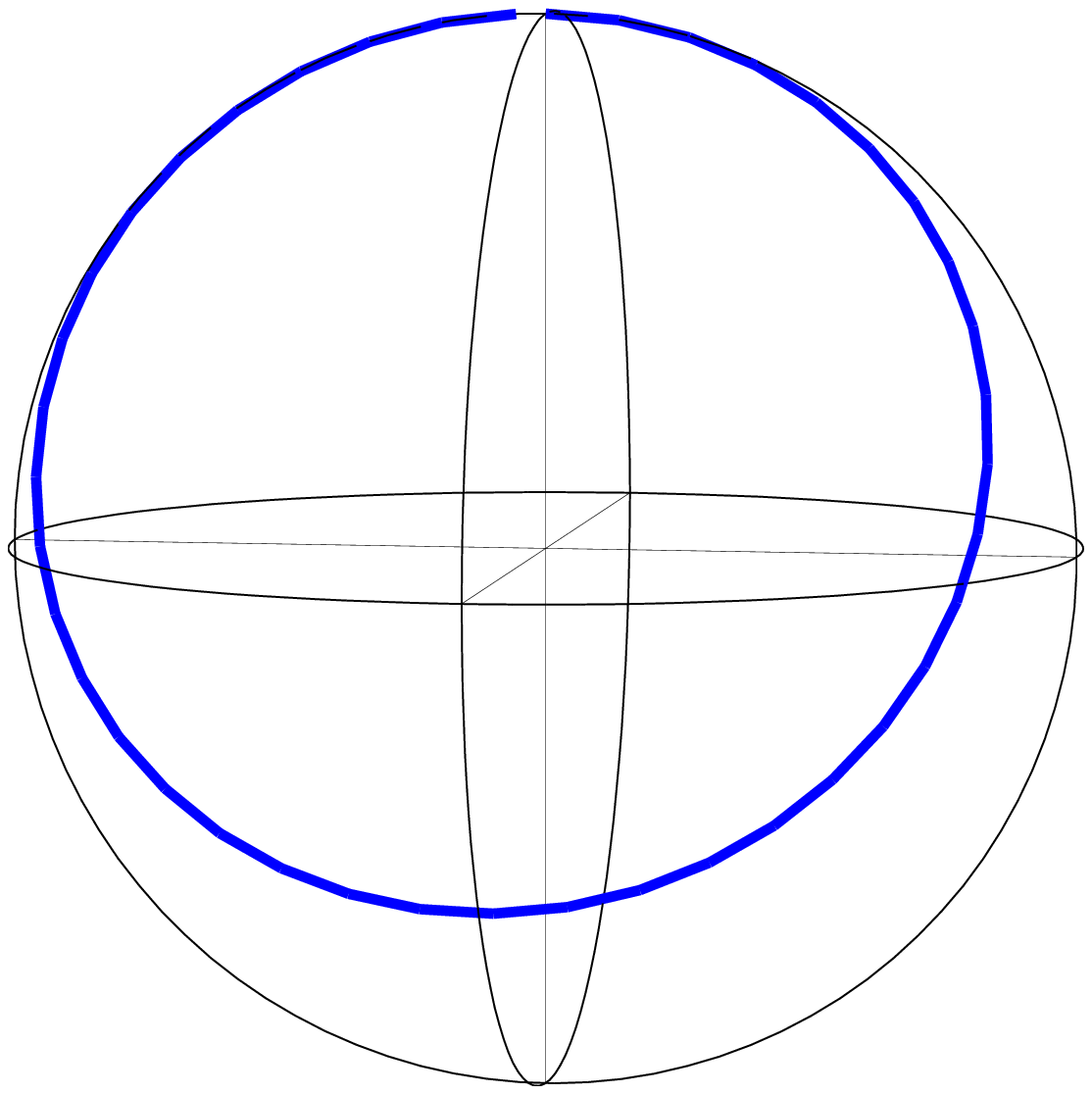} 
\includegraphics*[width=4cm]{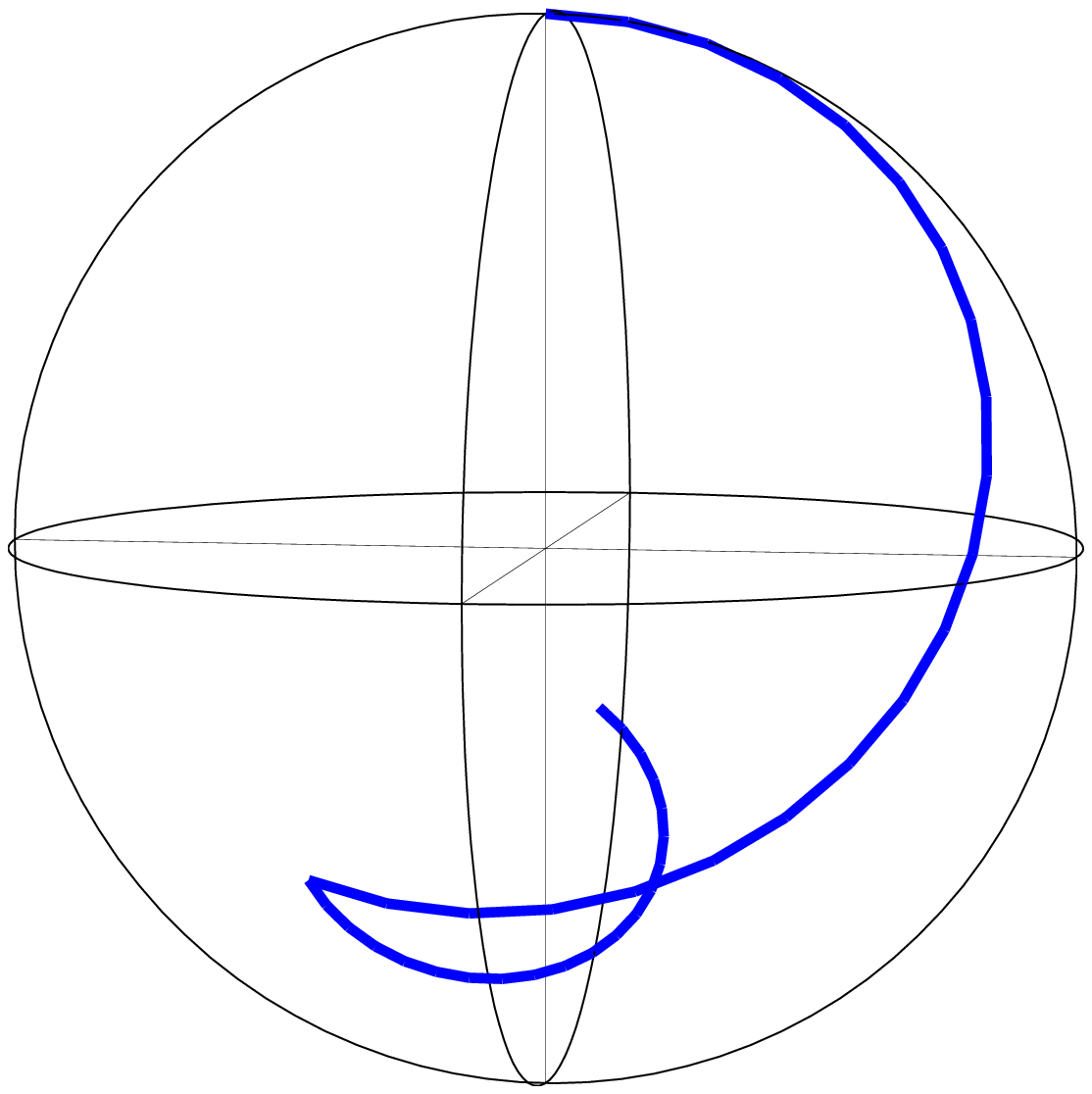} 
\vspace*{-2ex}
\ecen
\vspace*{-1.0ex}
\caption{Trajectory in the Bloch sphere of a qubit initially in 
$\ket{0}$, subject to two consecutive $180^\circ$ pulses, applied
off-resonance with $(\omega_0 - \omega_{rf} / \omega_1 = 0.5$.
(Left) 
If the two pulses are applied with the same phase ($X^2 X^2$), the
qubit is taken simply along a circular trajectory through $\ket{0}$, 
and reaches a point near $\ket{0}$; to be precise, the 
$50\%$ resonance offset makes the rotation angle
$\sqrt{(2^2 + 1^1)/2^2} = \sqrt{5/4}$ larger than $360^\circ$.
(Right) 
In contrast, if the two pulses are applied with opposite phases 
($X^2 \bar{X}^2$), the qubit is left far from $\ket{0}$.}
\label{fig:two180}
\efig
    
Evidently, quantum computing sequences are not as transparent as just
a train of $180^\circ$ pulses. Surprisingly, even throughout a quantum
computing sequence, the effect of RF inhomogeneities can be removed to 
a large extent~\cite{Vandersypen00a}, as illustrated in
Fig.~\ref{fig:grover_iteration}. After completion of a routine involving
the equivalent of about 1350 $90^\circ$ pulses, the measured amplitude was 
about $15\%$ of the full amplitude. Without removal of the effect of RF 
inhomogeneity, the signal would have been buried in the noise very rapidly.

\bfig
\bcen
\vspace*{1ex}
\includegraphics*[width=6cm]{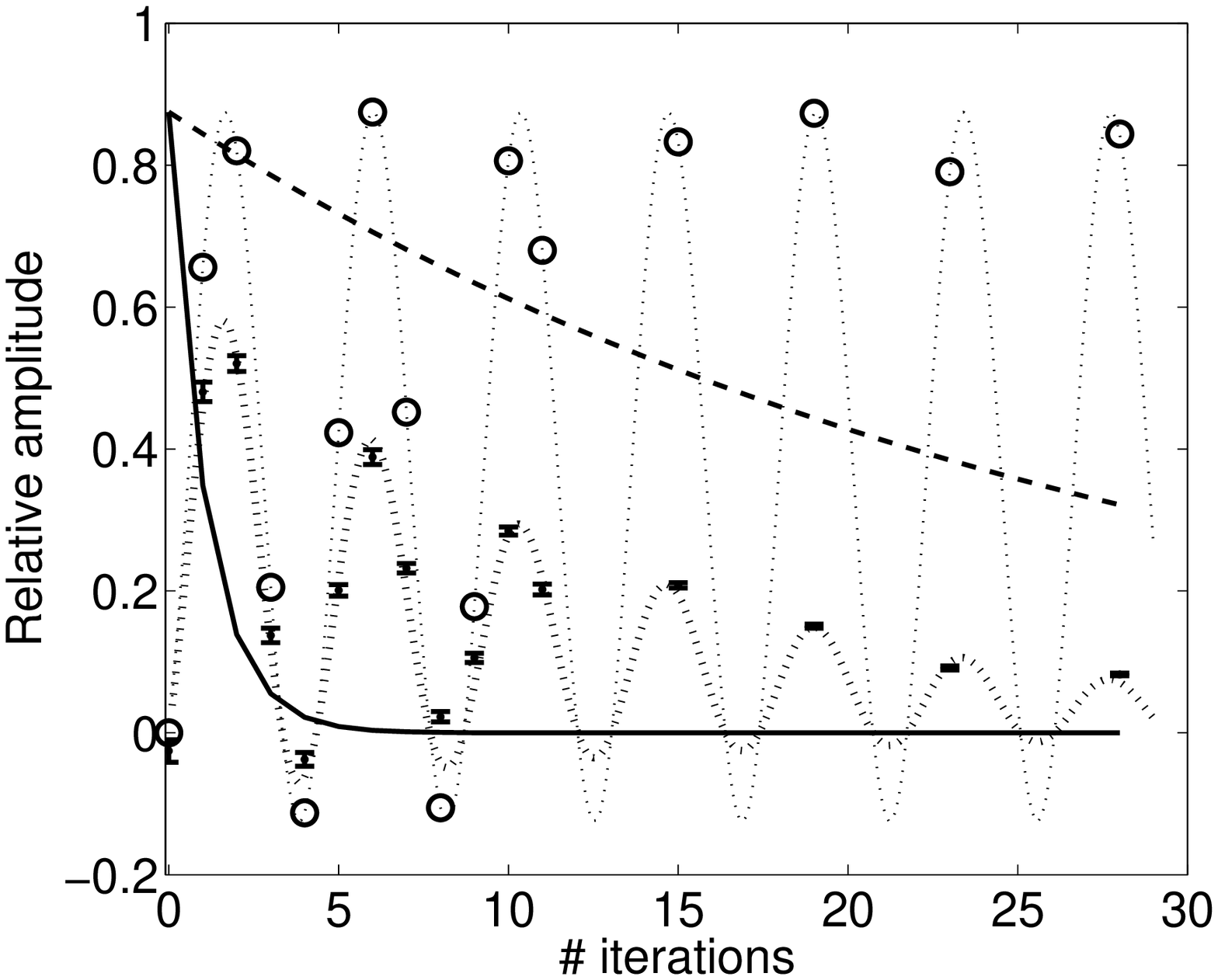} 
\vspace*{-2ex}
\ecen
\vspace*{-1.0ex}
\caption{Experimental (error bars) and ideal (circles) amplitude of
$\ket{\!\downarrow\uparrow\downarrow}$, as a function of the number of
iterations of a quantum search
algorithm~\cite{Nielsen00b,Vandersypen00a}, for three 
qubits, executed 
on $^{13}$CHFBr$_2$. Each 
iteration contains the equivalent of almost fifty $90^\circ$ pulses.
The dotted lines serve to guide the eye. Dashed line: the signal decay
for $^{13}$C due to decoherence, which represents a lower bound on the
decay rate. Solid line: the signal strength retained after applying a
continuous RF pulse of the same cumulative duration per search
iteration as the pulses in the actual experiment (averaged over 3
spins, measured up to 4 iterations and then extrapolated). Similar
observations have been reported in
Ref.~\protect\cite{Vandersypen99a}.}
\label{fig:grover_iteration}
\efig

This level of error cancellation was achieved partly due to a judicious
choice of the phases of the refocusing pulses.
Nevertheless, a more detailed description and understanding of the error
operators is needed in order to fully exploit the potential for error
cancellation in arbitrary pulse sequences.

%%%%%%%%%%%%%%%%%%%%%%%%%%%%%%%%%%%%%%%%
\subsubsection{Numerical optimization}
\label{sec:strongly_mod_pulses}

The composite pulses we discussed in the previous subsection are designed
to compensate for certain types of errors (mostly over- or underrotations
and frequency offsets), and work even when the exact Larmor frequencies,
spin-spin coupling strengths and the magnitude of the errors are unknown. 
This is the usual case in NMR spectroscopy. However, in quantum computing
experiments, detailed knowledge of the system Hamiltonian is usually 
available and can be used to tailor the composite pulses to the 
system specifics, taking the degree of quantum control one step further.

Following the notation of Ref.~\cite{Fortunato02a}, we
consider the concatenation of a number of rectangular pulses, each described
by four parameters: the pulse duration $\tau^m$, a constant amplitude
$\omega_1^m$, the transmitter frequency $\omega_{rf}^m$ and the
initial phase $\phi^m$, where $m$ indexes the pulse.
These parameters may be strongly modulated
from one pulse to the next\footnote{Jumps in the transmitter frequency 
can be conveniently realized with phase-ramping techniques; as discussed in 
Section~\ref{sec:phase_ramping}, this is done by phase shifting the raw RF 
excitation in fixed increments per time so a different RF frequency is 
obtained.} Via a numerical optimization procedure, the values of $\tau^m$, 
$\omega_1^m$, $\omega_{rf}^m$ and $\phi^m$ are chosen such that the 
resulting net unitary evolution $U_{net}$ is as close as possible to the 
ideal unitary transformation $U_{ideal}$, according to some fidelity 
measure (section~\ref{sec:fidelity}).

In practice, the number of time slices in the composite pulse is
increased starting from one, until a satisfactory solution is
found. While the fidelity function may have many local maxima and
finding the global maximum may therefore take a long time, suitable
algorithms such as the Nelder-Mead Simplex algorithm~\cite{Nelder65a} 
often succeed in
finding a reasonably good solution. Furthermore, the optimization
routine can incorporate penalties on high powers, large frequencies
and negative or very long time periods, in order to prevent the
algorithm from returning infeasible solutions.

Computation of $U_{net}$ uses the fact that the Hamiltonian during a
fixed-amplitude RF pulse can be made time independent by transforming
into a reference frame rotating at the transmitter frequency, as we
have seen in section~\ref{sec:rotating_frame}. We will call ${\cal
H}_{rot}^m$ the effective Hamiltonian in the frame rotating at
$\omega_{rf}^m$ during segment $m$.  Given that $\omega_{rf}^m$
may be different for every segment of the pulse, it is most
convenient to transform back to a common reference frame at the end of 
every time slice. This can be the frame of the raw RF frequency, or
the laboratory frame of the $n$-spin system. In the lab frame, the time 
evolution during segment $m$ is described by
\be
U^m = e^{-i\omega_{rf}^m \sum_{k=1}^{n} I_z^k \tau^m} 
e^{-i {\cal H}_{rot}^m \tau^m} \;.
\label{eq:cory_pulse}
\ee

Since all $U^m$ are expressed in the same reference frame, we can simply 
multiply them together to get $U_{net} = \prod_{m} U^m$, and compare the
result directly with $U_{ideal}$, expressed in the laboratory frame 
as well.

Two representative examples of composite pulses designed for 
spin-selective rotations in homonuclear 
spin systems are given in Fig.~\ref{fig:cory_waveform}.
The gate fidelity
(section~\ref{sec:gate_fidelity}) obtained with these two 
pulses is displayed in Fig.~\ref{fig:cory_profile}.
Naturally, the fidelity is close to
unity only near the resonance frequencies for which the gate was
designed to work.

\begin{figure}[htbp] 
\begin{center}
\includegraphics*[width=8cm]{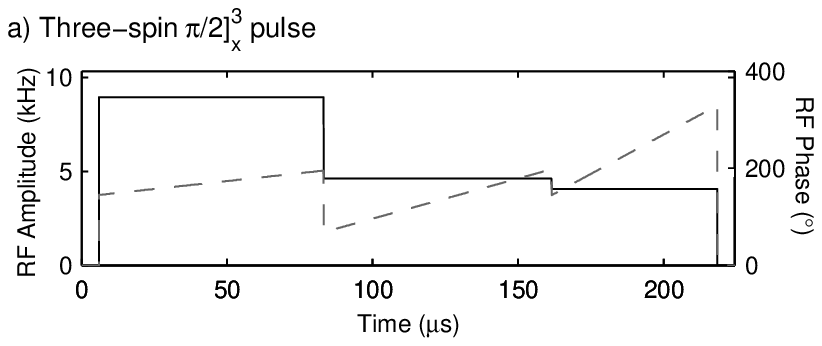} 
\includegraphics*[width=6cm]{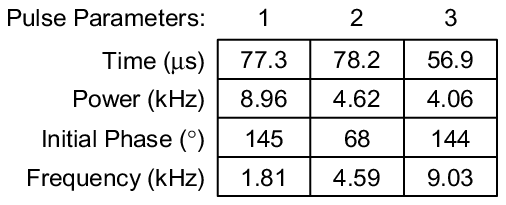} 
\includegraphics*[width=8cm]{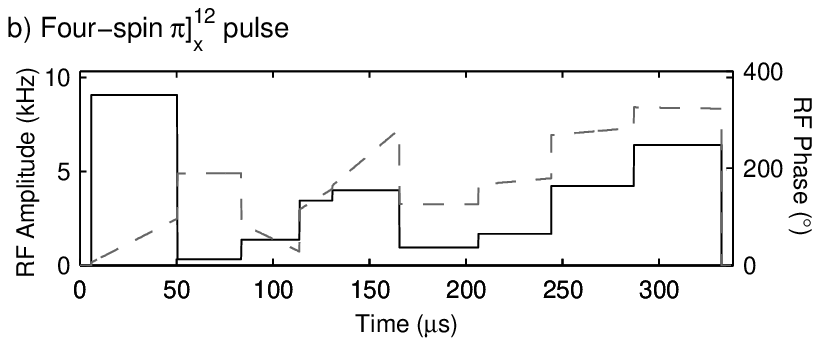} 
\includegraphics*[width=8cm]{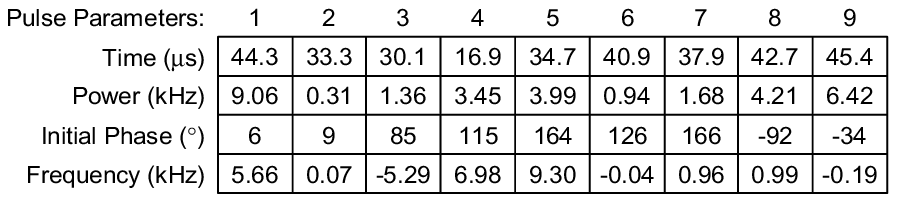} 
\end{center}
\caption{The ideal RF waveform for two example strongly modulated
pulses.  The solid (dashed) line is the amplitude (phase) of the
waveform. Details of the pulse parameters, as in
Eq.~\protect\ref{eq:cory_pulse}, are listed below each waveform. The 6
$\mu$s time interval with zero RF power before and after the composite
pulses is needed due to experimental implementation issues.  The
composite pulse in (a) performs a 90$^\circ$ rotation on one of the
$^{13}$C nuclei of $^{13}$C-labeled Alanine and the pulse in (b)
performs a simultaneous $180^\circ$ rotation on two $^{13}$C nuclei of
$^{13}$C-labeled Crotonic acid. Courtesy of D.G. Cory. Reproduced from
Ref.~\protect\cite{Fortunato02a}.}
\label{fig:cory_waveform}
\end{figure}

\begin{figure}[htbp]
\begin{center}
\includegraphics*[width=8cm]{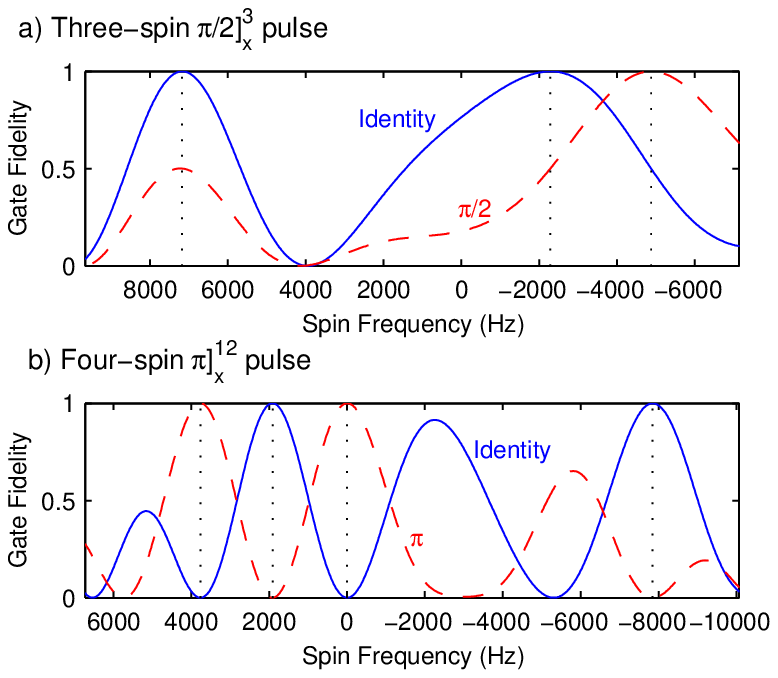} 
\end{center}
\vspace*{-2.0ex}
\caption{Gate fidelity of the two example pulses of
Fig.~\protect\ref{fig:cory_waveform} as the resonance frequency of a
test spin is varied. The solid (dashed) line is calculated with
identity (desired transformation) as the intended transformation. The
vertical dotted lines denote the actual chemical shifts for each
spin. Courtesy of D.G. Cory. Reproduced from
Ref.~\protect\cite{Fortunato02a}.}
\label{fig:cory_profile}
\end{figure}

Composite pulses can thus effectively generate accurate single- and
multiple-qubit Hamiltonians, using detailed knowledge of the system
Hamiltonian, and only limited knowledge about the errors.  Often,
however, full knowledge of the system parameters is not available, and
thus methods beyond composite pulses must be employed.

%%%%%%%%%%%%%%%%%%%%%%%%%%%%%%%%%%%%%%%%%%%%%%%%%%%%%%%%%%%%%%%%%%%%%%

\subsection{Average-Hamiltonian theory}
\label{sec:average_Ham}

The average-Hamiltonian formalism offers a versatile framework for
understanding how to effectively create or remove {\em arbitrary}
terms in the Hamiltonian by periodic perturbations, without requiring
full knowledge of the system dynamics.  The refocusing sequences
presented in section~\ref{sec:refocusing} and more general
multiple-pulse sequences designed to neutralize the effect of
dipole-dipole couplings can be explained within this framework.
Reduction of full dipole-dipole coupling given by
Eq.~\ref{eq:ham_dipole} to the simplified forms of
Eqs.~\ref{eq:ham_dipole2} and~\ref{eq:ham_dipole3} can also be
understood with average-Hamiltonian theory.

Following Ref.~\cite{Ernst87a}, we first introduce the Magnus 
expansion and then see how we can modify a time-independent Hamiltonian 
via a time-dependent perturbation. We use two concrete examples
to illustrate the concepts.

%%%%%%%%%%%%%%%%%%%%%%%%%%%%%%%%%%%

\subsubsection{The Magnus expansion}

The essence of average-Hamiltonian techniques is that the evolution $U(t)$ 
under a time-dependent Hamiltonian ${\cal H}(t)$ can be described by 
an effective evolution under a time-{\em in}\/dependent average Hamiltonian 
$\bar{\cal H}$, under two conditions~\cite{Haeberlen68a,Ernst87a}: (1) 
${\cal H}(t)$ is periodic and (2) the observation is stroboscopic and 
synchronized with the period $t_c$ of ${\cal H}(t)$.

We can then calculate $\bar{\cal H}$ exactly from 
\be
U(t_c) = \exp(-i \bar{\cal H} t_c) \;,
\label{eq:U_av_Ham}
\ee
by diagonalizing $U(t_c)$ and taking the logarithm of the resulting 
eigenvalues~\cite{Nielsen00b}.

In practice, it is often more convenient to compute $\bar{\cal H}$
approximately. Let us assume that ${\cal H}(t)$ is piecewise
constant (the analysis can be easily generalized to the case of
continuously changing Hamiltonians~\cite{Ernst87a}): 
${\cal H}(t)= {\cal H}_k$ for $\sum_0^{k-1} \tau_i < t <
\sum_0^k \tau_i$, and $t_c = \sum_0^n \tau_k$, so
\be
U(t_c) = \exp(-i {\cal H}_n \tau_n) \ldots \exp(-i {\cal H}_0 \tau_0) \,.
\label{eq:U_piecewise}
\ee
Repeated application of the Baker-Campbell-Hausdorff relation
\begin{eqnarray}
e^B e^A & = & \exp\left\{ A + B + \frac{1}{2}[B,A] \right. \nonumber\\
& & \left. + \frac{1}{12}\left([B,[B,A]] + [[B,A],A]\right) + \ldots \right\}
\end{eqnarray}
gives 
\be
\bar{\cal H} = 
\bar{\cal H}^{(0)} + \bar{\cal H}^{(1)} + \bar{\cal H}^{(2)} + \ldots \;,
\label{eq:magnus}
\ee
where
\begin{eqnarray}
\label{eq:magnus_discrete0}
\bar{\cal H}^{(0)} &=& \frac{1}{t_c} 
\left\{ {\cal H}_0 \tau_0 + \ldots + {\cal H}_n \tau_n \right\} \,,\\
\label{eq:magnus_discrete1}
\bar{\cal H}^{(1)} &=& \frac{-i}{2t_c} 
\left\{[{\cal H}_1 \tau_1,{\cal H}_0 \tau_0] \right. \nonumber\\
&&+ \left.[{\cal H}_2 \tau_2,{\cal H}_0 \tau_0] + 
          [{\cal H}_2 \tau_2,{\cal H}_1 \tau_1] + \ldots \right\} \;,
\end{eqnarray}
and so forth. 
This expansion, called the Magnus expansion~\cite{Magnus54a}, forms the basis 
of average-Hamiltonian theory.

%%%%%%%%%%%%%%%%%%%%%%%%%%%%%%%%%%%%%%%%
\subsubsection{Multiple-pulse decoupling}
\label{sec:mult_pulse_dec}

Let us consider a pulse sequence of $n$ infinitesimally short pulses $U_k$
separated by time intervals $\tau_k$ of free evolution under the system 
Hamiltonian ${\cal H}_0$, and such that $U_n \ldots U_2 U_1 = I$
(for pulses of finite length, the duration of the pulses must also be 
included in the average). The pulses correspond to basis transformations, 
and we can thus describe the system evolution via a sequence of time 
intervals $\tau_k$ of free evolution under the Hamiltonian 
$\tilde{\cal H}_0(k)$, with 
\begin{eqnarray}
\tilde{\cal H}_{0(0)} &=& {\cal H}_0 \,,
\label{eq:toggling_frame0}\\
\tilde{\cal H}_{0(1)} &=& U^{-1}_1 {\cal H}_0 U_1 \,,\\
\tilde{\cal H}_{0(2)} &=& U^{-1}_1 U_2^{-1} {\cal H}_0 U_2 U_1 \,,
\label{eq:toggling_frame2}
\end{eqnarray}
and so forth. Note that the order in which the $U_k$ are applied to
${\cal H}_0$ is reversed and that the $U_k$ themselves are reversed
as well. If we let $t_c = \sum_0^n \tau_k$, then the overall 
transformation $U(t_c)$ is given by
\be
U(t_c) = 
\exp(-i \tilde{\cal H}_{0(n)} \tau_n) \ldots 
\exp(-i \tilde{\cal H}_{0(0)} \tau_0) \,.
\ee
We can now use the Magnus expansion of Eq.~\ref{eq:magnus} and 
Eqs.~\ref{eq:magnus_discrete0}-\ref{eq:magnus_discrete1}, where we 
replace ${\cal H}_k$ by $\tilde{\cal H}_{0(k)}$, to obtain
the average Hamiltonian $\bar{\cal H}_0$ which describes the net
time evolution during $t_c$. The zeroth order average Hamiltonian 
is given by
\be
\bar{\cal H}_0^{(0)} = \frac{1}{t_c} \sum_{k=0}^n 
\tau_k U_1^{-1} \ldots U_k^{-1} {\cal H}_0 U_k \ldots U_1 \,.
\label{eq:zeroth_order}
\ee
The crux of average Hamiltonian theory is that by properly choosing
the pulse $U_k$, we can ensure that $\bar{\cal H}_0^{(0)}$ contain 
only the desired terms.

Sophisticated pulse sequences~\cite{Mehring83a} can also remove undesired 
contributions from the higher-order terms in the expansion, although this 
is generally harder since $\bar{\cal H}_0^{(1)}, \bar{\cal H}_0^{(2)} 
\ldots$ contain cross-terms between the various $\tilde{\cal H}_{0(k)}$. 
The commutators involved in the higher-order terms do become smaller for 
shorter cycle times, though, so fast cycles result in better averaging.

We also point out that pulse sequences which satisfy
\be
\tilde{\cal H}_{0(k)} = \tilde{\cal H}_{0(n-k)}
\ee
or equivalently
\be
U_{k+1} = U_{n-k}^\dagger
\ee
contain no contributions of odd orders to $\bar{\cal H}_0$,
\be
\bar{\cal H}_0^{(k)} = 0 \quad \mbox{for} \quad k=1,3,5,\ldots \,,
\ee
and thus perform significantly better than other sequences.

Let us now illustrate the operation of multiple-pulse decoupling via
two examples. First, the original multiple-pulse sequence for  
removal of dipole-dipole interactions is the {\sc wahuha}-4 
sequence~\cite{Waugh68a},
\be
%\tau \; X \tau \; \bar{Y} \; 2\tau \; Y \tau \bar{X} \tau \;,
\tau \bar{X} \; \tau Y \; 2\tau \; \bar{Y} \; \tau \; X \; \tau \;,
\ee
where the pulses are applied to all qubits involved, $\tau$ stands
for free evolution under the system Hamiltonian for a duration $\tau$, 
and the unitaries are ordered from right to left, as usual.
The pulses rotate the Zeeman terms in the Hamiltonian from $-\hat{z}$ to 
$-\hat{y}$, $-\hat{x}$, $-\hat{y}$, and back to $-\hat{z}$ (see 
Eqs.~\ref{eq:toggling_frame0}-\ref{eq:toggling_frame2}) for a duration 
$\tau$, $\tau$, $2\tau$, $\tau$ and $\tau$, respectively. The zeroth
order average Zeeman term is thus oriented along  
$-(\hat{x}+\hat{y}+\hat{z})$, with strength scaled down by a factor 
$1/\sqrt{3}$.  The dipolar Hamiltonian of Eq.~\ref{eq:ham_dipole2}
goes through the forms 
$[3 I_z^i I_z^j - I^i \cdot I^j]$, $[3 I_y^i I_y^j - I^i \cdot I^j]$ 
and $[3 I_x^i I_x^j - I^i \cdot I^j]$ for equal durations, and is thus zero
on average.

By selectively not pulsing specific qubits, it is also possible to 
reintroduce some of the couplings as desired. In Fig.~\ref{fig:refocusing3}, 
we already saw explicitly how to do this for $I_z^i I_z^j$ couplings.

A second example is an extension of the conventional spin-echo sequence 
(Section~\ref{sec:echo}) to three component spin-echoes\cite{Augustine97a}. In
conventional echo sequences, $180^\circ$ pulses about $\hat{x}$ or 
$\hat{y}$ remove the effect of a Hamiltonian of the form $c_z \sigma_z$.
Now we ask ourselves what sequence of pulses would freeze the evolution 
under a Hamiltonian of the form
\be
{\cal H} = c_x \sigma_x + c_y \sigma_y + c_z \sigma_z \,,
\ee 
where $c_x, c_y, c_z$ are arbitrary coefficients.
We can use Eq.~\ref{eq:zeroth_order} to verify that the 
sequence
\be
X^2 \; \tau \; \bar{X}^2 \; Y^2 \; \tau \; \bar{Y}^2 \; 
Z^2 \; \tau \; \bar{Z}^2 \; \tau \,,
\label{eq:3d_echo}
\ee
or equivalently, after simplification,
\be
X^2 \; \tau \; Z^{2} \; \tau \; X^{2} \; \tau \; \bar{Z}^2 \; \tau \,,
\ee
gives a zeroth order average Hamiltonian $\bar{\cal H}^{(0)} = 0$, and 
thus in effect corresponds to a three-component echo sequence.
Another way to show this is to note that
\begin{eqnarray}
X^2 \; {\cal H} \; \bar{X}^2 &=& 
   + c_x \sigma_x - c_y \sigma_y - c_z \sigma_z \,,\\
Y^2 \; {\cal H} \; \bar{Y}^2 &=& 
   - c_x \sigma_x + c_y \sigma_y - c_z \sigma_z \,,\\
Z^2 \; {\cal H} \; \bar{Z}^2 &=& 
   - c_x \sigma_x - c_y \sigma_y + c_z \sigma_z \,.
\end{eqnarray}
Clearly, ${\cal H} + X^2 \; {\cal H} \; \bar{X}^2 + 
Y^2 \; {\cal H} \; \bar{Y}^2 + Z^2 \; {\cal H} \; \bar{Z}^2 = 0$, and 
so the sequence of Eq.~\ref{eq:3d_echo}
gives, to zeroth order, no net evolution. Again, if $\tau$ is 
sufficiently short, the higher order contributions will be negligible.

%%%%%%%%%%%%%%%%%%%%%%%%%%%%%%%%%%%%%

\subsubsection{Reversing errors due to decoherence}
\label{sec:robust_decoherence}

Can we apply multiple-pulse sequences to reverse the effect of
interactions of a qubit with degrees of freedom in the environment? 
It is not clear a priori that this is possible: standard 
average-Hamiltonian theory assumes that we can manipulate both
interacting particles involved, for instance via RF pulses.
However, we have no control of degrees of freedom in the environment.

Remarkably, it is actually possible to remove the effect of unwanted
interactions with degrees of freedom in the environment, even when
applying operations to the system
only~\cite{Viola98a,Viola99a,Vitali99a,Duan99a}, provided the control
operations are applied faster than the fluctuations.  Knowledge about
the nature of the interactions can be applied to simplify the sequence
of decoupling operations, and such knowledge can even be experimentally
extracted~\cite{Byrd03a}, in part using a procedure known as {\em
process tomography}, described in Section~\ref{sec:process_tomo}.

If the fluctuations are faster than the accessible control operations,
errors can be corrected using quantum error
correction~\cite{Shor95a,Steane96a,Nielsen00b}, or they can prevented
by encoding the qubits in a subspace which is not affected by
decoherence~\cite{Lidar98a,Zanardi98a}. This is discussed further in
Section~\ref{sec:threshold}.

\section{Evaluation of quantum control}
\label{sec:evaluation}

The pulse control methods presented in the last two sections can have
impressive performance, but this is very much contingent on having an
accurate model of the system under control.  A variety of techniques
have been used in NMR to characterize the system dynamics and to 
evaluate the performance of control
sequences.  In this last section, we review some of these techniques,
beginning with a set of standard experiments to determine how quantum
a qubit system is, then proceeding to tomographic methods for fully
characterizing system dynamics, and concluding with fidelity metrics
for control, and implications these have for scalability to control
over large systems.

%%%%%%%%%%%%%%%%%%%%%%%%%%%%%%%%%%%%%%%%%%%%%%%%%%%%%%%%%%%%%%%%%%%%%%%%%%%%%
\subsection{Standard experiments}
\label{sec:standard}

In NMR spectroscopy as in atomic physics, a number of standard
experiments serve to test the quantum-mechanical behavior of a given
system, and to determine the extent of its isolation from the
environment (see Section~\ref{sec:decoherence}), in terms of its phase
coherence time $T_2$ and its energy relaxation time $T_1$, as well
as the decay time in the rotating frame $T_{1\rho}$.

%%%%%%%%%%%%%%%%%%%%%%%%%%%%%%%%%%%%%%%%
\subsubsection{Coherent oscillations driven by a resonant field}
\label{sec:rabi}

The dynamics of a single spin, driven resonantly by a coherent field,
were presented in sections~\ref{sec:rf} and~\ref{sec:rotating_frame}.
From Eq.~\ref{eq:spin_ideal_rot}, we have that in the ideal case
the RF field induces transitions from $\ket{0}$ to $\ket{1}$, where a 
qubit initially in $\ket{0}$ will be found in $\ket{1}$ after an RF 
pulse of duration $t_{pw}$ with probability
\be
   \mbox{Pr} [\ket{1}] = \sin^2 (\gamma B_1 t_{pw}/2) 
                       = \sin^2 (\omega_1 t_{pw}/2) 
\,,
\ee
The probability initially increases over time, until it reaches a
maximum $\mbox{Pr} [\ket{1}] = 1$ and then decreases again, by
stimulated emission, a cycle which keeps repeating itself.  

Such oscillations of a two-level quantum system driven by a resonant
field are known as Rabi oscillations~\cite{Rabi37a}, and the Rabi
frequency $\omega_1/2\pi$ is proportional to the amplitude of the
control field.  Observation of Rabi oscillations is usually accepted
as a signature of quantum coherent behavior.

In reality, the envelope of the Rabi oscillation signal is always
damped, due to decoherence as well as instrumental imperfections;
measurement of this decay time is useful, and known as a {\em
nutation} experiment.  In NMR, the Rabi decay time is often much 
shorter than the intrinsic phase randomization time constant $T_2$, due 
to the inhomogeneity of the RF field driving the Rabi oscillation,
across the macroscopic sample.  In other systems, 
the Rabi decay time may be longer than $T_2$, because (1) a long pulse 
can be seen as a concatenation of many $180^\circ$ pulses, which can 
have a refocusing effect (section~\ref{sec:refocusing}), and (2) the 
qubit is near $\pm \hat{z}$, where phase randomization has no effect, 
for roughly half the time during Rabi oscillations.

Coherent oscillations driven by a resonant field have been observed in
NMR and in many atomic systems for a long time.  Recently, however,
observations of such coherent dynamics have been made in other qubit
systems, including systems made from Josephson
junctions~\cite{Nakamura99a}, in molecular vibrational
states~\cite{Vala02a,Tesch02a}, and excitons in semiconductor quantum
dots~\cite{Stievater01a}.

%%%%%%%%%%%%%%%%%%%%%%%%%%%%%%%%%%%%%%%%
\subsubsection{Coherent oscillations initiated by a kick}
\label{sec:larmor}

A quantum system starting off in a state which is not an eigenstate of
the (static) system Hamiltonian, will precess about the quantization
axis of the system Hamiltonian, a motion known as Larmor precession
(e.g. Section~\ref{sec:singlespins}).  Such a situation could be realized 
by abruptly changing the system Hamiltonian, e.g. by suddingly applying 
a strong static field along $\hat{x}$ instead of along $\hat{z}$. 
Alternatively, and more realistically in NMR, Larmor precession can be
initiated by suddenly kicking the qubit out of the Hamiltonian eigenbasis. 
For a nuclear spin with Hamiltonian $- \hbar \omega_0 I_z$ 
(as in Eq.~\ref{eq:1spin_ham}), this is done by applying a $90^\circ$ RF 
pulse, causing a transition for instance from $\ket{0}$ to 
$(\ket{0} + \ket{1})/\sqrt{2}$, which initiates the time evolution
\be
	\ket{\psi(t)} = 
		\frac{e^{i \omega_0t/2} \ket{0} + 
		      e^{-i \omega_0t/2} \ket{1}}{\sqrt{2}} 
\,,
\ee
as illustrated in Fig.~\ref{fig:precession}. Like Rabi oscillations,
the observation of Larmor precession is also a signature of quantum
coherent behavior.
 
The Larmor precession is also damped, but in contrast to the Rabi
decay time, the Larmor decay time, termed $T_2^*$, is never longer
than $T_2$. Usually,
$T_2^* < T_2$; in particular, for NMR,
\be
	\frac{1}{T_2^*} = \frac{1}{T_2} + \frac{1}{T'_2}
\,,
\ee
where $T'_2$ is the dephasing time constant due to static magnetic field
inhomogeneities or other instrumental imperfections.

Larmor oscillations initiated by a kick have been observed recently in
a variety of systems, including those driven resonantly mentioned
earlier, and in addition a system of charges in coupled quantum
dots~\cite{Hayashi03a}.
The oscillations can be observed directly if the measurement basis
lies in the $\hat{x}-\hat{y}$ plane, as is the case in NMR. If the
measurement takes place along $\pm\hat{z}$, we must first change basis 
via a second $90^\circ$ pulse.

%%%%%%%%%%%%%%%%%%%%%%%%%%%%%%%%%%%%%%%%
\subsubsection{Ramsey Interferometry}
\label{sec:ramsey}

The double-pulse experiment 
\be X \; \tau \; X \;,
\ee
where time goes from right to left (as always, for unitary
transformations given in this article), and $\tau$ denotes a free
evolution period, under the evolution operation $e^{-i {\cal H}_{sys}
\tau / \hbar}$.  This is known as a Ramsey interference
experiment~\cite{Ramsey50a}.  Originally, this ``method of separated
oscillatory fields'' was applied to electronic states of molecular
beams traversing through two microwave excitation zones.  In NMR, two
pulses are involved, separated by a delay time $\tau$.  Ramsey
interference is most naturally described in the rotating frame of the
RF. If the qubit starts off along $\hat{z}$, the first $X$ pulse
rotates it to $-\hat{y}$.  Then the qubit precesses about $\hat{z}$
for a time $\tau$.  Finally the second $X$ pulse rotates the
$\pm\hat{y}$ component of the qubit state to $\mp\hat{z}$. Components
along $\pm\hat{x}$ at the end of the interval $\tau$ remain along
$\pm\hat{x}$ after the second $X$ pulse.

If only a single qubit is considered and the RF field is exactly 
on-resonance with the qubit precession, the qubit stays in place in 
the rotating frame during the time interval in between the two pulses, 
and the final state does not vary with $\tau$.
However, if the RF and the qubit are detuned in frequency by 
$\Delta \omega$, both the $\hat{x}$ and the $\hat{z}$ components of the
final state display a beating pattern as a function of 
$\Delta \omega \, \tau$, the so-called Ramsey fringes.
The decay time of the envelope of the Ramsey fringes is $T_2^*$, the 
same as that for Larmor precession.

For coupled qubits, the beating pattern contains information on the
coupling strengths. This fact forms the basis for two-dimensional 
correlation spectroscopy\cite{Jeener71a,Ernst87a}, a widely used range 
of two-pulse techniques for molecular structure determination.

%%%%%%%%%%%%%%%%%%%%%%%%%%%%%%%%%%%%%%%%
\subsubsection{Measurement of $T_2$}
\label{sec:echo}

The intrinsic $T_2$ time can be extracted in an experiment which 
is based on Larmor or Ramsey experiments. Certain imperfections
which cause the Ramsey or Larmor decay time $T_2^*$ to be smaller 
than $T_2$ can be removed by applying refocusing sequences.

The simplest instance of such a refocusing sequence consists of a
single $180^\circ$ pulse applied halfway during the time interval of 
free evolution initiated by an initial $90^\circ$ pulse. The entire 
sequence is thus
\be
%X - \frac{\tau}{2} - X^2 - \frac{\tau}{2} \;.
	\frac{\tau}{2} \; X^2 \; \frac{\tau}{2} \; X 
\,.
\ee
A second $X$ pulse should be added at 
the end if the measurement takes place in the $\pm\hat{z}$ basis.
In multi-spin systems, the pulses must be applied selectively
to one spin, in order to measure the $T_2$ of that spin.

The $X^2$ refocusing pulse removes not only simple 
scalar spin-spin couplings, as described in 
section~\ref{sec:refocusing}, but also undoes the effect of spatial
variations of the static magnetic field along $\hat{z}$. 
Such field inhomogeneities make spins in 
different regions of the sample progressively get out of phase with 
each other during the first time interval $\tau/2$. As a result, 
their magnetic moments cancel each other out and the NMR signal
vanishes. Provided the magnetic field variations are constant 
throughout the experiment, all the spins get exactly in-phase again 
(now along $+\hat{y}$)
by the end of the second time interval $\tau/2$, because of the 
$180^\circ$ refocusing pulse. As a result, the signal recovers, 
producing the well-known spin-echo. 
A generalization of this technique known as {\em three-component} 
refocusing (Section~\ref{sec:mult_pulse_dec}), undoes
effects from {\em any} static spin Hamiltonian terms.

The echo signal decays as a 
function of $\tau$, and the decay time constant is a measure of $T_2$.
However, terms in the Hamiltonian fluctuating on a timescale shorter 
than $\tau$ are not removed by a single refocusing pulse. Their effect
can still be removed if the fluctuations are slow compared to $\tau/n$
and a train of $n$ refocusing pulses is applied, each preceded and 
followed by a time interval $\tau/2n$ of free evolution. This so-called 
Carr-Purcell sequence~\cite{Carr54a},
\be
%\lbL
%X - \frac{\tau}{2n} - X^2 - \frac{\tau}{n} - \ldots X^2 - \frac{\tau}{n} 
%- X^2 - \frac{\tau}{2n} \rb \;,
\frac{\tau}{2n} \; X^2 \; \frac{\tau}{n} \; \ldots \; X^2 \; \frac{\tau}{n} 
\; X^2 \; \frac{\tau}{2n} \; X \;,
\ee
produces a first echo along $+\hat{y}$ after $\tau/n$, a second echo along
$-\hat{y}$ after $2\tau/n$, a third along $+\hat{y}$ after $3\tau/n$, and
so forth.
The magnitude of the echo signal decays exponentially throughout this
sequence and the echo signal left at the end of this sequence decreases
exponentially as a function of the total time $\tau$.  
To the extent that slow fluctuations in the Hamiltonian have been 
refocused, the decay time constant is the intrinsic $T_2$.

As we have seen in Section~\ref{sec:elem_pulse} of this review, small
but fixed errors in the pulse amplitude or duration may accumulate 
throughout a multiple-pulse sequence such as the Carr-Purcell sequence. 
However, if the phase of the refocusing pulses is shifted by $90^\circ$ 
with respect to the initial $90^\circ$ pulse, pulse length errors are 
compensated on even-numbered echoes and are thus not cumulative. In 
this sequence, 
\be
%X - \frac{\tau}{2n} - Y^2 - \frac{\tau}{n} \ldots - Y^2 - \frac{\tau}{n} 
%- Y^2 - \frac{\tau}{2n}\;,
\frac{\tau}{2n} \; Y^2 \; \frac{\tau}{n} \ldots Y^2 \; \frac{\tau}{n} 
\; Y^2 \; \frac{\tau}{2n} \; X\;,
\ee
known as the Carr-Purcell-Meiboom-Gill or CPMG 
sequence~\cite{Meiboom58a}, the echoes appear all along $-\hat{y}$.
Again, the decay time constant of the echo signal gives $T_2$.

Since $T_2$ indicates for how long a qubit can remain phase coherent,
it is usually called the coherence time, although the terms {\em phase
randomization time}, {\em dephasing time} and {\em transverse
relaxation time}, are also used. In NMR, $T_2$ is also known as the 
spin-spin relaxation time. In any case, $T_2$ is an important
number for evaluating the potential of quantum computers, as the ratio 
of $T_2$ over the typical duration of a quantum logic gate expresses 
the number of operations that can be completed coherently.

%%%%%%%%%%%%%%%%%%%%%%%%%%%%%%%%%%%%%%%%
\subsubsection{Measurement of $T_1$}
\label{sec:recovery}

Energy exchange with the environment makes a qubit which is out of
equilibrium gradually return to thermal equilibrium. In thermal
equilibrium, the qubit is in a statistical mixture of $\ket{0}$ and
$\ket{1}$, with probabilities set by the temperature and the energy
difference between $\ket{0}$ and $\ket{1}$. The time constant of this
equilibration process, $T_1$, is often called the energy relaxation
time, the longitudinal relaxation time or simply the relaxation time.
In NMR, $T_1$ is often termed the spin-lattice relaxation time.

Two standard experiments for measuring $T_1$ are inversion recovery
and saturation recovery. The sequence for the inversion recovery
experiment is 
\be
X \; \tau \; X^2 \;.
\ee 
The $180^\circ$ pulse
inverts the $\ket{0}$ and $\ket{1}$ probabilities, then during time
$\tau$, relaxation takes place, and finally a $90^\circ$ read-out
pulse is applied if necessary (i.e. when the measurement basis is in 
the $\hat{x}-\hat{y}$ plane).
In saturation recovery, a strong RF field is applied
for a long enough time such that it saturates the qubit transition and
equalizes the $\ket{0}$ and $\ket{1}$ probabilities. As in inversion
recovery, the original $\ket{0}$ and $\ket{1}$ populations are
altered, and we can monitor the populations return to their
equilibrium value as a function of $\tau$. The time constant of this
equilibration process is $T_1$.

Note that both the inversion recovery and saturation recovery
experiments bring the qubit out of equilibrium, but to a state which
has no coherence. As a result, phase randomization does not affect the
equilibration process --- we measure purely the effect of energy
exchange with the bath. In contrast, Ramsey and spin-echo experiments
for measuring $T_2^*$ and $T_2$ pick up contributions from phase
randomization both without and with energy exchange with the bath. If
energy exchange dominates phase randomization, the measured $T_2$ is
$2 T_1$.

The relevance of $T_1$ is twofold. First, it sets an upper bound for
$T_2$ and second, it tell us how much time we have to perform a
measurement in the $\{ \ket{0},\ket{1} \}$ basis. Phase randomization 
does not change the $\ket{0}$ and $\ket{1}$ probabilities, so $T_2$ 
is irrelevant during such a measurement. 
In many cases, $T_1 \gg T_2$, in which case we have more time to 
measure than to perform coherent operations.

%%%%%%%%%%%%%%%%%%%%%%%%
\subsubsection{Measurement of $T_{1\rho}$}
\label{sec:T1rho}

A third decay time useful to characterize the degree of isolation 
between a qubit and the environment is $T_{1\rho}$. This time constant
can be measured via a technique called {\em spin-locking}, where
the spin is first rotated into the $\hat{x}-\hat{y}$ plane, say
by a $Y$ pulse, and next continuous irradiation is applied, 
phase shifted by $90^\circ$ with respect to the pulse, so it is
aligned with the spin state (along the $\hat{x}$ axis):
\be
R_x(\mbox{continuous}) \; Y \;.
\ee
The continuous irradiation along $\hat{x}$ locks the spin to the
$\hat{x}$ axis, in the following sense. Whenever the spin starts to 
diverge from the $\hat{x}$ axis due to interactions with the
environment, the RF field rotates it to the opposite side of the 
$\hat{x}$ axis within a time $\pi/\omega_1$. Provided that the spin 
is still moving in the same direction after this time, it will thus 
return to the $\hat{x}$ axis (note that spin-locking thus also
inhibits evolution due to $J$-couplings and moderate frequency 
offsets). Only if the spin moves in the
opposite direction after $\pi/\omega_1$, it continues to depart
from the $\hat{x}$ axis due to the rotation by the RF field. So the 
amplitude along $\hat{x}$ decays, and the decay time constant 
is termed $T_{1\rho}$, 
known in NMR as the spin-lattice relaxation time in the rotating 
frame.

We see thus that, whereas $T_2$ is governed by low-frequency 
fluctuations in the environment and $T_1$ depends on fluctuations 
at the Larmor frequency, the decay during spin-locking arises from 
fluctuations at the Rabi frequency used during spin-locking.
The spin-locking experiment thus gives additional information
on the spectral density of the interactions with the environment.

%%%%%%%%%%%%%%%%%%%%%%%%%%%%%%%%%%%%%%%%%%%%%%%%%%%%%%%%%%%%%%%%%%%%%%%%%%%%%%
\subsection{Measurement of quantum states and gates}

The standard experiments presented in the previous section give only
partial information on the system dynamics. Here we show that in fact
the {\em full relaxation superoperator} can be determined
systematically by a procedure known as {\em process tomography}, which
builds upon {\em state tomography}, as described below.

%%%%%%%%%%%%%%%%%%%%%%%%%%%%%%%%%%%%%%%%
\subsubsection{Quantum state tomography}
\label{sec:state_tomo}

The density matrix $\rho$ completely describes our knowledge of 
the state of a system.
Measurement of the density matrix is therefore extremely helpful when 
testing or claiming the preparation of specific quantum states.

One-time measurement of each of $n$ qubits, in a given basis of $2^n$
states $\ket{m}$, gives only very little information on $\rho$.  All
that can be inferred from a measurement outcome $m$ is that
$\mbox{Pr}[\ket{m}] \neq 0$.

Repeated measurements of $n$ qubits, each time prepared in the same
state and measured in the same basis, reveals the probability 
distribution for the measurement basis states,
\be
	\mbox{Pr}[\ket{m}] = \bra{m} \rho \ket{m}
		= \mbox{Tr}(\rho \ket{m}\bra{m})= \mbox{Tr}(\rho M) 
\,,
\ee
where $M$ is an observable or measurement operator.  If we repeatedly 
measure each
qubit in the $\{\ket{0},\ket{1}\}$ basis, we thus obtain all the 
diagonal entries of $\rho$, $\rho_{ii}$.

Quantum state tomography~\cite{Chuang98b,Chuang98a,Chuang98c} is a
method which allows {\em all} the elements of the density
matrix $\rho$ to be determined.  This method consists of repeating the
measurement of the same state in various measurement bases, until all
the elements of $\rho$ can be determined, by solving a set of linear
equations. In practice, it is often more convenient to first rotate the 
qubits via a unitary transformation and then perform the measurement 
in a fixed basis. This is equivalent to measuring in different basis,
since
\be
	\mbox{Tr}\,[\rho (U M U^\dagger)] = 
		\mbox{Tr}\,[(U^\dagger \rho U) M] 
\,.
\ee

Specifically, we can expand the density matrix of a single qubit
$\rho$ as
\be
\left[ \begin{array}{cc} \rho_{00} & \rho_{01} \\ 
                         \rho_{10} & \rho_{11} \end{array} \right]
 = 
	\rho_{00} \ket{0}\bra{0} + \rho_{01} \ket{0}\bra{1}
	+ \rho_{10} \ket{1}\bra{0} + \rho_{11} \ket{1}\bra{1} 
\,.
\ee
Measurements of a single qubit in the $\{\ket{0},\ket{1}\}$ basis give
us $\rho_{00}$ and $\rho_{11}=1-\rho_{00}$. However, after changing
basis via a $90^\circ$ rotation about $\hat{x}$, transforming $\rho$
to $X \rho X^\dagger$, the measurement gives access to
$\mbox{Im}(\rho_{10}) = -\mbox{Im}(\rho_{01})$. Similarly, measurement 
after transformation by $Y$ reveals $\mbox{Re}(\rho_{10}) 
= \mbox{Re}(\rho_{01})$. Thus, by measuring the qubit
state first directly, then measuring the same state again after an $X$
read-out pulse, and then again after a $Y$ read-out pulse, we can
reconstruct $\rho$ completely.

Similarly, for $n$ qubits, we can expand $\rho$ as
\be
\rho = 
\sum_{i=0}^{2^n-1} \sum_{j=0}^{2^n-1} \rho_{ij} \ket{i}\bra{j} \,,
\label{eq:expand_rho_n}
\ee
and choose a set of basis changes which gives access to all $4^n-1$
degrees of freedom in $\rho$.

However, it is much easier to find a suitable set of basis changes 
if we use the Pauli expansion of $\rho$ instead of 
Eq.~\ref{eq:expand_rho_n}. The Pauli expansion for a single-qubit state is
\be
\rho = c_0 \sigma_0 + c_1 \sigma_1 + c_2 \sigma_2 + c_3 \sigma_3 \,,
\label{eq:pauli_1qubit}
\ee
where $c_0=1$ for normalization, and we use $\sigma_0=I/2$, 
$\sigma_1=\sigma_x/2$, $\sigma_2=\sigma_y/2$, $\sigma_3=\sigma_z/2$. 
Measurement in the computational basis, described by the observables
$\sigma_0 \pm \sigma_3$, gives us $\mbox{Pr}(\ket{0}) = (c_0 + c_3)/2$, 
and $\mbox{Pr}(\ket{1}) = (c_0 - c_3)/2$ so we can extract $c_3$.
Since
\begin{eqnarray}
X \rho X^\dagger &=& 
c_0 \sigma_0 + c_1 \sigma_1 - c_3 \sigma_2 + c_2 \sigma_3 \\
Y \rho Y^\dagger &=& 
c_0 \sigma_0 + c_3 \sigma_1 + c_2 \sigma_2 - c_1 \sigma_3 \,,
\end{eqnarray}
we indeed obtain $(c_0 \pm c_2)/2$ after applying $X$ and 
$(c_0 \mp c_1)/2$ after using $Y$.

For $n$ qubits, Eq.~\ref{eq:pauli_1qubit} generalizes to
\be
\rho = 
\sum_{i=0}^3 \sum_{j=0}^3 \ldots \sum_{k=0}^3 \; c_{ij \ldots k} \;
\sigma_i \otimes \sigma_j \otimes \ldots \sigma_k \,,
\label{eq:pauli_nqubits}
\ee
where $c_{00\ldots0} = 1$. Measurement in the computational basis is
described by observables of the form
\be
(\sigma_0 \pm \sigma_3) \otimes (\sigma_0 \pm \sigma_3) \otimes
\ldots (\sigma_0 \pm \sigma_3) \;,
\ee
and returns the probabilities
\be
\sum_{i,j,\ldots k \in \{0,3\}} \pm \, \frac{c_{ij\ldots k}}{2^n}  \,.
\ee
For example, for two qubits, these are
\begin{eqnarray}
\mbox{Pr}(\ket{00}) &=& (c_{00} + c_{03} + c_{30} + c_{33})/4 
\label{eq:Pr00}\\
\mbox{Pr}(\ket{01}) &=& (c_{00} - c_{03} + c_{30} - c_{33})/4 \\
\mbox{Pr}(\ket{10}) &=& (c_{00} + c_{03} - c_{30} - c_{33})/4 \\
\mbox{Pr}(\ket{11}) &=& (c_{00} - c_{03} - c_{30} + c_{33})/4 \,.
\label{eq:Pr11}
\end{eqnarray}
After measurement of the four $\mbox{Pr}[\ket{m}]$, we can solve 
for $c_{03}, c_{30}, c_{33}$ from this overdetermined set of linear
equations. Again, we can determine the other $c_{ij\ldots k}$
by transformation of the corresponding $\sigma_{ij\ldots k}$ to 
an observable, for instance
\be
X_1 Y_2 (\sigma_0 + \sigma_2) \otimes (\sigma_0 + \sigma_1) 
X_1^\dagger Y_2^\dagger
=
(\sigma_0 + \sigma_3) \otimes (\sigma_0 - \sigma_3) 
\;.
\ee

We end this discussion of state tomography with three additional 
comments:

First, in order to obtain all the basis state probabilities such as in
Eqs.~\ref{eq:Pr00}-\ref{eq:Pr11}, we must each time read out all the
qubits. If it is only possible to read out any one single qubit in
each experiment, we obtain $n$ bit-wise probabilities instead of $2^n$
basis state probabilities, giving spin-spin correlations. The
measurement operators are then of the form
\be
2^{n-1} [\sigma_0 \otimes \sigma_0 \otimes (\sigma_0 \pm \sigma_3) 
\otimes \ldots \otimes \sigma_0]
\ee
and we measure probabilities
\be
\frac{1}{2} (c_{0\ldots 000\ldots 0} \pm c_{0\ldots 030\ldots 0})
\ee
It is now no longer possible to rotate arbitrary components of $\rho$ 
into observable positions using just single-qubit rotations. Two-qubit 
gates are necessary to obtain all $c_{ij\ldots k}$.

Second, the measurement basis need obviously not be the computational 
basis.
In NMR experiments, for instance, the single-qubit measurement operator 
can be written as $-i \sigma_1 - \sigma_2$. For two coupled spins, the 
measurement operators are
\begin{eqnarray}
2(-i \sigma_1 - \sigma_2) &\otimes & (\sigma_0 \pm \sigma_3) \\
(\sigma_0 \pm \sigma_3) &\otimes & 2(-i \sigma_1 - \sigma_2) \,,
\label{eq:nmr_observable}
\end{eqnarray}
and so forth. Since NMR experiments are normally performed on a large
ensemble of molecules, the expectation value of the observables
can be read out by acquiring a single spectrum. The four operators in 
Eq.~\ref{eq:nmr_observable} correspond to the four lines in the 
spectrum of a two-spin system (two doublets). Phase sensitive 
detection permits us to separately record the real and imaginary 
component of each spectral line and distinguish $\sigma_1$ and
$\sigma_2$ contributions. 

Third, errors in the gates used for changing basis during
state tomography, lead to a measured density matrix which differs 
from the actual state of the system. If the errors are known and can be 
modeled accurately, they can be incorporated in the state tomography 
procedure and the actual state can nevertheless be determined 
accurately. 

Quantum state tomography has been experimentally implemented in many
atomic systems, notably the early work mapping out photon
states~\cite{Smithey93a} and vibrational cat states of trapped
atoms~\cite{Meekhof96a}.  Recently, it has become a common tool used to
evaluate NMR states~\cite{Chuang98b,Chuang98c}, states of optical
photon qubits~\cite{Thew02a}, and even vibrational states of
molecules~\cite{Skovsen03a}.

%%%%%%%%%%%%%%%%%%%%%%%%%%%%%%%%%%%%%%%%
\subsubsection{Quantum process tomography}
\label{sec:process_tomo}

Now that we know how to experimentally determine the state of a quantum 
system, it is only a short step to the characterization of a quantum 
process, such as a quantum logic gate, communication channel, storage 
device and so forth. In general, let us consider a quantum mechanical 
black box whose input may be an arbitrary quantum state, and whose
output is the result of the internal dynamics of the black box, as well 
as interactions to the outside world. Then can we ascertain the transfer 
function of this black box?

The answer is yes~\cite{Chuang97a,Poyatos97a,Dariano01a,Boulant02a}. 
The outline of the procedure is to first determine the output state
of the black box for a set of input states which form a basis for the 
system Hilbert space, and then to use the fact that quantum mechanics
is linear to compute the entire transfer function from this finite 
set of input-output pairs. 

An arbitrary quantum state transformation is a linear map ${\cal E}$,
\be
	\rho \mapsto \frac{{\cal E}(\rho)}{\mbox{Tr}({\cal E}(\rho))} 
\,,
\ee
where we can express ${\cal E}(\rho)$ in the operator-sum 
representation or Kraus representation~\cite{Kraus83a,Nielsen00b}
(an alternative to the superoperator formalism widely used in 
NMR~\cite{Ernst87a}),
\be
	{\cal E}(\rho) = \sum_k A_k \rho A_k^\dagger 
\,.
\label{eq:opsumrep}
\ee
The $A_k$ are operators acting on the system alone, yet ${\cal E}$ 
completely describes the possible state changes of the system, 
including unitary operations, generalized measurements and decoherence 
(for trace-preserving processes, $\sum_i A_k^\dagger A_k = 1$). The 
expansion of Eq.~\ref{eq:opsumrep} is in general not unique. 
In fact, we can always describe ${\cal E}$ using a fixed set of
operators $\tilde{A_k}$ which form a basis for the set of operators on 
the state space, so that~\cite{Chuang97a}
\be
	{\cal E}(\rho) = 
		\sum_{p,q} \chi_{pq} \tilde{A}_p \rho \tilde{A}_q^\dagger 
\,,
\label{eq:fixed_opsumrep}
\ee
where $\chi_{pq}$ is a positive Hermitian matrix. Since the 
$\tilde{A_k}$ are fixed, ${\cal E}$ is completely described by $\chi$. 
In general, $\chi$ will contain $16^n - 4^n$ independent real parameters, 
where $n$ is the number of qubits. 

In order to determine $\chi$ experimentally, we choose a basis of $4^n$ 
linearly independent density matrices $\rho_j$ which span the system 
Hilbert space, and determine ${\cal E}(\rho_j)$ for each $j$. 
We can then write down a set of linear equations of the form of 
Eq.~\ref{eq:fixed_opsumrep}, where we plug in the 
measurement outcomes ${\cal E}(\rho_j)$ and solve for the $\chi_{mn}$.

The most convenient choice for the $\rho_j$ depends on the
implementation of the qubits and on the observables, as was the case
for state tomography. Clearly, the effort needed to perform quantum
process tomography increases even more rapidly with the number of
qubits than quantum state tomography. This procedure has
experimentally been used only for one and two qubit
NMR~\cite{Childs01a,Boulant02a} and optical
photon~\cite{Mitchell03a,Altepeter03a} systems.

The operation elements $A_k$ in the operator sum representation of
Eq.~\ref{eq:opsumrep} can describe arbitrary quantum operations, but
among these a select subset are useful to identify.  For example, when
$A_k$ is a unitary matrix, that corresponds to perfect, closed system
Hamiltonian evolution.  Phase damping ($T_2$) is described by
\be
		A_0 = \mattwoc{1}{0}{0}{\sqrt{\gamma}}
~~~~~~~~	A_1 = \mattwoc{0}{0}{0}{\sqrt{1-\gamma}}
\,,
\ee
and amplitude damping ($T_1$) by
\be
		A_0 = \mattwoc{1}{0}{0}{\sqrt{\gamma}}
~~~~~~~~	A_1 = \mattwoc{0}{\sqrt{1-\gamma}}{0}{0}
\,,
\ee
where $\gamma \sim e^{-t/\tau}$ parameterizes the strength of the
damping, for time $t$, in terms of a time constant $\tau$.  These, and
other relaxation parameters~\cite{Nielsen00b}, can be obtained by
process tomography.

Such results can, in turn, be useful for approximate numerical
simulation of relaxation and decoherence processes in spin systems.
Phase damping and energy relaxation can be simulated in alternation
with unitary evolution under the system and control Hamiltonian,
taking sufficiently short time slices to obtain good approximation of
true dynamics.  This permits a $n$ spins system to be simulated using
$n 4^n$ steps, compared to $16^n$ for fully general quantum
operations.  Experimental results have shown this method to be
predictive of the system dynamics throughout sequences containing
hundreds of RF pulses~\cite{Vandersypen01a,Vandersypen01c}.

%%%%%%%%%%%%%%%%%%%%%%%%%%%%%%%%%%%%%%%%%%%%%%%%%%%%%%%%%%%%%%%%%%%%%%%%%%%%%%
\subsection{Fidelity of quantum states and gates}
\label{sec:fidelity}

The methods of the previous section give us full knowlegde of the system
state and dynamics, but sometimes a more succinct measure for comparing
theoretical expectations with experimental measurements is useful.
These are given by quantum state and gate fidelities.

%%%%%%%%%%%%%%%%%%%%%%%%%%%%%%%%%%%%%%%%
\subsubsection{Quantum state fidelity}
\label{sec:state_fidelity}

One elementary goal of quantum control is to create some pure state
$|\psi\>$.  However, suppose the final output is instead the pure
state $|\phi\>$.  Does $|\phi\>$ represent $|\psi\>$ with high
fidelity?  

Classically, the fidelity of two probability distributions $\{p_x\}$
and $\{q_x\}$ is given by $F(p_x,q_x) = \sum_x \sqrt{p_x q_x}$; when
they are equal, the fidelity is one.  The analogous quantum measure of
fidelity for two pure states $|\psi\>$ and $|\phi\>$ is
\be
	F(|\psi\>,|\phi\>) = |\<\phi|\psi\>|
\,,
\label{eq:pure_state_fidelity}
\ee
which is simply the absolute value of the overlap between the two states.

More generally, the output state of a control sequence is often
described by a density matrix $\rho$; this is useful because density
matrices can describe classical statistical mixtures of quantum
states, arising from decoherence processes, for example.  The fidelity
between a pure state $|\psi\>$ and a mixed state $\rho$ is
\be
	F(|\psi\>,\rho) = \sqrt{\<\psi|\rho|\psi\>}
\,,
\label{eq:pure_mixed_fidelity}
\ee
which reduces to Eq.(\ref{eq:pure_state_fidelity}) when $\rho =
|\phi\>\<\phi|$.  

The most general case is the fidelity between two density matrices,
$\rho$ and $\sigma$, which is defined as~\cite{Nielsen00b}
\be
	F(\sigma,\rho) \equiv {\rm Tr} \sqrt{\sqrt{\sigma} \rho \sqrt{\sigma}}
\,,
\ee
and despite the apparent asymmetry in this expression, it is actually
symmetric in $\rho$ and $\sigma$, and furthermore, reduces properly to
Eq.(\ref{eq:pure_mixed_fidelity}) when one density matrix is
pure.

Note that in the literature, sometimes the square of
Eq.(\ref{eq:pure_state_fidelity}) is defined as the
fidelity~\cite{Bowdrey02a}; that departs from the usual classical
definition for fidelity, but is convenient because
$F(|\psi\>,|\phi\>)^2$ can be interpreted as the probability that
a system in the state $|\phi\>$ is found to be in the state $|\psi\>$
when measured in the $\{\ket{\psi},\ket{\psi_\perp}\}$ basis.  Such
probabilities are meaningful in the accuracy thresholds discussed in
Section~\ref{sec:threshold}.

Other metrics for comparing two states have been used to quantify 
the relative error between theoretical and experimental states, such
as the simple two-norm~\cite{Vandersypen00a} and other
expressions~\cite{Fortunato02a}.  These were used
because the diagonal elements of the density matrix were suppressed;
such metrics are inferior to the fidelity measure, which should be
used when possible, due to its direct connection to quantum
information measures and fault tolerance theorems.

It is worthwhile to consider a specific example relating control
precision to state fidelity.  Suppose we desire $\ket{\psi} = |1\>$, 
but obtain $\ket{\phi} = R_y(\pi+\epsilon) |0\> = 
-\sin{\frac{\epsilon}{2}} \ket{0} + \cos{\frac{\epsilon}{2}} \ket{1}
\approx -{\epsilon}/{2} \ket{0} + (1 - {\epsilon^2}/{8}) \ket{1}$.
The resulting error probability is $1-|\<\phi|\psi\>|^2 = 
\epsilon^2/4$. This example makes the point that that
for small rotation angle errors $\epsilon$, the gate failure 
probability goes as $\epsilon^2$.

%%%%%%%%%%%%%%%%%%%%%%%%%%%%%%%%%%%%%%%%
\subsubsection{Quantum gate fidelity}
\label{sec:gate_fidelity}

A more complex goal of quantum control is to accomplish a desired
quantum operation.  Perhaps the most common scenario is one in which
the desired operation $U$ is a unitary transform on a single qubit,
whereas the actual transform accomplished is some quantum operation
${\cal E}$ (given in the operator sum representation).

A natural means to evaluate control precision is through the {\em
average gate fidelity}
\bea
	\bar{F}({\cal E},U) &=& \int F(|\psi\>,U^\dagger {\cal
					 E}(|\psi\>\<\psi|) U )^2 \,d\psi
\label{eq:gatefavg}
\\
	&=& \int 
		 {\<\psi| U^\dagger {\cal
			E}(|\psi\>\<\psi|) U |\psi\>} \,d\psi
\,,
\eea
where the integral is over the uniform (Haar) measure $d\psi$ on the
Hilbert space of the system.  For a single qubit, this formula can be
reduced to a simple expression~\cite{Nielsen02a,Bowdrey02a},
\be
	\bar{F}({\cal E},U) = 
		{
		\frac{1}{2} + \frac{1}{12} \sum_{k=\{1,2,3\}}
			{\rm Tr}\left( 
			U \sigma_k U^\dagger {\cal E}(\sigma_k)\right)
		}
\,,
\ee
where $\sigma_k$ are the three Pauli matrices.  Similar simple
formulas can be obtained for higher dimensional
systems~\cite{Nielsen02a}.  Note that by convention~\cite{Nielsen00b},
the average gate fidelity is defined such that it goes as the {\em
square} of the usual state fidelity; thus, it can be interpreted as a
probability.

A more difficult quantity to calculate is the {\em minimum} gate
fidelity,
\be
	F({\cal E},U) = {\rm min}_{|\psi\>} F\left(U|\psi\>,
				\cal{E}(|\psi\>\<\psi|)\right)
\,;
\ee
the square of this quantity gives the worst-case gate failure
probability that is relevant for fault-tolerance threshold theorems.

%%%%%%%%%%%%%%%%%%%%%%%%%%%%%%%%%%%%%%%%%%%%%%%%%%%%%%%%%%%%%%%%%%%%%%%%%%%%%%
\subsection{Evaluating Scalability}
\label{sec:threshold}

This article has been concerned with the control of complex systems
composed of multiple distinct physical pieces.  Given some degree of 
control over a few such pieces, how controllable is a {\em very} large 
quantum system composed from many pieces? 
Normally, one would expect that a system composed of
unreliable pieces would itself be unreliable, and that the overall
probability of failure increases rapidly with the number of pieces.
Unexpectedly, however,
arbitrarily reliable quantum systems can be built from unreliable
parts as long as certain criteria are met.

The main criterion for being able to construct a reliable system is
that the probability of error per operation $p$ be below the
``accuracy
threshold,''~\cite{Aharonov97a,Kitaev97b,Knill98c,Gottesman97a,Preskill98b}
$p_{th}$.  When $p<p_{th}$ is satisfied, a quantum error correction
circuit (for instance as demonstrated by NMR
experiments~\cite{Cory98a,Leung99a,Knill01b}) can be constructed using
the unreliable components; this circuit performs computations on
encoded qubits, such that a net decrease in error is achieved even
when error correction itself is done with the faulty gates.  

The probability of failure per operation $p$, must of course be
defined, and this is done in terms of the fidelity metrics discussed
in the previous section, which incorporate decoherence (e.g. $T_1$,
$T_2$, gate times, etc.) and control imperfections.  Thus, for
example, $p$ is bounded from above by the gate fidelity $p \leq 1-
F({\cal E},U)^2$. 

Remarkably, no reliable resources need be utilized for the fault
tolerant construction.  Through $k$ levels of recursive application of
error correction, the device error $p$ can be reduced to $p^{2^k}$,
using physical resources (space, time, and energy) which scale as
$d^k$ for some constant $d$.  Thus, a small increase in resources
exponentially reduces the overall error.  Many assumptions are made in
obtaining $p_{th}$, such as the availability of local, fast, parallel
classical control resources, but the generally accepted theoretical
optimal value of $p_{th}$ is about $10^{-4}$~\cite{Knill98a,Knill98c},
with optimistic estimates ranging as high as $10^{-3}$ with additional
restrictions~\cite{Steane02a}.  As we have seen at the end of 
section~\ref{sec:state_fidelity}, this implies that for
instance rotation angles must be precise to order 
$\sim \sqrt{10^{-4}}=10^{-2}$. In principle, $p_{th}$ can be
experimentally measured, for example, by implementing a recursive
error correction circuit and testing its probability of failure, but
this has not yet been accomplished.

The fault tolerance threshold $p_{th}$, and its relative value
compared with state and gate fidelities give a crisp criterion for
system scalability for specific implementations.  Modern classical
systems are robust mainly because component failures can be
controlled; similarly, the future of control over quantum systems
hinges on our ability to evaluate $p_{th}$ and to build components
which fail with probability $p<p_{th}$.

\section{Discussion and conclusions}

In this review, we have presented a diverse set of tools intended to
compensate for undesired or uncontrolled terms in the Hamiltonian of 
coupled qubits, as well as for instrumental limitations. 
These tools are most powerful and easiest to design when all the 
terms in the system Hamiltonian commute with each other, and the
control terms can be much stronger than the system Hamiltonian.
The common theme of the control techniques 
is careful tailoring of the amplitude, phase and frequency of 
the time-dependent terms in the Hamiltonian, whether in the form of 
shaped pulses, composite pulses or multiple-pulse sequences. 
We now discuss the effectiveness and applicability of these 
advanced control techniques, which points at where they could be 
used in other quantum systems.

%%%%%%%%%%%%%%%%%%%%%%%%%%%%%%%%%%%%%%%%

{\bf Pulse shaping} is particularly attractive because of the modular 
and scalable design approach. Amplitude profiles are selected 
from a library of standard or specially designed shapes in order to 
minimize cross-talk (frequency-selective pulses) and coupling 
effects (self-refocusing pulses). Robustness to experimental 
imperfections can also be considered in the choice of pulse shape.
Once suitable amplitude profiles have been chosen, the pulse lengths 
are set as short as possible while maintaining qubit-selectivity. The 
same amplitude profiles and pulse lengths are then used throughout 
the pulse sequence.

Remaining cross-talk effects can be further reduced at a small cost
(quadratic in the number of qubits). 
From the amplitude profiles and pulse lengths, unintended phase shifts 
produced by single RF pulses as well as off-resonance effects during 
simultaneous pulses can be precomputed, once for every pair of 
qubits. 

The main disadvantage of the standard pulse shaping techniques is 
that often the coupled evolution during the pulses (in particular
$90^\circ$ pulses or simultaneous pulses) cannot be completely frozen.
The remaining coupled evolution can be 
unwound to a large extent during the time-intervals before and after 
the pulse, but such reversal is never perfect because the RF terms in 
the Hamiltonian, $I_x^i$ and $I_y^i$, do not commute with the coupling 
terms, $I_z^i I_z^j$. Furthermore, shaped pulses must often be quite 
long in order to remain spin-selective, which means that decoherence 
has more effect. This problem evidently gets worse as the Larmor
frequencies of the spins approach each other.

Nevertheless, the combination of pulse shaping and phase ramping 
techniques has been very successful in practice. It has enabled the 
implementation of the most complex sequences of operations realized 
to date, acting on up to seven nuclear spins.\\

%%%%%%%%%%%%%%%%%%%%%%%%%%%%%%%%%%%%%%%%

{\bf Composite pulses} have proven to be a versatile tool in NMR
spectroscopy, mostly for compensating systematic errors such as 
RF field strength variations and frequency off-sets. Another 
useful application of composite pulses is the effective creation 
of unitary operators which are otherwise not or not easily 
accessible. A good example is the composite $\hat{z}$ rotation,
created from a sequence of $\hat{x}$ and $\hat{y}$ rotations.

Even so, the use of (hard) composite pulses in NMR quantum 
computing experiments has been limited so far. Their main drawback 
is that in multi-spin homonuclear molecules, single-frequency but 
high-power and rectangular pulses will rotate spins in a large 
frequency window, about an axis and over angles which depend on RF 
field strength and the respective resonance off-sets. This 
severely limits straightforward application of hard composite 
pulses in homonuclear spin systems.

Nevertheless, it is in principle possible to take advantage of the 
differences in resonance off-sets in order to rotate one spin while 
the other spins undergo no net rotation. Such effective frequency 
selectivity despite the use of hard pulses was demonstrated in a 
quantum computation on a homonuclear two-spin system, first using 
single hard pulses~\cite{Jones99b}, and later using composite hard 
pulses~\cite{Cummins00a}.

The same idea underlies the operation of composite pulses tailored
to achieve any rotation of one or more spins about independent axis, 
using detailed knowledge of the system Hamiltonian. Furthermore, 
short, high-power pulses can be used, so the effect of decoherence 
is reduced compared to the case of the long, low-power shaped pulses.  
Even more attractive here is the fact that {\em all} the coupling 
terms can be effectively frozen, and that other types of cross-talk, 
such as Bloch-Siegert effects, are automatically taken care of, 
sunlike the case of shaped pulses.

The main disadvantage of such strongly modulated composite pulses is
that the time needed to find near-optimal pulse parameters increases
exponentially with the number of qubits $n$, as it involves computing 
unitary matrices of size $2^n$ by $2^n$. Nevertheless, for small numbers 
of qubits, this technique can be very useful.\\

%%%%%%%%%%%%%%%%%%%%%%%%%%%%%%%%%%%%%%%%

{\bf Average-Hamiltonian} techniques underlie the operation of 
widely used multiple-pulse refocusing sequences. In the context
of liquid NMR quantum computing, couplings are of the form 
$I_z^i I_z^j$ and refocusing sequences consist simply of a 
train of $180^\circ$ pulses. Such refocusing sequences are an
essential ingredient of all NMR quantum computing experiments 
involving more than two spins.

More complex decoupling sequences exist to remove the effect of
coupling Hamiltonians of a different form, as is the case of
solid-state NMR and many other qubit implementations. Even 
errors arising from interactions with the environment, i.e.
decoherence, can be removed using multiple-pulse sequences.

In all cases, the refocusing operations (e.g. the $180^\circ$ 
pulses) must be fast compared to the fluctuations they are 
intended to cancel, and they must also be repeated at a rate 
faster than the fluctuations.\\

%%%%%%%%%%%%%%%%%%%%%%

{\bf Perspective ---} 
In NMR quantum computing experiments on heteronuclear spin systems, 
where short, high-power RF pulses were used, errors in the time 
evolution were usually dominated by various instrumental limitations. 
Most experiments on homonuclear systems, in contrast, made use of long, 
low-power pulses, and were limited by cross-talk and coupling effects. 
As the pulse techniques for coping with limitations of the Hamiltonian 
and instrumentation became more advanced, the field reached the point 
where errors due to imperfect quantum control were smaller than errors 
caused by decoherence.

Reaching this point in many-qubit systems must be a prime objective for 
any implementation of quantum computers, along with reduction of 
decoherence itself. Quantum information and computation theory offers
a common language which can facilitate transfer and translation of the 
techniques for coherent control of coupled nuclear spins to other 
fields of physics. Such cross-fertilization has already started, in 
systems as diverse as trapped ions~\cite{Gulde03a}, excitons in quantum 
dots~\cite{Chen01a} and Cooper pair boxes~\cite{Collin04a},
and is likely to accelerate the 
progress towards the elusive goal of complete control over quantum systems.

%%%%%%%%%%%%%%%%%%%%%%%%%%%%%%%%%%%%%%%%%%%%%%%%%%%%%%%%%%%%%%%%%%%%%%%%%%%%%

\bibliographystyle{unsrt}
\bibliography{nmrqcrmp}

\end{document}